\documentclass[aps,prb,twocolumn,notitlepage,amsfonts,citeautoscript,superscriptaddress,showpacs]{revtex4-1}

\usepackage{natbib}
\usepackage{listings}
\usepackage{color}
\definecolor{dkgreen}{rgb}{0,0.6,0}
\definecolor{gray}{rgb}{0.5,0.5,0.5}
\definecolor{mauve}{rgb}{0.58,0,0.82}

\usepackage{amsmath,amssymb,mathrsfs}
\usepackage{latexsym}
\usepackage{graphicx} 
\usepackage{grffile}
\usepackage{epstopdf}
\usepackage{graphicx,epstopdf,color}
\usepackage{amsfonts}
\usepackage{amsmath,amssymb,mathrsfs}
\usepackage{hyperref}
\usepackage{color}
\usepackage{url}
\usepackage{soul}
\usepackage{bm}

\newcommand{\textalert}[1]{}

\newcommand{\eps}{\varepsilon}

\newcommand{\Real}[1]{\text{Re}\left(#1\right)}

\newcommand{\ed}{\eps_\text{d}} 
\newcommand{\om}{\omega} 
\newcommand{\wdr}{\om_\text{d}} 
\newcommand{\SFN}{S} 
\newcommand{\sfn}{s} 
\newcommand{\PFN}{P} 

\newcommand{\ssq}{a} 
\newcommand{\ssc}{c} 
\newcommand{\ssqc}{\textit{ac}}

\newcommand{\bwq}{\bar{\om}_a}
\newcommand{\bwc}{\bar{\om}_c}
\newcommand{\bXq}{\hat{\bar{X}}_\ssq}
\newcommand{\bXc}{\hat{\bar{X}}_\ssc}
\newcommand{\bYq}{\hat{\bar{Y}}_\ssq}
\newcommand{\bYc}{\hat{\bar{Y}}_\ssc}

\newcommand{\bq}{\hat{a}}
\newcommand{\nq}{\hat{n}_\ssq}
\newcommand{\wq}{\om_\ssq}
\newcommand{\kq}{\kappa_\ssq}
\newcommand{\bc}{\hat{c}}
\newcommand{\nc}{\hat{n}_\ssc}
\newcommand{\wc}{\om_\ssc}
\newcommand{\kc}{\kappa_\ssc}
\newcommand{\Bop}{\hat{B}_k}

\newcommand{\Xq}{\hat{X}_\ssq}
\newcommand{\Xc}{\hat{X}_\ssc}
\newcommand{\Yq}{\hat{Y}_\ssq}
\newcommand{\Yc}{\hat{Y}_\ssc}

\newcommand{\uqq}{u_\textit{aa}}
\newcommand{\uqc}{u_\textit{ac}}
\newcommand{\ucq}{u_\textit{ca}}
\newcommand{\ucc}{u_\textit{cc}}
\newcommand{\vqq}{v_\textit{aa}}
\newcommand{\vqc}{v_\textit{ac}}
\newcommand{\vcq}{v_\textit{ca}}
\newcommand{\vcc}{v_\textit{cc}}

\newcommand{\half}{\frac{1}{2}}

\usepackage{xr}

\graphicspath{{fig/}{fig/fig_rrc/}}

\usepackage[caption=false]{subfig}
\usepackage{epstopdf}
\newcommand{\HO}{\hat{\mathcal{H}}}

\begin{document}
\title{Lifetime renormalization of driven \\weakly anharmonic superconducting qubits: II. The readout problem}
\author{Alexandru Petrescu}
\affiliation{Institut Quantique and D\'epartement de Physique, Universit\'e de Sherbrooke, Sherbrooke, Qu\'ebec, J1K 2R1, Canada}
\affiliation{Department of Electrical Engineering, Princeton University, Princeton, New Jersey, 08544}
\author{Moein Malekakhlagh}
\affiliation{Department of Electrical Engineering, Princeton University, Princeton, New Jersey, 08544}
\author{Hakan E. T\"ureci}
\affiliation{Department of Electrical Engineering, Princeton University, Princeton, New Jersey, 08544}
\date{\today}
\begin{abstract}
Recent experiments in superconducting qubit systems have shown an unexpectedly strong dependence of the qubit relaxation rate on the readout drive power. This phenomenon limits the maximum measurement strength and thus the achievable readout speed and fidelity. We address this problem here and provide a plausible mechanism for drive-power dependence of relaxation rates. To this end we introduce a two-parameter perturbative expansion in qubit anharmonicity and the drive amplitude through a unitary transformation technique introduced in Part I. This approach naturally reveals number non-conserving terms in the Josephson potential as a fundamental mechanism through which applied microwave drives can activate additional relaxation mechanisms. We present our results in terms of an effective master equation with renormalized state- and drive-dependent transition frequency and relaxation rates. Comparison of numerical results from this effective master equation to those obtained from a Lindblad master equation which only includes number-conserving terms (\textit{i.e.} Kerr interactions) shows that number non-conserving terms can lead to significant drive-power dependence of the relaxation rates. The systematic expansion technique introduced here is of general applicability to obtaining effective master equations for driven-dissipative quantum systems that contain weakly non-linear degrees of freedom.
\end{abstract}
\maketitle

\section{Introduction}
\label{Sec:Intro}
The dispersive interaction between a qubit and a cavity forms the basis for qubit state measurement widely employed in superconducting quantum circuits. As predicted by the Jaynes-Cummings model of this interaction  \cite{blais_et_al_2004}, each qubit state induces a different shift on the effective resonance frequency of the readout cavity \cite{wallraff_et_al_2004}. By monitoring this shift with a microwave probe-pulse, the qubit state can be accurately measured. The rapid and high-fidelity application of qubit state readout is widely recognized to be a critical component in the implementation of current quantum computing algorithms. The fidelity of this protocol is predicated on the dominance of certain number-conserving terms in the effective qubit evolution under the action of the probe-pulse that is quasi-resonant with the readout cavity. This dynamical regime, sometimes referred to as the linear dispersive regime, is generally expected to prevail for cavity photon occupations well below the ``critical photon number" $n_\text{crit} = \Delta^2/4g^2$, where $\Delta = \wc - \wq$ is the detuning between the  cavity ($\wc$) and qubit ($\wq$) resonance frequencies and $g$ is the vacuum Rabi frequency characterizing the coupling strength \cite{blais_et_al_2004, boissonneault_et_al_2009}. For present systems based on transmon qubits \cite{koch_et_al_2007}, this number is typically $n_\text{crit} \ge 25$.

\begin{figure}[t!]
  \includegraphics[width=0.65\linewidth]{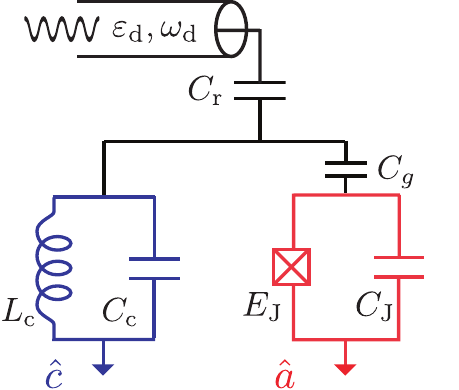}
  \caption{Schematics of the readout setup. A Josephson junction qubit (mode $\bq$) is capacitively coupled to a waveguide through a readout cavity (mode $\bc$). A microwave drive (amplitude $\ed$, frequency $\wdr$) is applied through the waveguide.  \label{Fig:Circuits} }
\end{figure}

Recent experimental analysis \cite{mundhada_et_al_2016, minev_et_al_2019} indicates that $T_1$ relaxation time may decrease by as much as a factor of two for relatively small cavity photon occupations $\bar{n}_\ssc \sim 5$. Understanding the plausible fundamental mechanism behind this observation is one of the goals of this paper. 

It should come as no surprise that in a coherently driven nonlinear system the validity of perturbation theory in Hamiltonian parameters (such as $g/\Delta$) requires some care regarding the nature of the qubit nonlinearity. Early work carefully analyzed the so-called ``nonlinear dispersive regime" of operation and the systematic corrections to the frequencies and dissipation rates \cite{boissonneault_et_al_2009} within the Jaynes-Cummings framework, suitable for qubits with a strong anharmonicity, such as the Cooper pair box or quantronium qubit \cite{makhlin_et_al_2001,bouchiat_et_al_1998,vion_et_al_2002,lehnert_et_al_2003}. This approach predicts \cite{boissonneault_et_al_2009, sete_et_al_2014} that in the absence of any dephasing noise, the relaxation rate ($1/T_1$) of the qubit {\it decreases} with drive strength. The presence of a dephasing noise, on the other hand, is found to lead to an {\it increase} of the relaxation rate with the drive strength. This ``dressed dephasing hypothesis'' does seem to agree with some experimental data that found an increase in the relaxation rate with the drive-strength \cite{slichter_et_al_2012}, but does not seem to correctly capture the effective temperature of the qubit in the steady-state in experiments conducted on 3D transmon qubits \cite{mundhada_et_al_2016}. The question therefore arises whether accurate modeling of the Josephson nonlinearity of the qubit changes any of these predictions in a qualitative way. We address this question here building on the technique of unitary transformations established in Ref.~[\onlinecite{malekakhlagh_et_al_2018}], hereafter called Part I.

Here we derive an effective master equation (EME) for a weakly anharmonic qubit driven by a coherent microwave tone. We consider the situation typical of dispersive readout, where the weakly anharmonic qubit is coupled to a single-mode resonator, which in turn is connected capacitively to a semi-infinite transmission line [see Fig.~\ref{Fig:Circuits}]. Extending the formalism developed in Part I to the coherently driven case, we provide analytical expressions for effective system frequencies, as well as relaxation and excitation rates that depend on drive parameters. Through a two-parameter expansion in the weak Josephson anharmonicity and the drive strength, we show that at lowest order the system unitary dynamics is governed by a multi-mode Kerr Hamiltonian \cite{Nigg_BlackBox_2012}, as found in Part I, but with drive-adapted parameters. The renormalization of relaxation rates can only be captured by retaining the number non-conserving terms in the Josephson potential. One important finding is that drive-activated correlated qubit-cavity relaxation processes are dominantly responsible for  large renormalizations of the qubit relaxation rates. The formalism presented here is the time-dependent generalization of that in Part I, and the results reduce to those obtained in Part I in the limit of zero drive strength.

There are several conclusions that can be drawn from our results regarding driven Josephson junctions. Here we consider solely the electromagnetic fluctuations of the infinite transmission line at zero temperature as a source of relaxation (and excitation, when mixed with the coherent drive tone, as we show). We find that the lowest order impact of the drive is to {\it increase} the relaxation rate of a dispersively coupled qubit. This is in contrast to earlier findings \cite{boissonneault_et_al_2009, sete_et_al_2014} that the relaxation rate decreases with the drive strength in the absence of dephasing sources. The reason can be traced back to the two-level approximation to the Josephson nonlinearity that underlies the Jaynes-Cummings (and the Rabi) model. From the point of view of the anharmonicity, the Josephson nonlinearity is a softening potential, while the two-level truncation is the extreme case of a hardening potential. In terms of the parameter $\epsilon = \sqrt{2 E_\text{C} / E_\text{J}}$ of a Josephson potential, the two-level truncation corresponds to $\epsilon<0$, a principally unphysical limit. This has the additional consequence that the impact of the drive is effective already at lower excitation powers than previously foreseen, with important implications for optimization of readout protocols. Finally, the impact of a radiative bath through which the drive is incident is found to also lead to excitation of the qubit in proportion to the drive strength, even at zero temperature. 
   
Any initialization, computation and readout operation on superconducting circuits involves microwave drives. Our results indicate that the accurate modeling of the Josephson potential of qubits in such circuits is critical as the demand for high-fidelity operations is pushed to its limits. Methods to deal with this challenge may be based on purely numerical schemes. Indeed in recent years, it has become necessary to better model strongly driven Josephson circuits, in a variety of applications: parametric schemes for engineering effective nonlinearities \cite{zhang_et_al_2019,frattini_et_al_2018,sivak_et_al_2019}, high-power readout schemes \cite{ginossar_et_al_2010,reed_et_al_2010}, as well as the driven-dissipative stabilization of states confined to a given quantum manifold, such as cat states \cite{leghtas_et_al_2015,vlastakis_et_al_2013,puri_et_al_2017}, as well as implementations of parametric two-qubit gates \cite{rigetti_devoret_2010,chow_et_al_2013,mckay_et_al_2016,magesan_et_al_2018,sheldon_et_al_2016,caldwell_et_al_2018,didier_et_al_2018,reagor_et_al_2018}. The initial evaluation of the effectiveness of the two-level system approximation for modeling high-power dynamics \cite{bishop_et_al_2010} has been addressed in Ref.~\onlinecite{boissonneault_et_al_2010}. More recently the Floquet master equation \cite{grifoni_haenggi_1998} has been successful in describing the escape of certain strongly driven Josephson circuits into states unconfined by the cosine potential \cite{lescanne_et_al_2019,verney_et_al_2019}. Earlier theoretical and experimental work also points to the role of counter-rotating terms in explaining the unexpectedly high susceptibility of certain Josephson circuits to excitation in certain power bands \cite{sank_et_al_2016}. 
 
The pursuit of deriving effective generators for the evolution of open systems has a long history which can be traced back to the projection-operator formalism of Feshbach \cite{feshbach_1958}. Most of these schemes rely on numerical methods to extract the low-frequency dynamics generated by linear operators of the Lindblad class \cite{kessler_2012,reiter_sorensen_2012,mirrahimi_rouchon_2009}. A similar method has been applied to obtain effective dynamics on reduced manifolds using quantum stochastic differential equations \cite{bouten_et_al_2008,bouten_silberfarb_2008,tezak_et_al_2017}. An important aspect of the approach presented here is that one obtains explicit drive-dependent renormalizations of both frequencies and relaxation rates because of the inclusion of number-nonconserving terms. Underlying our method is a series of unitary Schrieffer-Wolff transformations \cite{schrieffer_wolff_1966} that remove number-nonconserving terms order-by-order from the system Hamiltonian, but dress the interactions of the system with its environment.

\begin{figure*}[t!]
  \includegraphics[width=\textwidth]{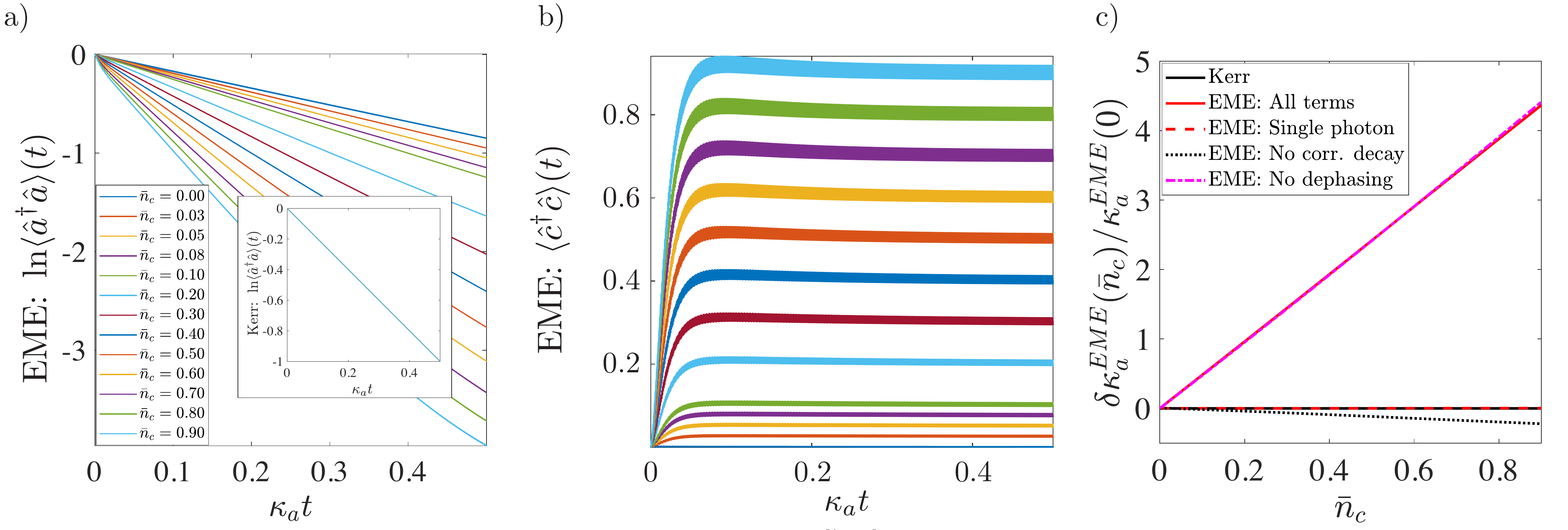}
  \caption{\label{Fig:EMEQ1M1} Master equation simulations for our model of dispersive readout [see Fig.~\ref{Fig:Circuits}]. a) EME solution: The natural logarithm of the qubit occupation number $\langle \bq^\dagger \bq \rangle$ as a function of time, for different values of the drive power. As the drive strength is increased, the relaxation rate of the qubit increases linearly as a function of the cavity steady state population. Inset: The Kerr-only master equation predicts no drive- or nonlinearity-induced renormalization of the qubit relaxation rate.  b) The drive strength is adjusted such that the cavity has a  mean steady state population $\bar{n}_\ssc$   c) Qubit relaxation rate [Eq.~(\ref{Eq:Fitted})] extracted from a number of numerical simulations: Kerr theory, EME with all terms included, as well as EME in which a subset of the terms are included only (see text for complete discussion).} 
\end{figure*}

The remainder of the paper is organized as follows. Section~\ref{Sec:Model} introduces the model for the quantum circuit consisting of a qubit coupled to a cavity, which is the standard setup for the dispersive readout scheme, and outlines the main steps of the perturbation theory in weak anharmonicity and weak drive used to obtain corrections to frequency and decay rate. We apply the EME to understand dispersive readout in Sec.~\ref{Sec:PT}. The EME is analytically derived and then numerically simulated. The main outcome of this section is our prediction for the renormalizations of qubit transition frequency and zero-temperature relaxation rate in the presence of a driven cavity at a steady-state population $\bar{n}_\ssc$. Finally, we summarize our results in Sec.~\ref{Sec:Summary}. We have opted to relegate many details to appendices in an effort to improve clarity. Each of the appendices will be pointed to in the main sections of the paper when necessary.

\section{Model and main result}
\label{Sec:Model}

In this section, we present the model and the steps towards obtaining the EME. The system under consideration is a superconducting transmon qubit \cite{koch_et_al_2007} capacitively coupled to a cavity, which is an idealization of the circuitry typically used for the dispersive readout \cite{blais_et_al_2004,Wallraff_Strong_2004}. The dynamics of the system (subscript ``s'') coupled to the waveguide (``bath'', subscript ``b'') follows from the full Hamiltonian:
\begin{equation}
  \HO = \HO_{\text{s}} + \HO_\text{d}(t) + \HO_{\text{sb}} + \HO_{\text{b}}.
\end{equation}
We have approximated the circuit as an oscillator characterized by inductance $L_\ssc$ and capacitance $C_\ssc$ [see Fig.~\ref{Fig:Circuits}], resulting in the oscillator frequency $\bwc = \sqrt{L_\ssc C_\ssc}$. The simplification of the superconducting cavity to a single mode is done for transparency of results. The techniques presented in this work can be easily generalized to a multi-mode setup, starting from the exact electromagnetic modeling of the system \cite{Malekakhlagh_NonMarkovian_2016}. The coupling capacitance between the qubit and the cavity is denoted by $C_g$. The transmon qubit is defined by the Josephson and Coulomb charging energies, respectively, denoted by $E_\text{J}$ and $E_\text{C} = e^2/(2C_\text{J})$, where $C_\text{J}$ is the capacitance across the Josephson junction. This leads to the qubit transition frequency\cite{koch_et_al_2007} $\bwq \approx \sqrt{8 E_\text{J} E_\text{C}}$ in the limit of high anharmonicity $E_\text{J}/E_\text{C} \gg 1$. Upon quantizing\cite{Devoret_Quantum_1995} the circuit of Fig.~\ref{Fig:Circuits}, we arrive at the following system Hamiltonian, which was the starting point of Part I: 
\begin{equation}
\label{Eq:Hs}
\hat{\mathcal{H}}_\text{s} = \frac{\bwq}{4}\left[\bYq^2-\frac{2}{\epsilon}\cos\left(\sqrt{\epsilon}\bXq\right)\right]+\frac{\bwc}{4}\left(\bXc^2+\bYc^2\right)+g\bYq\bYc,
\end{equation}
consisting of a part describing the transmon qubit, one describing the linear superconducting cavity, and finally a coupling term between the two. With our conventions, the commutator of the phase and charge quadratures contains an additional factor of two: $[\hat{\bar{X}}_{\ssq,\ssc}, \hat{\bar{Y}}_{\ssq,\ssc}] = 2i$ [see App.~\ref{App:Conventions} for an explanation of our conventions].

The energy scale $g$ denotes the capacitive qubit-cavity coupling strength, and it can be related to the coupling capacitance $C_g$. Finally, an important dimensionless quantity appearing in the qubit Hamiltonian $\hat{\mathcal{H}}_\ssq$ is the anharmonicity parameter 
\begin{equation}
  \epsilon = \sqrt{2 E_\text{C}/E_\text{J}},
\end{equation} 
which will form the basis of the perturbative expansion for weak anharmonicity $\epsilon \ll 1$, which coincides with the regime of operation for transmon qubits \cite{koch_et_al_2007}. Note that the quadratures in Eq.~(\ref{Eq:Hs}) are standard phase and number operators scaled by factors of $\sqrt{\epsilon}$ (see App.~\ref{App:Conventions}). Switching to the dimensionless quadratures $\hat{\bar{X}}_{\ssq,\ssc}$ etc. in the expression of the system Hamiltonian~(\ref{Eq:Hs}) will permit a perturbative expansion in powers of $\epsilon$.

We assume that the drive term $\HO_\text{d}(t)$ acts on the bare charge quadrature, as depicted in Fig.~\ref{Fig:Circuits}. The driven Hamiltonian, time-periodic with a period $2\pi / \wdr$, introduces additional complexity in the derivation of EMEs as compared to the undriven case treated in Part I. In the EMEs derived below, the strength of the drive, $\ed$, and the anharmonicity parameter, $\epsilon$, will be treated on equal footing as small parameters in a perturbative expansion that leads to effective driven-dissipative dynamics.  To organize this double perturbative expansion, we first need to switch from the bare mode basis to the normal mode basis, whose advantage is that the linear theory becomes diagonal in Fock space. This involves re-expressing the bare quadratures in Eq.~(\ref{Eq:Hs}) in terms of normal mode quadratures 
\begin{equation}
\{\bXq,\bYq,\bXc,\bYc\} \to \{\Xq,\Yq,\Xc,\Yc\},
\end{equation}
at the expense of introducing hybridization coefficients [see App.~\ref{App:Conventions}]. 

We now turn to the description of the drive and relaxation mechanisms in this normal mode basis. In this work, we think of both drive and relaxation as being facilitated by capacitively coupling the bare cavity mode to the waveguide, see Fig.~\ref{Fig:Circuits}. Thus the drive is distributed to the normal modes as follows
\begin{equation}
\hat{\mathcal{H}}_\text{d}=\ed (\vcq \Yq + \vcc \Yc) \sin(\wdr t).
\label{eqn:Model-Hd in normpic}
\end{equation}
We assume that there is no intrinsic decay rate for the bare qubit oscillator, \textit{i.e.} that relaxation is only induced on the qubit by coupling it to the open cavity. This is the situation of pure radiative decay known as the Purcell effect. The effect of higher harmonics of the cavity can be addressed using the theoretical framework introduced in Ref.~[\onlinecite{Malekakhlagh_Cutoff_2017}].

In the normal mode basis, the system-bath coupling arises from the capacitive coupling of the system to the waveguide:
\begin{eqnarray}
  \hat{\mathcal{H}}_{\text{sb}} &=& \left(\vcq \Yq + \vcc \Yc \right)  \hat{Y}_\text{b}, \label{Eq:HSB}
\end{eqnarray}
where 
\begin{equation}
  \hat{Y}_\text{b}=\sum_k g_{k}  \left( -i \hat{B}_{k} + i\hat{B}_{k}^\dag \right)
\end{equation} 
is the noise operator to which the bare cavity quadrature couples, and the continuum of bath modes is described by bosonic creation and annihilation operators obeying the commutation relation $[ \Bop, \Bop^\dag ]=1$ for each index $k$, governed by the linear Hamiltonian $\hat{\mathcal{H}}_\text{b} = \sum_k \om_{k} \hat{B}_{k}^\dag \hat{B}_{k}$.

In order to prepare a perturbative expansion in the two parameters $\ed$ and $\epsilon$, we may now bring $\hat{\mathcal{H}}_\text{s} + \hat{\mathcal{H}}_\text{d}(t)$ to a new form in which the anharmonicity and the drive appear on equal footing. This is achieved by a displacement transformation that removes the terms which are linear in the quadratures (App.~\ref{Ap:Shift}). Upon performing this transformation, we denote the resulting Hamiltonian $\hat{\mathcal{H}}_\text{s} + \hat{\mathcal{H}}_\text{d}(t) \to \hat{\mathcal{H}}_{\text{s}}(t)$, in which the drive terms appear as follows: 
\begin{eqnarray}
&&\hat{\mathcal{H}}_{\text{s}}(t) =\wq\left(\bq^{\dag}\bq+\half\right)+\wc\left(\bc^{\dag}\bc+\half\right)\;\;\;\;\;\; \label{Eq:DispH}
 \\
&&  +\frac{\bwq}{2} \sum_{n=2}^{\infty} \frac{(-\epsilon)^{n-1}}{(2n)!} \left[\uqq \bq + \uqc \bc + \eta_x e^{-i\wdr t} + \text{H.c.} \right]^{2n}. \nonumber 
\end{eqnarray}
The Hamiltonian of Eq.~(\ref{Eq:DispH}) is the starting point for the analysis of an arbitrary linearly-driven weakly anharmonic two-mode circuit. The form above is general: the displacement parameter $\eta_x$ and the hybridization coefficients $u_{ij}$ would take different forms for different types of linear coupling between the drives, the qubit, and the cavity.

We now proceed to illustrate the distinct role of number non-conserving terms in the renormalization of relaxation rates. We show that we can perform a unitary transformation on the system Hamiltonian that removes number non-conserving terms up to a desired order $\epsilon^n$ in the Hamiltonian. Because $\hat{\mathcal{H}}_{\text{s}}(t)$ is time dependent, the condition that a unitary transformation preserve the dynamics of the Schr\"odinger equation needs to be formulated in terms of the Floquet Hamiltonian, which differs from the Hamiltonian through the addition of the energy operator $-i\partial_t$ \cite{sambe_1973}:
\begin{eqnarray}
\hat{\mathcal{H}}_{\text{s},\text{eff}}(t) - i\partial_t =  e^{-\hat{G}(t)} \left[ \hat{\mathcal{H}}_\text{s}(t) - i\partial_t  \right] e^{\hat{G}(t)}. \label{Eq:FloquetUGHUG}
\end{eqnarray}
The antihermitian generator $\hat{G}(t)$ is time-dependent and it is defined by the condition that the \textit{effective} Hamiltonian, $\hat{\mathcal{H}}_{\text{s},\text{eff}}(t)$, contains no number-nonconserving terms up to some order in $\epsilon$. The generator can be found order by order upon an expansion in powers of the anharmonicity, $\hat{G}(t) = \epsilon \hat{G}_4(t) + \epsilon^2 \hat{G}_6(t) + \ldots$, through a hierarchical set of operator-valued ordinary differential equations, which are derived in App.~\ref{Sec:Hierarchy}. 

In this article, we present the solution for the generator $\hat{G}_4(t)$ that cancels the number-nonconserving terms of the Josephson nonlinearity up to linear order $\epsilon$. To this end we expand the system Hamiltonian in powers of the anharmonicity, to wit
\begin{equation}
  \hat{\mathcal{H}}_{\text{s}}(t) = \hat{\mathcal{H}}_{\text{2}} - \epsilon \hat{\mathcal{H}}_4(t)  + \epsilon^2 \hat{\mathcal{H}}_6(t) + \ldots, \label{Eq:HstExpansionMain}
\end{equation}
and decompose each operator $\HO_{2n}(t) = \hat{\mathcal{S}}_{2n}(t) + \hat{\mathcal{N}}_{2n}(t)$ into a sum of two normal-ordered operators. These are the number-conserving and number-nonconserving terms, respectively.  The condition for the generator can be written to lowest order in the anharmonicity $\epsilon$ in the compact form of a differential equation [see App.~\ref{Sec:Hierarchy}]:
\begin{eqnarray}
  - i \dot{\hat{G}}_4(t) + \left[ \hat{\mathcal{H}}_2, \hat{G}_4(t) \right] = \hat{\mathcal{N}}_4(t),\label{Eq:G4tODEMain}
\end{eqnarray}
with initial condition $\left[ \hat{\mathcal{H}}_2, \hat{G}_4(0) \right] = \hat{\mathcal{N}}_4(0)$, where $\hat{\mathcal{N}}_4(t)$ contains the number-nonconserving terms arising from the normal-ordered expression of the fourth power in the expansion of the Josephson nonlinearity in Eq.~(\ref{Eq:DispH}). The key point here is that there is a major simplification of the operator-valued ordinary differential Eq.~(\ref{Eq:G4tODEMain}) if one expands $\hat{G}_4(t)$ as the sum of all possible normal-ordered ``monomials'' $\bq^{\dagger m} \bq^n \bc^{\dagger p} \bc^q$, which are merely many-body operators consisting of powers of creation and annihilation operators of the two normal modes.
By virtue of the following property of the bosonic algebra, 
\begin{equation}
  [\bq^\dagger \bq, \bq^{\dagger m} \bq^n] = (m-n) \bq^{\dagger m} \bq^n,
\end{equation}
with an analogous form for $\bc$, one can turn Eq.~(\ref{Eq:G4tODEMain}) into collection of \textit{uncoupled} ordinary differential equations for the complex-valued coefficients of these monomials in the expansion of the generator. Therefore, the generator $\hat{G}_4(t)$ is analytically tractable and closed-form expressions can be written down for the simplest examples (see App.~\ref{Ap:SWCorr-DrQu} for a one-mode theory), while computer algebra \cite{zitko_2011} can be used for the general situation encountered in the problem of dispersive readout. 

Once the generator is determined, the first effect of this transformation is that number-nonconserving terms have been removed to order $\epsilon$ from the effective Hamiltonian. The latter takes a Kerr form, containing interactions up to quadratic order in the number operators counting photons in the two normal modes corresponding to qubit and cavity, and terms at most linear in the anharmonicity $\epsilon$: $\HO_{\text{s},\text{eff}} = \HO_2 - \epsilon \hat{\mathcal{S}}_4(t)$. Secondly, the action of the generator $\hat{G}(t)$ on the system-bath Hamiltonian yields corrected system operators coupling to the bath noise operator $\hat{Y}_\text{b}$. In the Born-Markov approximation \cite{breuer_petruccione_2002}, this leads to the EME in the Lindblad form:
\begin{eqnarray}
  \dot{\hat{\rho}}(t) = -i \left[ \HO_{\text{s},\text{eff}}(t), \hat{\rho}(t) \right] + \sum_{j}2 \kappa(\om_j) \mathcal{D}\left[ \hat{C}_{\text{eff}}(\om_j) \right] \hat{\rho}(t),\label{Eq:EMESumIntro} \nonumber \\ \;
\end{eqnarray}
where $\hat{C}(\om_j)$ are renormalized system collapse operators defined at a set of frequencies $\{\om_j\}$, which are linear combinations involving integer multiples of the normal mode and the drive frequencies, and the dissipator superoperators are defined as usual, $\mathcal{D}[\hat{C}](\bullet)=\hat{C}(\bullet)\hat{C}^{\dag}-1/2\{\hat{C}^{\dag}\hat{C},(\bullet)\}$. 

Note that we have performed the Born-Markov and secular approximations \textit{after} the application of two unitary transformations on the full Hamiltonian describing the system and its environment: the first, a displacement transformation into the frame rotating at the drive frequency, and, the second, a Schrieffer-Wolff transformation that eliminated the number-nonconserving terms. This was the essential step that allowed us to derive drive- and anharmonicity-corrected dissipators. This point underlies the derivation of the EME in App.~\ref{App:EME}.

We are now in a position to summarize our main result. For the readout problem, where the drive is nearly-resonant with the cavity, there are two dominant contributions entering the EME, arising from the following two dissipators:
\begin{eqnarray}
\hat{C}(\wq)\approx&&-i\Bigg[\vcq-\frac{\epsilon}{8}\left(\frac{\bwq}{\wq}\vcq\uqq^2 - 4\frac{\bwq\wq}{\wc^2-\wq^2}\vcc\uqc\uqq\right) \nonumber \\ 
&&\;\;\;\;\;\;\;\times\left(\uqq^2+\uqc^2+\uqq^2 \nq +2\uqc^2 \nc + 2|\eta_x|^2\right)\Bigg]\bq \nonumber \\
&& - \frac{i \epsilon}{2} \frac{\bwq}{\wc} \vcq \uqc \uqq^2 \frac{\wdr}{\wc-\wdr} \bq \left(\eta_x^*\bc - \eta_x  \bc^\dag\right),  \label{Eq:CqEffApprox} 
\end{eqnarray}
where $\eta_x$ is a complex number arising from the displacement to the frame rotating with the drive, $\nq = \bq^\dagger \bq$ and $\nc=\bc^\dagger \bc$, and
\begin{eqnarray}
\hat{C}(\wc)\approx&&-i\Bigg[\vcc-\frac{\epsilon}{8}\left(\frac{\bwq}{\wc}\vcc\uqc^2 - 4\frac{\bwq\wc}{\wq^2-\wc^2}\vcq\uqq\uqc\right) \nonumber \\ 
&&\;\;\;\;\;\;\;\times\left(\uqc^2+\uqq^2+\uqc^2 \nc +2\uqq^2 \nq + 2|\eta_x|^2\right)\Bigg]\bc \nonumber \\ 
&& - i\frac{\epsilon}{8} \frac{\bwq \wdr}{\wc}  \vcc \uqc \uqq^2  \frac{\eta_x}{\wdr-\wc} \nq. \nonumber \\ 
 \label{Eq:CcEffApprox}
\end{eqnarray}
These collapse operators derive from the coupling of the bare cavity to the environment, $\HO_\text{sb}$, dressed (to lowest order in $\epsilon$) by the number-nonconserving terms of the Josephson anharmonicity [see Sec.~\ref{Sec:PT}]. Note that, in addition to scalars rescaling the annihilation operators $\bq$ and $\bc$, there are other contributions which become important in the presence of drive, such as a qubit dephasing term $\bq^\dagger \bq$ appearing in the cavity dissipator, as well as a correlated cavity-qubit relaxation $\bq \bc$ and qubit-cavity conversion $\bq \bc^\dagger$. The correlated decay processes are responsible for stark renormalizations of the qubit relaxation rates, as illustrated in Fig.~\ref{Fig:EMEQ1M1}, which summarizes the numerical results hinging on the EME fully developed in Sec.~\ref{Sec:PT}.

The rates associated with the collapse operators~(\ref{Eq:CqEffApprox}) and~(\ref{Eq:CcEffApprox}) correspond to transitions at or nearly at the qubit and cavity normal mode frequencies, respectively:
\begin{eqnarray}
\kq = \kappa(\wq) =\half \SFN (\wq),\; \kc = \kappa(\wc) = \half \SFN (\wc). \label{Eq:RatesQ1M1}
\end{eqnarray}
In defining rates above, we needed the bilateral power spectral density corresponding to the bosonic bath described by $\HO_\text{b}$, defined as the Fourier transform of the finite-temperature two-point correlation function: 
\begin{equation}
  \SFN (\om) = \lim_{T \to 0}\int_{-\infty}^\infty d\tau \, e^{-i\om \tau}  \text{Tr} \left[ \frac{e^{- \hat{\mathcal{H}}_\text{b}/ k_\text{B} T}}{Z_\text{b}(T)} \hat{Y}_{\text{b}}(\tau) \, \hat{Y}_{\text{b}}(0) \right], \label{Eq:SFN}
\end{equation} 
where the bath partition function is
\begin{equation}
  Z_\text{b}(T) =  \text{Tr} \left[ e^{- \hat{\mathcal{H}}_\text{b}/ k_\text{B} T} \right]
\end{equation}
and the bath modes are assumed to be in thermal equilibrium at zero temperature obeying Bose-Einstein statistics: $\text{Tr}_\text{b} \left\{ \Bop \hat{B}_l^\dag \right\} \equiv \delta_{kl} (1+n_k)$, and $\text{Tr}_\text{b} \left\{ \Bop^\dag \hat{B}_l \right\} \equiv \delta_{kl} n_k$; $n_k =\left[e^{\om_k/(k_B T)} - 1\right]^{-1}$ is the value of the Bose-Einstein distribution at energy $\om_k$ and temperature $T$. 

We conclude our presentation of the model and the main steps towards obtaining the EME by reiterating the main property underlying the derivation of the EME to lowest order in $\epsilon$: Corrections to the eigenfrequencies are captured by the number-conserving terms in $\hat{\mathcal{H}}_{\text{s},\text{eff}}$, whereas the renormalized dissipators in~(\ref{Eq:EMESumIntro}) arise from the number-nonconserving terms of the Josephson nonlinearity. Correlated processes between the qubit and the cavity in the presence of drive can result in a significant drive-dependent renormalization of the qubit relaxation rates.

\section{Effective Master Equation for the readout problem}
\label{Sec:PT}

In this section we carry out the program outlined in Sec.~\ref{Sec:Model} for the EME describing dispersive readout. We develop the pertubation theory for a weakly anharmonic qubit coupled to an open driven resonator, shown schematically in Fig.~\ref{Fig:Circuits}. We are confining ourselves to the analysis of the enhancement of the Purcell effect in the presence of drive and anharmonicity.

For a pedagogical application of the method, we point the reader to App.~\ref{Ap:SWCorr-DrQu} where we consider a one-mode theory of a weakly driven, weakly anharmonic qubit coupled to an infinite waveguide, which yields the effective dressing of the qubit decay rate and frequency. That toy problem contains all the essential ingredients of the methodology to derive the EME and sets up the stage for the readout problem treated in this section.

The remainder of this section is organized as follows. Subsection~\ref{SubSec:DerivEME} contains the derivation of the EME for dispersive readout. Equations~(\ref{Eq:EMESum}) and~(\ref{Eq:CqEff}) contain the main results, with approximate forms applicable to the typical scenario for dispersive readout, when the drive is close-to-resonant with the cavity normal mode frequency, obtained in Eqs.~(\ref{Eq:CqEffApproxSec}) and~(\ref{Eq:CcEffApproxSec}). The reader interested in the numerical results directly could skip to Subsec.~\ref{SubSec:NumerEME}, where the EME numerical simulations are discussed, with numerical results summarized in Fig.~\ref{Fig:EMEQ1M1}.

\subsection{Derivation of EME}
\label{SubSec:DerivEME}
With number-nonconserving terms removed from the driven system Hamiltonian $\hat{\mathcal{H}}_{\text{s}}$, their effect carries over to two different quantities appearing in the dynamical equations. First, applying the unitary transformation derived from the condition above to the system-bath coupling yields a renormalized system quadrature coupling to the bath [cf. Eq.~(\ref{Eq:HSB})]:
\begin{eqnarray}
  \hat{\mathcal{H}}_{\text{sb}} \to e^{-\hat{G}(t)}\hat{\mathcal{H}}_{\text{sb}} e^{\hat{G}(t)} = \hat{\mathcal{H}}_{\text{sb}} + \epsilon \left[ \hat{\mathcal{H}}_{\text{sb}}, \hat{G}_4(t) \right] + O(\epsilon^2). \nonumber\\
\;
\end{eqnarray}
Secondly, the unitary must be applied to the system reduced density matrix, which becomes
\begin{eqnarray}
  \hat{\rho}_{\text{s}}(t) \to e^{-\hat{G}(t)}  \hat{\rho}_{\text{s}}(t) e^{\hat{G}(t)} =  \hat{\rho}_{\text{s}}(t) + \epsilon \left[  \hat{\rho}_{\text{s}}(t), \hat{G}_4(t) \right] + O(\epsilon^2). \nonumber\\ \;
\end{eqnarray}
We show in this section that, among the many terms that correct the quadratures, there will be a simple rescaling of the qubit and cavity collapse operators leading to the enhancement of relaxation rates.  

The Hamiltonian describing the setup of Fig.~\ref{Fig:Circuits}, which is an idealization of the circuit used in dispersive readout schemes, is:
\begin{eqnarray}
  \HO = \HO_\text{s}(t) + \HO_\text{b} + \HO_{\text{sb}},
\end{eqnarray} 
where $\HO_\text{s}(t)$ is the displaced system Hamiltonian introduced in Eq.~(\ref{Eq:DispH}) truncated after the linear order in the anharmonicity $\epsilon$, 
\begin{eqnarray}
\hat{\mathcal{H}}_{\text{s}}(t) &=&\wq\left(\bq^{\dag}\bq+\half\right)+\wc\left(\bc^{\dag}\bc+\half\right) \label{Eq:DispHRep} \;\;\;\;\;\; \\
&&  -\frac{\epsilon\bwq}{48}\left(\uqq \bq +\uqc \bc + \eta_x e^{-i\wdr t} + \text{H.c.} \right)^4. \nonumber 
\end{eqnarray}
The system-bath coupling $\HO_\text{sb}$ was expressed already in Eq.~(\ref{Eq:HSB}) and $\HO_\text{b}$ is the Hamiltonian describing the bath modes. Note that although only the bare cavity was driven, now both the qubit-like and the cavity-like normal modes are subjected to the drive due to hybridization:
\begin{eqnarray}
  \eta_x = \uqq \eta_{\ssq,x} + \uqc \eta_{\ssc,x}.  
\end{eqnarray}
The coherent parts corresponding to each normal mode are given by 
\begin{eqnarray} 
  \eta_{\ssq,x} &=& \frac{\vcq \ed (\wdr+i\kq)}{\wq^2-(\wdr+i\kq)^2}, \nonumber  \\ \eta_{\ssc,x}&=& \frac{\vcc\ed (\wdr + i\kc)}{\wc^2 - (\wdr + i\kc)^2}.
\end{eqnarray}
These are the amplitudes of the displacement of the phase quadrature for the two normal modes $\bq,\bc$. Note that these expressions depend explicitly on the relaxation rates and they are obtained from the linear theory [for a derivation, see App.~\ref{Ap:ShiftME}]. That is, if the anharmonicity were turned off, $\epsilon=0$, then the steady state population of the cavity would be 
\begin{equation}
  \bar{n}_\ssc = |(\eta_{\ssc,x} + i \eta_{\ssc,y})/2|^2, \label{Eq:nbarc}
\end{equation} 
where $\eta_{\ssc,y} = -i \wc/(\wdr + i\kc) \eta_{\ssc,x}$ is the corresponding amplitude of the displacement of the charge quadrature. Note that since the hybridization between the cavity and the qubit is typically taken to be weak, the dressed cavity is only weakly nonlinear, and therefore we can use Eq.~(\ref{Eq:nbarc}) as a very good estimate of the actual numerical steady state population.

We now follow the same program as in the previous section to find the generator $\hat{G}(t)$ to lowest order in $\epsilon$ that removes the number-nonconserving terms of the nonlinear potential of Eq.~(\ref{Eq:DispHRep}), according to the general condition~(\ref{Eq:FloquetUGHUG}). The generator $\hat{G}_4(t)$ has been obtained by analogy to the one-mode theory [App.~\ref{Ap:SWCorr-DrQu}] using computer algebra \cite{zitko_2011}.  

The number-conserving terms of the quartic nonlinearity amount to the following contributions
\begin{eqnarray}\;
\epsilon \hat{\mathcal{S}}_4(t) =  \lambda_\ssq(t) \nq  + \lambda_\ssc(t) \nc + \chi_\ssqc \nq \nc + \alpha_\ssq \nq^2  + \alpha_\ssc \nc^2 \label{Eq:S4Text}, \nonumber \\ \;
\end{eqnarray}
with
\begin{eqnarray}
  \lambda_\ssq(t)&=& \epsilon \frac{ \bwq}{8}  \uqq^2 \left[4\Real{\eta_x^2 e^{2 i \wdr t}} + 4|\eta_x|^2 +\uqq^2 + 2\uqc^2\right], \nonumber \\
  \lambda_\ssc(t)&=& \epsilon \frac{ \bwq}{8} \uqc^2 \left[4\Real{\eta_x^2 e^{2 i \wdr t}} + 4|\eta_x|^2 +\uqc^2+2\uqq^2\right],  \nonumber \\
  \chi_\ssqc &=& \epsilon \frac{ \bwq}{4} \uqc^2 \uqq^2, \;\; \alpha_\ssq = \epsilon \frac{ \bwq}{8} \uqq^4, \; \; \alpha_\ssc = \epsilon \frac{ \bwq}{8} \uqc^4. \label{Eq:DefinitionsEffectiveHam}
\end{eqnarray}
These terms enter the effective Hamiltonian:
\begin{eqnarray}
  \HO_{\text{s},\text{eff}} (t) = [\wq - \lambda_\ssq(t)] \nq + [\wc - \lambda_\ssc(t)] \nc \nonumber \\
  - \chi_\ssqc \nq \nc - \alpha_\ssq \nq^2  -  \alpha_\ssc \nc^2.
\end{eqnarray}
This form includes AC Stark shift contributions on the first row, and cross-Kerr, and self-Kerr contributions, respectively, on the second row. On the one hand, $\HO_{\text{s},\text{eff}}(t)$ is the quantum non-demolition Hamiltonian required for dispersive measurement in circuit QED. On the other hand, the explicit form above shows that, at linear order in $\epsilon$, the qubit transition frequencies acquire a dependence on the qubit and cavity states as well as on the drive power.

Next, we address the system-bath coupling in order to categorize all the possible relaxation processes induced by the number non-conserving terms. For this, as before, we calculate the corrections to the dressed system quadratures $\Yq$ and $\Yc$ which enter the system-bath couplings, Eq.~(\ref{Eq:HSB}). These quadratures transform according to
\begin{eqnarray}
  \Yq &\to& \Yq + \epsilon \left[ \Yq, \hat{G}_4(t) \right] + O(\epsilon^2), \nonumber \\
  \Yc &\to& \Yc + \epsilon \left[ \Yc, \hat{G}_4(t) \right] + O(\epsilon^2).
\end{eqnarray}
We focus first on the corrections to the qubit quadrature, \textit{i.e.} $\left[ \Yq, \hat{G}_4(t) \right]$, which will induce corrections to qubit relaxation. The resulting expressions are lengthy; they can be found in App.~\ref{Ap:Tables} (Tables~\ref{Tab:CoeffsQuadQ}, \ref{Tab:CoeffsQuadC}, and \ref{Tab:CoeffsQuadQC} for qubit-only, cavity-only and mixed processes, respectively). The results for the corrected cavity quadrature, $\left[ \Yc, \hat{G}_4(t) \right]$, can be found by applying the following transformation to the three tables: $\wq \leftrightarrow \wc$, $\uqq \leftrightarrow \uqc$, $\vqq \leftrightarrow \vqc$, and $\bq \leftrightarrow \bc$, while leaving $\bwq$ intact. 

To derive the EME, we next express the renormalized qubit quadrature in the interaction picture with respect to $\HO_{\text{s},\text{eff}}(t) + \HO_\text{b}$. This amounts to a sum of operators effecting transitions between the states of the effective Hamiltonian, multiplied by phase factors oscillating at the transition frequency [for a detailed derivation, see App.~\ref{App:EME}]:
\begin{eqnarray}
&&  e^{i \int_0^t dt' \HO_{\text{s},\text{eff}}(t')}  \left\{ \Yq + \epsilon \left[ \Yq, \hat{G}_4\right] \right\} e^{-i \int_0^t dt' \HO_{\text{s},\text{eff}}(t')} \nonumber \\ 
&&\;\;\;\;\;\;\;\;\;\equiv  \sum_j \hat{C}(\om_j) e^{i \om_j t},
\end{eqnarray}
where $j$ indexes a discrete set of frequencies $\{\om_1,\om_2,...\}$ which are linear combinations of $\om_{\text{d}},\wq,$ and $\wc$. $\hat{C}(\om_j)$ are operators at most linear in $\epsilon$, which will enter the dissipators of the EME, according to the prescription:  
\begin{eqnarray}
  \hat{C}(\om_j) e^{i\om_j t} \to 2 \kappa(\om_j) \mathcal{D}\left[ \hat{C}(\om_j) \right],
\end{eqnarray}
where $2\kappa(\om_j) = \SFN(\om_j)$. To order $\epsilon$, the effective collapse operator for the qubit is: 
\begin{widetext}
\begin{eqnarray}
\hat{C}(\wq)=&&-i\Bigg[\vcq-\frac{\epsilon}{8}\left(\frac{\bwq}{\wq}\vcq\uqq^2 - 4\frac{\bwq\wq}{\wc^2-\wq^2}\vcc\uqc\uqq\right) \left(\uqq^2+\uqc^2+\uqq^2 \nq +2\uqc^2 \nc + 2|\eta_x|^2\right)\Bigg]\bq \nonumber \\
&& - i\frac{\epsilon}{8} \frac{\bwq \wdr}{\wq} \vcq  \uqq^2 \left[ \frac{\eta_x^2 }{\wdr+\wq} + \frac{\eta_x^{*2}}{\wdr-\wq} \right] \bq^\dag \nonumber \\
&& +\frac{i \epsilon}{2} \frac{\bwq \wdr}{\wc-\wq} \vcq \uqq \uqc \left[\frac{\eta_x^2}{2\wdr+(\wc-\wq)}  + \frac{\eta_x^{*2}}{2\wdr-(\wc-\wq)}     \right]    \bc \nonumber \\
&& -\frac{i \epsilon}{2} \frac{\bwq \wdr}{\wc+\wq} \vcq \uqq \uqc \left[\frac{\eta_x^{*2}}{2\wdr+(\wc+\wq)}  + \frac{\eta_x^2}{2\wdr-(\wc+\wq)}     \right]    \bc^\dagger \nonumber \\
&& - i\frac{\epsilon}{2} \frac{\bwq \wdr}{\wq}  \vcq \uqq^3  \left[  \frac{\eta_x^*}{\wdr+\wq} + \frac{\eta_x}{\wdr-\wq} \right] \nq - i\frac{\epsilon}{8} \frac{\bwq \wdr}{\wq}  \vcq \uqq \uqc^2  \left[  \frac{\eta_x^*}{\wdr+\wq} + \frac{\eta_x}{\wdr-\wq} \right] \nc \nonumber \\ 
&& +i\frac{\epsilon}{4} \frac{\bwq \wdr}{\wq} \vcq \uqq^3 \left[ \frac{\eta_x^*}{\wdr-\wq} + \frac{\eta_x}{\wdr+\wq} \right]   \bq^2  - i\frac{\epsilon}{4} \frac{\bwq \wdr}{3\wq} \vcq \uqq^3 \left[ \frac{\eta_x}{\wdr-3\wq} + \frac{\eta_x^* }{\wdr+3\wq} \right]   \bq^{\dagger 2} \nonumber \\
&& +i\frac{\epsilon}{4} \frac{\bwq \wdr}{2\wc-\wq} \vcq \uqq \uqc^2 \left[ \frac{\eta_x}{\wdr+ (2\wc - \wq)} + \frac{\eta_x^*}{\wdr-(2\wc - \wq)} \right] \bc^2 \nonumber \\
&& -i\frac{\epsilon}{4} \frac{\bwq \wdr}{2\wc+\wq} \vcq \uqq \uqc^2 \left[ \frac{\eta_x^*}{\wdr+(2\wc + \wq)} + \frac{\eta_x}{\wdr-(2\wc + \wq)} \right] \bc^{\dagger 2} \nonumber \\
&& + \frac{i \epsilon}{2} \frac{\bwq \wdr}{\wc} \vcq \uqc \uqq^2 \left[ \frac{\eta_x^*}{\wdr-\wc} + \frac{\eta_x}{\wdr+\wc} \right] \bq \bc - \frac{i \epsilon}{2} \frac{\bwq \wdr}{\wc} \vcq \uqc \uqq^2 \left[ \frac{\eta_x}{\wdr-\wc} + \frac{\eta_x^*}{\wdr+\wc} \right] \bq \bc^\dag \nonumber \\
&& + \frac{i \epsilon}{2} \frac{\bwq \wdr}{\wc-2\wq} \vcq \uqc \uqq^2  \left[ \frac{\eta_x}{\wdr+(\wc-2\wq)} + \frac{\eta_x^*}{\wdr-(\wc-2\wq)} \right] \bq^\dag \bc \nonumber \\
&& - \frac{i \epsilon}{2} \frac{\bwq \wdr}{\wc+2\wq} \vcq \uqc \uqq^2  \left[ \frac{\eta_x^*}{\wdr+(\wc+2\wq)} + \frac{\eta_x}{\wdr-(\wc+2\wq)} \right] \bq^\dag \bc^\dag.
 \label{Eq:CqEff}
\end{eqnarray}
\end{widetext}
One can determine the effective collapse operator for the cavity normal mode, $\hat{C}(\wc)$, by replacing $\bq \leftrightarrow \bc$, $\wq \leftrightarrow \wc$, $\uqq \leftrightarrow \uqc$, $\ucc \leftrightarrow \ucq$, and $\vcq \leftrightarrow \vcc$, while $\bwq$ remains fixed.


We note that there are other single-photon contributions resulting in dissipators at frequencies different from $\wc$ and $\wq$. Nonetheless, these contributions are order $\epsilon^2$ in the EME, and we therefore neglect them. The collapse operators derived above enter the EME for the qubit coupled to the resonator:
\begin{eqnarray}
  \dot{\hat{\rho}}(t) = -i \left[ \HO_{\text{s},\text{eff}}(t), \hat{\rho}(t) \right] + \sum_{j=\ssq,\ssc}2 \kappa(\om_j) \mathcal{D}\left[ \hat{C}(\om_j) \right] \hat{\rho}(t).\label{Eq:EMESum} \nonumber \\ \;
\end{eqnarray}
State-dependent relaxation rates can be obtained  as before from the Fock-state representation of the EME, which we omit here for brevity.

\begin{figure}[t!]
  \includegraphics[width=\linewidth]{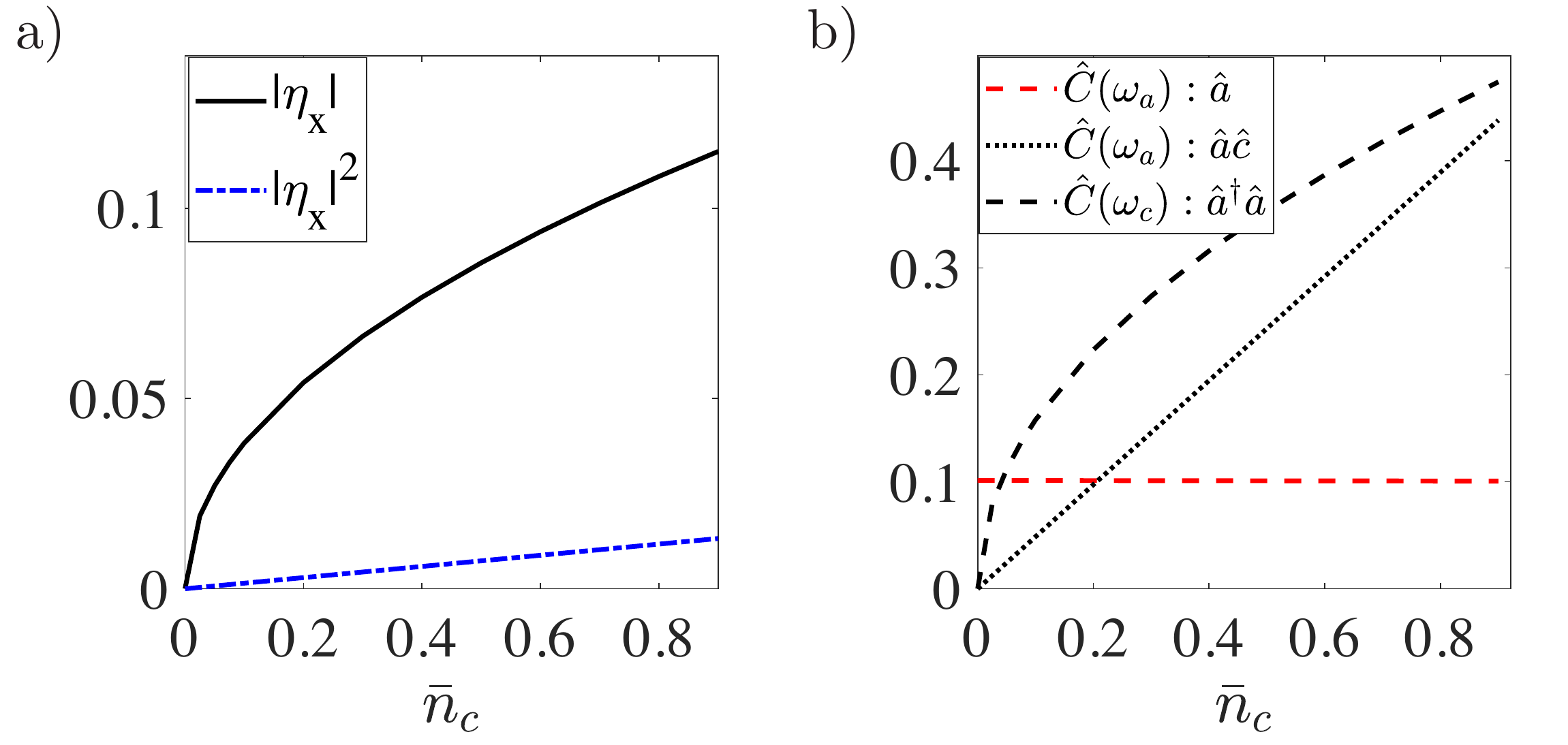}
  \caption{\label{Fig:EMEQ1M1Mag}Magnitudes of a) $\eta_{x}$ as a function of $\bar{n}_\ssc$ for the parameters chosen for the EME simulation (see text); b) for the same parameters, the magnitudes of the most significant terms in the EME.}
\end{figure}

We now turn to the analysis of the various contributions entering the EME. We can simplify the expressions and distill an analytical interpretation of the numerical results for the parameter regime chosen. By direct calculation, we have obtained that the leading contributions to the dissipators of Eq.~(\ref{Eq:EMESum}) are as follows. For the qubit dissipator, there is the dressed single-photon relaxation in the operator $\bq$, along with a correlated relaxation process $\bq(\bc-\bc^\dagger)$ which is large when the drive is nearly resonant with the cavity 
\begin{eqnarray}
\hat{C}(\wq)\approx&&-i\Bigg[\vcq-\frac{\epsilon}{8}\left(\frac{\bwq}{\wq}\vcq\uqq^2 - 4\frac{\bwq\wq}{\wc^2-\wq^2}\vcc\uqc\uqq\right) \nonumber \\ 
&&\;\;\;\;\;\;\;\times\left(\uqq^2+\uqc^2+\uqq^2 \nq +2\uqc^2 \nc + 2|\eta_x|^2\right)\Bigg]\bq \nonumber \\
&& - \frac{i \epsilon}{2} \frac{\bwq}{\wc} \vcq \uqc \uqq^2 \frac{\wdr}{\wc-\wdr} \bq \left(\eta_x^*\bc - \eta_x  \bc^\dag\right).  \label{Eq:CqEffApproxSec} 
\end{eqnarray}
Turning to the cavity dissipator, there are two leading contributions, one corresponding to single photon decay via $\bc$, and one corresponding to qubit dephasing via $\bq^\dagger \bq$: 
\begin{eqnarray}
\hat{C}(\wc)\approx&&-i\Bigg[\vcc-\frac{\epsilon}{8}\left(\frac{\bwq}{\wc}\vcc\uqc^2 - 4\frac{\bwq\wc}{\wq^2-\wc^2}\vcq\uqq\uqc\right) \nonumber \\ 
&&\;\;\;\;\;\;\;\times\left(\uqc^2+\uqq^2+\uqc^2 \nc +2\uqq^2 \nq + 2|\eta_x|^2\right)\Bigg]\bc \nonumber \\ 
&& - i\frac{\epsilon}{8} \frac{\bwq \wdr}{\wc}  \vcc \uqc \uqq^2  \frac{\eta_x}{\wdr-\wc} \nq. \nonumber \\ 
 \label{Eq:CcEffApproxSec}
\end{eqnarray}
Subleading corrections from the remaining terms in Eq.~(\ref{Eq:CqEff}) are at least two orders of magnitude smaller for the parameters chosen.  In  the next subsection we provide numerical estimates for the relative sizes of these contributions in the EME.

\subsection{Numerical results}
\label{SubSec:NumerEME}
Let us now turn to our numerical results based on Eq.~(\ref{Eq:EMESum}), shown in Fig.~\ref{Fig:EMEQ1M1}. Our aim is to illustrate qubit relaxation in the presence of a steady state population in the cavity. This imposes certain constraints on the numerical parameters for the simulation of EME. We have chosen (in rescaled units where ``1'' corresponds to $10$ GHz for typical experiments)
\begin{eqnarray}
 \bwq = 0.77 \pi, \; \bwc = \pi,\; g = 0.025  \pi, \nonumber \\
\end{eqnarray}
for the bare qubit and cavity frequencies, and qubit-cavity coupling $g$, respectively, amounting to $n_\text{crit} = \left[\Delta/(2g)\right]^2 \approx 21$
and hence the following ratio of quality factors of the dressed qubit and cavity:
\begin{equation}
  \frac{Q_\ssq}{Q_\ssc} = \frac{\wq}{\wc} \frac{\kc}{\kq} \approx 51.5. \label{Eq:QFactorRatio}
\end{equation}
This choice for the bare Q-factors guarantees that the population $\langle \bc^\dagger \bc \rangle(t)$ relaxes to the steady state value, with a mean population $\bar{n}_c$, markedly faster than the qubit population. Additionally, we have chosen the anharmonicity parameter $\epsilon=0.1$ which corresponds to $E_\text{C}/E_\text{J} = 1/200$. The drive frequency is detuned from the cavity frequency at half of the value of the Kerr interaction between cavity and qubit, which is the typical situation for dispersive readout \cite{blais_et_al_2004,Wallraff_Strong_2004}:
\begin{eqnarray}
  \wdr = \wc - \chi_\ssqc / 2,
\end{eqnarray}
with $\chi_{\ssqc} = \epsilon\wq \uqq^2 \uqc^2/2 \approx 1.7 \times 10^{-3} \bwc$. Moreover, the initial state corresponds to one photon in the hybridized qubit mode, and the vacuum state for the cavity, that is
\begin{equation}
  \hat{\rho}(0) = |1_\ssq 0_\ssc \rangle \langle 1_\ssq 0_\ssc |.
\end{equation}
By virtue of our choices of Q-factors in Eq.~(\ref{Eq:QFactorRatio}), the population of the qubit, which is in the excited state at the beginning of the simulation according to the initial density matrix $\hat{\rho}(0)$, will relax slowly in the presence of a relatively rapidly stabilizing steady-state population of the cavity, $\bar{n}_\ssc$.

Note that it is not typical for dispersive readout that $\kappa_\ssc \approx 10^{-2} \pi$ is overwhelmingly large compared to the dispersive shift $\chi_{\ssqc}$. Working at low quality factors is imposed by the necessity of simulations to be performed in a reasonable amount of time. This is the consequence of not performing rotating-wave approximation resulting in widely different timescales. However, as our expressions show, we expect the EME to correct the relaxation rates multiplicatively: that is, an order of magnitude decrease of the cavity relaxation rate $\kc$ is expected to result in an order of magnitude decrease in the corrections predicted by the EME. This is why we present our relaxation rates rescaled by the bare relaxation rates instead of absolute units.

We plot the expectation value of the photon number operator corresponding to the hybridized qubit, $\bq^\dag \bq$, and extract the leading exponential decay in its time-dependence. Figure~\ref{Fig:EMEQ1M1}\textit{a}) shows this time dependence for variable drive strength, parametrized by the mean steady-state population of the cavity $\bar{n}_\ssc$ [plotted in Fig.~\ref{Fig:EMEQ1M1}\textit{b)}]. The leading dependence of $\langle\bq^\dag\bq\rangle$ is exponential, and the rate of decay as a function of time increases visibly as a function of drive power. To extract the relaxation rate of the qubit, $\kq^{\text{EME}}$, numerically, as a function of $\bar{n}_\ssc$, we assume the following form for the transient qubit population:
\begin{equation}
  \langle \bq^\dagger \bq \rangle (t) = e^{-2 \kq^{\text{EME}} t} + ...,
\end{equation} 
where the ellipsis contains subleading oscillatory terms (negligible for our parameter choices). The result of this fit is summarized in Fig.~\ref{Fig:EMEQ1M1}\textit{c)}, where the relaxation rate obtained from fitting the EME curves of Fig.~\ref{Fig:EMEQ1M1}a) is plotted versus $\bar{n}_\ssc$:
\begin{equation}
  \frac{\delta\kq^{\text{EME}}(\bar{n}_\ssc)}{\kq^\text{EME}(0)} \equiv \frac{\kq^{\text{EME}}(\bar{n}_\ssc) - \kq^\text{EME}(0)}{\kq^\text{EME}(0)}. \label{Eq:Fitted}
\end{equation}
For the left-hand side of Eq.~(\ref{Eq:Fitted}), we obtain a monotonically increasing correction to the qubit relaxation rate, with almost-linear behavior at low cavity photon number [solid red Fig.~\ref{Fig:EMEQ1M1}\textit{c})]. This increase is primarily due to the nearly-resonant behavior of the correlated decay term in Eq.~(\ref{Eq:CqEffApproxSec}).

Note that, since the hybridization between the qubit mode and the cavity is weak, the EME dynamics closely reproduces the steady state population of the cavity predicted by the linear theory. This is illustrated, for example, by the cavity population, plotted as a function of time and drive strength in Fig.~\ref{Fig:EMEQ1M1}\textit{b}).  A comparison of the relaxation dynamics of the cavity and qubit populations in the first two panels of Fig.~\ref{Fig:EMEQ1M1} reveals that the cavity population relaxes on a time scale which is markedly shorter than the interval of transient exponential decay of the qubit mode.

To illustrate the essential role of number-nonconserving terms, we consider for comparison a Kerr-theory master equation simulation, which exhibits no visible renormalization of the relaxation rates [see inset of Fig.~\ref{Fig:EMEQ1M1}\textit{a})]. This theory retains the number-conserving terms of the Josephson nonlinearity up to quartic order in the undriven Hamiltonian, plus the drive:
\begin{equation}
  \hat{\mathcal{H}}_\text{s,Kerr}(t) = \hat{\mathcal{H}}_\text{s,Kerr} + \hat{\mathcal{H}}_\text{d}(t),
\end{equation}
where
\begin{eqnarray}
  \HO_{\text{s},\text{Kerr}} = [\wq -  \lambda_\ssq^{(0)}] \nq + [\wc -  \lambda_\ssc^{(0)}] \nc \nonumber \\
  -  \chi_\ssqc \nq \nc -  \alpha_\ssq \nq^2  -  \alpha_\ssc \nc^2.
\end{eqnarray}
The frequency shifts amount to
\begin{eqnarray}
  \lambda_\ssq^{(0)}&=& \frac{ \bwq}{8}  \uqq^2 \left[\uqq^2 + 2\uqc^2\right], \nonumber \\
  \lambda_\ssc^{(0)}&=& \frac{ \bwq}{8} \uqc^2 \left[ \uqc^2+2\uqq^2\right],   
\end{eqnarray}
and $\chi_{\ssqc}$, $\alpha_\ssq$, and $\alpha_\ssc$ have been defined in Eq.~(\ref{Eq:DefinitionsEffectiveHam}). 

This driven Kerr Hamiltonian would form the basis of an oversimplified theory in which the rotating-wave approximation has been performed at the level of the Hamiltonian without considering renormalization effects onto dissipators. The associated master equation amounts to adding dissipators $\mathcal{D}[\bq]$ and $\mathcal{D}[\bc]$, thus neglecting the essential contributions to the dissipators from the Josephson nonlinearity and from the drive term:
\begin{eqnarray}
    \dot{\hat{\rho}}(t) = -i \left[ \HO_{\text{s},\text{Kerr}}(t), \hat{\rho}(t) \right] + 2 \kappa(\wc) \mathcal{D}\left[ \bc \right] \hat{\rho}(t)  \nonumber \\
+ 2 \kappa(\wq) \mathcal{D}\left[ \bq \right] \hat{\rho}(t).\label{Eq:KerrQ1M0}
\end{eqnarray}
As shown in the inset of Fig.~\ref{Fig:EMEQ1M1}\textit{a}), there is no renormalization of the decay rate in a Kerr-only master equation simulation.

In Figure~\ref{Fig:EMEQ1M1Mag} we investigate the sizes of the various terms entering the Eqs.~(\ref{Eq:CqEffApproxSec}) and~(\ref{Eq:CcEffApproxSec}). We first note that the drive term $|\eta_x|$, which is proportional to $\sqrt{\bar{n}_\ssc}$, reaches $\approx 10^{-1}$ at $\bar{n}_\ssc = 1.0$, which verifies our condition that the drive should cause only a small deviation on the phase quadrature [Fig.~\ref{Fig:EMEQ1M1Mag}\textit{a})]. Figure~\ref{Fig:EMEQ1M1Mag}\textit{b}) shows the leading contributions in the dissipators, as a function of drive power. The absolute value of the coefficient of the single photon dissipator, normalized by $\vcq$, has almost no renormalization as a function of drive (dashed red curve). However, this value differs from $\vcq$, which would be the amplitude of this term in a purely linear theory. Two contributions control the dressing of the dissipators as a function of drive: the correlated decay $\bq \bc$ in $\hat{C}(\wq)$ (black dotted line), and the photon dephasing term $\bq^\dagger \bq$ in $\hat{C}(\wc)$ (dot-dashed magenta line). 

To further illustrate the effects of these contributions, we have devised EME numerical simulations containing subsets of the terms [Fig.~\ref{Fig:EMEQ1M1}\textit{c})]. The correlated decay $\bq \bc$ in $\hat{C}(\wq)$ seems to be responsible for most of the renormalization of relaxation rates in the presence of drives, as shown by EME simulations where this term is omitted (black dotted line). Moreover, the omission of the dephasing term $\bq^\dagger \bq$ in $\hat{C}(\wc)$ leaves the EME result largely unaffected (see dot-dashed magenta) curve. Finally, we note that the Kerr simulation (solid black line) and an EME simulation retaining only the single-photon terms (red-dashed line) both predict negligible renormalization of the qubit relaxation rate as a function of drive.

\begin{figure}
  \includegraphics[width=\linewidth]{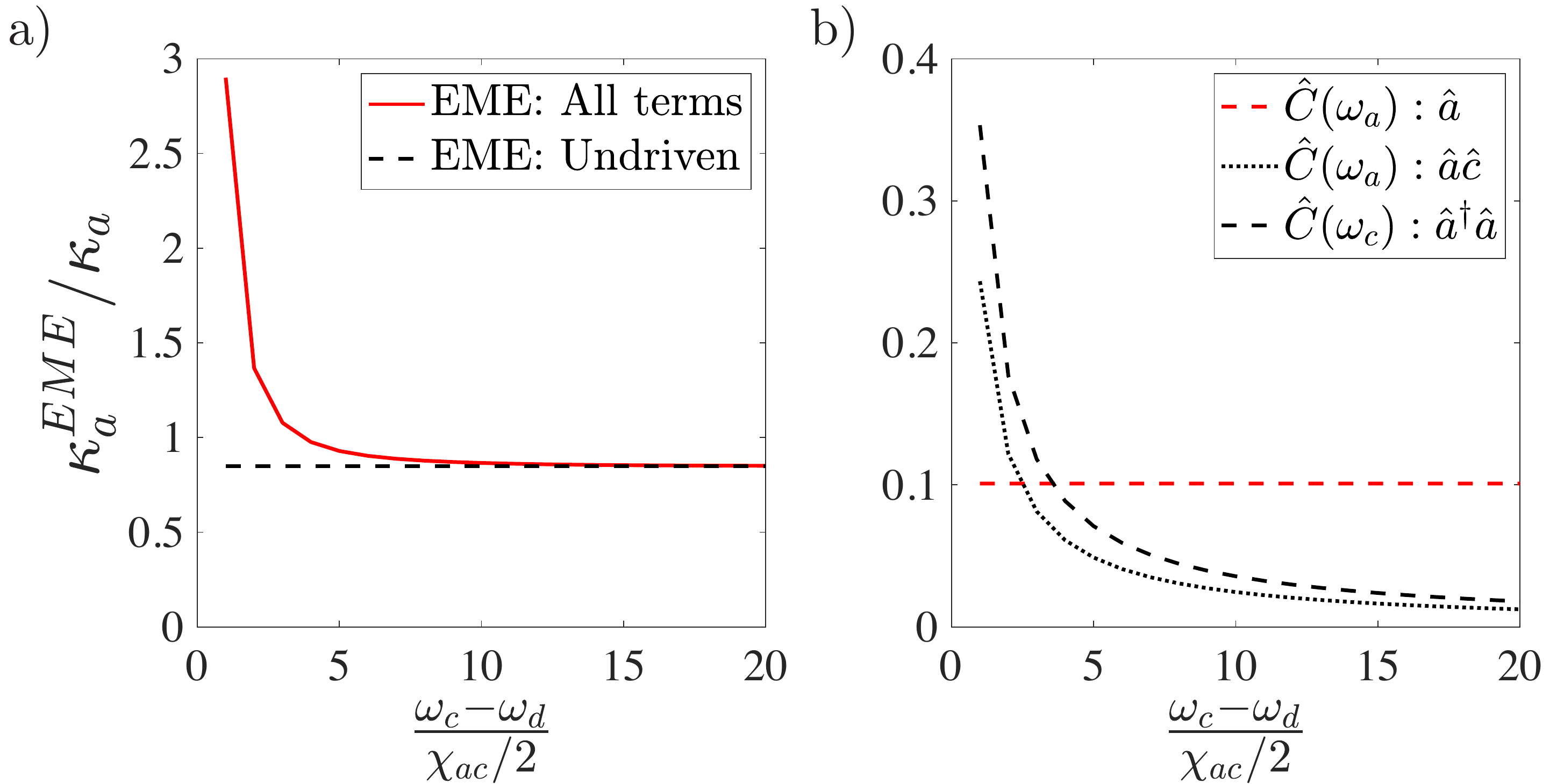}
  \caption{EME results versus drive frequency, at $\bar{n}_\ssc = 0.5$ steady-state cavity photons. a) The relaxation rate obtained from the EME (solid red) exhibits a large renormalization only in the vicinity of the cavity resonance. At large drive-cavity detuning $\wc - \wdr$ (ten times the cross-Kerr energy scale $\chi_{\text{ac}}$) there is almost no correction from the drive-induced terms, as exemplified by a comparison with the EME result for an undriven system (black dashed line). b) This is consistent with the coefficients of the most resonant contributions in the EME decaying algebraically with the detuning. \label{Fig:EMEQ1M1VsVd}}
\end{figure}

Before summarizing, we would like to add a new wrinkle. We have seen that the correction from the drive-induced contributions in the EME is dominated by almost-resonant contributions $\propto 1/(\wc - \wdr)$. In a second set of numerical simulations performed with the same parameters ($\bwq / \bwc = 0.77$, $g/\bwc = 0.025$), we have varied the drive frequency in the interval $[\wdr-10\chi_{\ssq\ssc}, \wdr - \chi_{\ssq\ssc}/2]$ while keeping the cavity steady-state population $\bar{n}_\ssc$ fixed at a reference value of $0.5$ photons. Our results are summarized in Fig.~\ref{Fig:EMEQ1M1VsVd}. The relaxation rate obtained from the EME only shows a markedly large renormalization close to the cavity frequency $\wc$ and decays rapidly as the drive frequency is shifted. When the drive is detuned to around $10 \chi_{\ssq\ssc}$ under the cavity frequency, there is very little renormalization discernible from the drive-induced terms, and the rate obtained from the EME matches to good approximation that corresponding to the EME of the undriven theory [Fig.~\ref{Fig:EMEQ1M1VsVd}a)]. The value of the relaxation rate $\kq^\text{EME}$ predicted by the EME for the undriven case is smaller than $\kq$, as already shown in Part I. Overall, these results are consistent with our understanding of the fact that the coefficients of the drive-induced corrections to the EME decay algebraically with the detuning of the readout drive [Fig.~\ref{Fig:EMEQ1M1VsVd}]. This suggests that there is a marked sensitivity of the renormalization of the decay rate of the qubit as a function of the detuning between the readout drive and the cavity. 

To summarize, it appears that in the driven qubit-cavity system, for a choice of parameters inspired by the setup for dispersive readout, the renormalization of the qubit relaxation rate is primarily driven by nearly-resonant, correlated decay processes corresponding to one photon leaking out of each normal mode, $\bq \bc$. These processes appear as drive-activated corrections to the qubit dissipator. As the drive is detuned from the cavity normal mode frequency, the strength of the terms in the dissipators corresponding to these processes decays inversely proportional to the detuning between the readout drive and the cavity frequency.

\section{Summary}
\label{Sec:Summary}
To conclude, we have argued that the relaxation rate and the transition frequency of a driven, weakly anharmonic, superconducting qubit depend strongly on drive power. We have arrived at these conclusions by devising a perturbation theory in the weak nonlinearity and in the strength of the drive. We have shown that, to lowest order, the effect arises from the interplay of number-nonconserving terms in the nonlinear Hamiltonian with the drive, and that the lowest-order contributions of the Josephson potential, the quartic terms, predict significant corrections to qubit dynamics. Moreover, through full numerical simulation of the EME, we have quantitatively confirmed our qualitative analytical predictions. The theory presented here can be adapted to a wide range of experimental parameters. A quantitative comparison to current experiments would necessitate the inclusion of the effects of finite temperature and pure dephasing \cite{boissonneault_et_al_2009} which is the subject of future work. We expect that these refinements will only bring quantitative corrections to the results presented here, with the qualitative picture conveyed in this work, in particular the net increase of the qubit relaxation rate with drive in the dispersive readout setup, remaining intact.

More generally, our results shed light on the importance of number-nonconserving terms in the theoretical description of driven nonlinear systems. In the limit of zero drive, number-nonconserving terms correspond to the counter-rotating terms of the Hamiltonian, which are frequently neglected in current theories of transmon qubit systems \cite{Nigg_BlackBox_2012,solgun_et_al_2019}. We have shown that, while number-conserving terms dress frequencies to lowest order in the strength of anharmonicity, $\epsilon$, it is the number-nonconserving terms that actually correct the collapse operators, ultimately leading to a $\epsilon-$order corrections to the qubit relaxation rate. These are linear in the mean cavity photon occupation in the steady state, for small photon numbers. This is the central finding of our work.

\section{Acknowledgements}
We acknowledge useful discussions with Alexandre Blais, Michel Devoret, S. M. Girvin, Zlatko Minev, Shantanu Mundhada, Ioan M. Pop, Shyam Shankar, and Yaxing Zhang. A.P. acknowledges funding by the Institut Quantique Postdoctoral Fellowship at the Universit\'e de Sherbrooke. This research was supported by the US Department of Energy, Office of Basic Energy Sciences, Division of Materials Sciences and Engineering under Award No. DE-SC0016011.

\appendix
\section{Notation conventions}
\label{App:Conventions}
Our variables are rescaled from the ones conventionally used in the literature. If $\hat{\varphi_j}$ and $\hat{n}_j$, $j=\ssq,\ssc$, are the canonically conjugate superconducting phase and Cooper pair number operators, then they are related to the operators introduced above as follows:
\begin{eqnarray}
  \hat{X}_j = \frac{1}{\sqrt{\epsilon}} \hat{\varphi}_j, \hat{Y}_j = 2 \sqrt{\epsilon} \hat{n}_j.
\end{eqnarray}
These conventions allow us write the harmonic part of the Hamiltonian in a form that is symmetric with respect to an interchange of the quadratures. Secondly, it allows us to keep the dependence on the anharmonicity parameter $\epsilon$ explicit and outside of the operators.

To organize our double expansion in the drive amplitude and in the anharmonicity, we first needed to switch from the bare mode basis to the normal mode basis, that is: 
\begin{eqnarray}
  \bXq &=& \uqc \Xc + \uqq \Xq, \nonumber \\
  \bXc &=& \ucc \Xc + \ucq \Xq, \nonumber \\
  \bYq &=& \vqc \Yc + \vqq \Yq, \nonumber \\
  \bYc &=& \vcc \Yc + \vcq \Yq. 
\end{eqnarray}
When expressed with respect to the normal mode quadratures on the right hand side of the equations above, the linear Hamiltonian becomes diagonal in the Fock basis. Equivalently, the coupling $g$ has been absorbed along with other details of the linear Hamiltonian into the normal mode coefficients $u_{ij}$, with $i,j \in \{ \ssq,\ssc\}$.  The explicit dependence of these variables on the parameters of the Hamiltonian can be found in Part I. Note, however, that the remaining nonlinear part of the Hamiltonian mixes the normal modes.

The normal-mode quadratures are related to the bosonic creation and annihilation operators as follows:
\begin{eqnarray}
  \Xq = \bq + \bq^\dagger, \; \Yq = -i(\bq - \bq^\dagger), \nonumber \\
  \Xc = \bc + \bc^\dagger, \; \Yc = -i(\bc - \bc^\dagger),
\end{eqnarray}
and the commutator of the quadratures is $[\Xq,\Yq]= 2i$ if $[\bq,\bq^\dagger]=1$ etc.

\section{Displacement transformation without rotating wave approximation}
\label{Ap:Shift}
\textalert{Need to make changes here with a drive on the $\bYc$ quadrature. This will change the form of the displacement.}
This Appendix follows closely the derivation in Ref.~[\onlinecite{Velasco-Martinez_Unitary_2014}] in order to generate a unitary transformation that removes the coherent part of a continuous wave drive on a harmonic oscillator [App.~\ref{Ap:ShiftSE} below]. The second subsection, App.~\ref{Ap:ShiftME}, generalizes this derivation to the case of a harmonic oscillator coupled to a harmonic bath, leading to the formulae used in the main text. 

\subsection{Displacement transformation on the Schr\"odinger equation}
\label{Ap:ShiftSE}
Consider the driven harmonic oscillator described by
\begin{eqnarray}
  \hat{\mathcal{H}}_\text{s}(t)   &=& \hat{\mathcal{H}}_0 + \hat{\mathcal{H}}_\text{d}(t), \nonumber \\
  \hat{\mathcal{H}}_0 &=& \frac{\wq}{4}\left( \Xq^2 + \Yq^2\right),  \nonumber \\
  \hat{\mathcal{H}}_\text{d}(t) &=& \ed \Yq \sin(\wdr t). \label{Eq:DrivenSHO}
\end{eqnarray}
where the canonical commutator between the two quadratures is $[ \Xq,\Yq ] = 2i$. The problem is to find a unitary transformation
\begin{equation}
  \hat{U}(t) = e^{ i \Xq \frac{y_\ssq(t)}{2}} e^{ - i \Yq \frac{x_\ssq(t)}{2}} e^{ - i S(t)},  \label{Eq:Ut}
\end{equation}
with $x_\ssq(t), y_\ssq(t)$ and $S(t)$ three real-valued functions of time, such that
\begin{equation}
  \hat{U}^\dag(t) \left[\hat{\mathcal{H}}_\text{s}(t) - i\partial_t\right] \hat{U}(t) = \frac{\wq}{4} \left( \Xq^2 + \Yq^2 \right) - i \partial_t. \label{Eq:CondU}
\end{equation}

In other words, such a unitary transformation appropriately displaces the two quadratures $\Xq$ and $\Yq$ in order to remove the time-dependent drive term. The task is to find $x_\ssq(t), y_\ssq(t)$ and $S(t)$ satisfying the condition~(\ref{Eq:CondU}). 

The canonical commutation relation implies that $\Yq$ generates translations for $\Xq$ and \textit{vice versa}
\begin{eqnarray}
  e^{   i \Yq \frac{x_\ssq}{2}} \Xq e^{ - i \Yq \frac{x_\ssq}{2}} &=& \Xq + x_\ssq, \nonumber \\
  e^{ - i \Xq \frac{y_\ssq}{2}} \Yq e^{   i \Xq \frac{y_\ssq}{2}} &=& \Yq + y_\ssq. \label{Eq:ShiftsQuadratures}
\end{eqnarray}
Consequently
\begin{eqnarray}
  \hat{U}^\dag(t) \hat{\mathcal{H}}_\text{s}(t) \hat{U}(t) = \frac{\wq}{4} \left[ (\Xq + x_\ssq)^2 + (\Yq + y_\ssq)^2 \right] \nonumber \\+ \ed (\Yq + y_\ssq) \sin(\wdr t). 
\end{eqnarray}

The energy operator transforms according to
\begin{eqnarray}
  \hat{U}^\dag(t) ( - i \partial_t ) \hat{U}(t) = -i\partial_t + \frac{\dot{y_\ssq}}{2} \left( \Xq + x_\ssq \right) - \frac{\dot{x_\ssq}}{2} \Yq - \dot{S}. \nonumber \\ \;
\end{eqnarray}
This follows from an application of the chain rule in $\hat{U}^\dag(t) (-i \partial_t) \hat{U}(t) f(t)$ where $f$ is an arbitrary differentiable complex-valued function of time. 

Then the Floquet Hamiltonian transforms under the action of the unitary transformation $\hat{U}(t)$ as follows:
\begin{eqnarray}
  \hat{U}^\dag(t) \left[ \hat{\mathcal{H}}_\text{s}(t) - i \partial_t \right] \hat{U}(t) = \frac{\wq}{4} \left( \Xq^2 + \Yq^2 \right) - i \partial_t \nonumber \\
  + \left[  \frac{\wq}{2} x_\ssq + \frac{\dot{y_\ssq}}{2} \right] \Xq + \left[  \frac{\wq}{2} y_\ssq + \ed \sin(\wdr t)  - \frac{\dot{x_\ssq}}{2} \right] \Yq \nonumber \\
  + \frac{\wq}{4} \left(x_\ssq^2 + y_\ssq^2 \right) + \ed y_\ssq \sin(\wdr t) + \frac{\dot{y_\ssq}}{2} x_\ssq - \dot{S}.
\end{eqnarray}

In order to satisfy Eq.~(\ref{Eq:CondU}), we ask that the coefficient of the quadrature $\Xq$, the coefficient of the quadrature $\Yq$, and the coefficient of the time-dependent c-number in $\hat{U}^\dag \left[ \hat{\mathcal{H}}_\text{s}(t) - i \partial_t \right] \hat{U}(t)$ vanish, respectively:
\begin{eqnarray}
  \dot{y_\ssq} &=& -\wq x_\ssq, \nonumber \\
  \dot{x_\ssq} &=& \wq y_\ssq + 2 \ed \sin(\wdr t), \nonumber \\
  \dot{S} &=& L(t) \equiv + \frac{\wq}{4} \left(x_\ssq^2 + y_\ssq^2 \right) + \ed y_\ssq \sin(\wdr t) + \frac{\dot{y_\ssq}}{2} x_\ssq. \label{Eq:ClassEOMS}
\end{eqnarray}
The first two are classical equations of motion for the quadratures. $S$ corresponds to the action, whereas $L$ is the Lagrangian, defined here as the Legendre transform of the classical Hamiltonian. The Euler-Lagrange equations corresponding to $L$ are the first two rows of~(\ref{Eq:ClassEOMS}).

The $x_\ssq$ quadrature obeys the equation
\begin{equation}
  \ddot{x_\ssq} + \wq^2 x_\ssq - 2 \ed \wdr \cos(\wdr t) = 0.
\end{equation} 
Plugging in an Ansatz of the particular form that oscillates at the drive frequency $x_\ssq(t) = 2 \eta_{\ssq,x} \cos(\wdr t)$, we find
\begin{equation}
   \eta_{\ssq,x} = \frac{\ed \wdr}{\wq^2 - \wdr^2}.
   \label{Eq:EtadGen}
\end{equation}
Linear combinations of the solutions to the homogeneous equation ($\ed = 0$) can be added in order to enforce any boundary values for $x_\ssq(0), y_\ssq(0)$.  

We note that the form derived above in Eq.~(\ref{Eq:EtadGen}) becomes divergent if the drive is resonant with the mode frequency $\wq$. This is impractical for our application to dispersive readout, where the drive is close-to-resonant with the cavity frequency. One solution to circumvent this problem is to consider the effect of dissipation. If the oscillator had a relaxation rate $\kq$, then the formula above translates to:
\begin{equation}
  \eta_{\ssq,x} = \frac{\ed (\wdr + i \kq)}{\wq^2 - (\wdr + i \kq)^2}.
\end{equation}
For a full derivation of this form, which is adjusted for dissipation, and is divergence-free, the reader can refer to the next subsection.

\subsection{Displacement transformation on the full master equation}
\label{Ap:ShiftME}
A limitation of the transformation performed in App.~\ref{Ap:Shift} is that the effect of the bath is not included in the displacement. For consistency, in a numerical simulation, the transformation of App.~\ref{Ap:Shift} would need to be applied to the system operator in the system-bath coupling, leading to dissipators of displaced collapse operators. It turns out there is a simpler way to account for the effect of the bath by removing the drive terms directly at the level of the Lindblad master equation.
 
To this end, we consider a generalization of the transformation $\hat{U}(t)$ introduced in App.~\ref{Ap:Shift}, and apply it to the reduced density matrix:
\begin{equation}
  \hat{\rho}'( t ) = \hat{V}^\dagger(t) \hat{\rho}(t) \hat{V}(t),
\end{equation}
where $\hat{V}(t)$ has the same form as $\hat{U}(t)$ in Eq.~(\ref{Eq:Ut}),
\begin{equation}
  \hat{V}(t) = e^{ i \Xq \frac{y_\ssq(t)}{2} } e^{ - i \Yq \frac{x_\ssq(t)}{2}} e^{ - i S(t) }, 
\end{equation}
The problem is to find complex $x_\ssq(t)$, $y_\ssq(t)$, $S(t)$ such that the drive term is eliminated from the master equation altogether, not merely from the Hamiltonian as in the previous subsection. 
At the end of this section, we will prove that there actually exist \textit{real} $x_\ssq(t)$, $y_\ssq(t)$, $S(t)$ and hence unitary $\hat{V}(t)$ satisfying the condition above. For now, let us relax this assumption and find the necessary conditions for non-unitary, but invertible, $\hat{V}(t)$ such that the drive term is removed from the Lindblad master equation. We  denote the inverse of $\hat{V}(t)$ by $\hat{W}(t)$:
\begin{eqnarray}
  \hat{W}(t) \hat{V}(t) &=& \hat{V}(t) \hat{W}(t) = 1, \nonumber \\
  \hat{W}^\dagger(t) \hat{V}^\dagger(t) &=& \hat{V}^\dagger(t) \hat{W}^\dagger(t) = 1.
\end{eqnarray}

We now need to express the Lindblad master equation can be expressed in terms of the non-Hermitian Hamiltonian 
\begin{eqnarray}
  \hat{\mathcal{H}}_{\text{s}} &=& \frac{\wq - i \kq}{4}\left( \Xq^2 + \Yq^2 \right) + \ed \Yq \sin(\wdr t), \nonumber \\
  \dot{\hat{\rho}}(t) &=& \frac{1}{i} \left[ \hat{\mathcal{H}}_{\text{s}} \hat{\rho}(t) -  \hat{\rho}(t) \hat{\mathcal{H}}_{\text{s}}^\dagger(t) \right] + 2 \kq \bq \hat{\rho} \bq^\dag. \nonumber
\end{eqnarray}
The first step is to find the equation obeyed by $\hat{\rho}'(t)$. To this end, we may write $\hat{\rho}(t) = \hat{W}^\dagger(t) \hat{\rho}'(t) \hat{W}(t)$ and take the time derivative
\begin{equation}
  \dot{\hat{\rho}}(t) = \dot{\hat{W}}^\dagger(t) \hat{\rho}'(t) \hat{W}(t) + \hat{W}^\dagger(t) \dot{\hat{\rho}}'(t) \hat{W}(t) + \hat{W}^\dagger(t) \hat{\rho}'(t) \dot{\hat{W}}(t),
\end{equation}
then apply $\hat{V}^\dagger(t)$ to the left and $\hat{V}(t)$ to the right, which yields
\begin{eqnarray}
  \hat{V}^\dagger(t) \dot{\hat{\rho}}(t) \hat{V}(t) &=& \hat{V}^\dagger(t)\dot{\hat{W}}^\dagger(t) \hat{\rho}'(t) +  \dot{\hat{\rho}}'(t) +  \hat{\rho}'(t) \dot{\hat{W}}(t) \hat{V}(t) \nonumber \\
  &=& \hat{V}^\dagger(t)\dot{\hat{W}}^\dagger(t) \hat{\rho}'(t) - \hat{\rho}'(t) \hat{W}(t) \dot{\hat{V}}(t) +  \dot{\hat{\rho}}'(t). \nonumber \\ \;
\end{eqnarray}
Then 
\begin{eqnarray}
  \dot{\hat{\rho}}'(t) &=& \hat{V}^\dagger(t) 
  \left\{ \frac{1}{i} \left[ \hat{\mathcal{H}}_{\text{s}} \hat{\rho}(t) -  \hat{\rho}(t) \hat{\mathcal{H}}_{\text{s}}^\dagger(t) \right] + 2 \kq \bq \hat{\rho} \bq^\dag  \right\} \hat{V}(t) \nonumber \\
  && - \hat{V}^\dagger(t) \dot{\hat{W}}^\dagger(t) \hat{\rho}' + \hat{\rho}' \dot{\hat{W}}(t) \hat{V}(t) \nonumber \\
  &=& -i \left[ \hat{V}^\dagger \hat{\mathcal{H}}_{\text{s}} \hat{W}^\dagger \hat{V}^\dagger \hat{\rho} \hat{V} - \hat{V}^\dagger \hat{\rho} \hat{V} \hat{W} \hat{\mathcal{H}}_{\text{s}} \hat{V}\right]  \nonumber \\ 
  && \;\;\;\;\;\;\;\;+ 2 \kq \hat{V}^\dagger \bq \hat{W}^\dagger \hat{V}^\dagger \hat{\rho} \hat{V} \hat{W} \bq^\dagger \hat{V}   \nonumber \\
  && - \hat{V}^\dagger(t) \dot{\hat{W}}^\dagger(t) \hat{\rho}' + \hat{\rho}' \dot{\hat{W}}(t) \hat{V}(t) \nonumber \\ 
  &=& - i \left[ \hat{V}^\dagger \left(\hat{\mathcal{H}}_{\text{s}} - i\partial_t\right) \hat{W}^\dagger \hat{\rho}' - \hat{\rho}' \hat{W} \left(\hat{\mathcal{H}}^\dagger_{\text{s}} - i \partial_t \right) \hat{V}\right] \nonumber \\
  &&  + 2 \kq \left(\hat{V}^\dagger \bq \hat{W}^\dagger\right)  \hat{\rho}' \left(\hat{W} \bq^\dagger \hat{V}\right). \label{Eq:SHEOMRhoPrime}
\end{eqnarray}
We may now use
\begin{eqnarray}
  \hat{V}^\dagger \bq \hat{W}^\dagger &=& \frac{\Xq + i \Yq}{2} + \frac{x_\ssq^* + i y_\ssq^*}{2} \equiv \bq + \bar{a}, \nonumber \\
  \hat{W} \bq^\dagger \hat{V} &=&  \frac{\Xq - i \Yq}{2} + \frac{x_\ssq - i y_\ssq}{2} \equiv \bq^\dagger + \bar{a}^*,
\end{eqnarray}
to recast the last term of~(\ref{Eq:SHEOMRhoPrime}) in the form
\begin{eqnarray}
  && 2 \kq \left(\hat{V}^\dagger \bq \hat{W}^\dagger\right)  \hat{\rho}' \left(\hat{W} \bq^\dagger \hat{V}\right)  \\ && = 2 \kq   \left( \frac{\Xq + i \Yq}{2} + \frac{x_\ssq^* + i y_\ssq*}{2} \right) \hat{\rho}' \nonumber \\
  && \;\;\;\; \times \left( \frac{\Xq - i \Yq}{2} + \frac{x_\ssq - i y_\ssq}{2} \right)   \nonumber \\
  && = 2 \kq \bq\hat{\rho}' \bq^\dag + 2 \kq \bar{a} \hat{\rho}' \frac{\Xq - i \Yq}{2}  \nonumber \\ && \;\;\;\;\;\;\;+ 2 \kq \bar{a}^* \frac{\Xq + i \Yq}{2} \hat{\rho}' + 2 \kq |\bar{a}|^2 \hat{\rho}'. \nonumber \\ \; \label{Eq:TransformedCollapse}
\end{eqnarray}
Additionally, 
\begin{eqnarray}
  \hat{V}^\dagger(t) \left[ \hat{\mathcal{H}}_{\text{s}}(t) - i \partial_t \right] \hat{W}^\dagger(t) = \frac{\wq - i \kq}{4} \left( \Xq^2 + \Yq^2 \right)  \nonumber \\
  + \left[  \frac{\wq - i \kq}{2} x_\ssq^* + \frac{\dot{y_\ssq}^*}{2} \right] \Xq \nonumber \\
+ \left[  \frac{\wq - i\kq}{2} y_\ssq^* - \frac{\dot{x_\ssq}^*}{2} + \ed \sin(\wdr t) \right] \Yq \nonumber \\
  + \frac{\wq - i \kq}{4} \left[\left(x_\ssq^*\right)^2 + \left(y_\ssq^*\right)^2  \right]  \nonumber \\ + \ed y_\ssq^* \sin(\wdr t) + \frac{\dot{y_\ssq}^*}{2} x_\ssq^* - \dot{S}^*. \nonumber \\ \; \label{Eq:TransformedHefFlo}
\end{eqnarray} 
Note that the equation for $  \hat{W}(t) \left[ \hat{\mathcal{H}}_{\text{s}}^\dagger(t) - i \partial_t \right] \hat{V}(t)$ is the Hermitian conjugate of~(\ref{Eq:TransformedHefFlo}). In addition to~(\ref{Eq:TransformedHefFlo}) we must keep track of the second term and the fourth term in the expression of the transformed collapse operator, Eq.~(\ref{Eq:TransformedCollapse}). This yields the following three equations that need to be satisfied for the coefficients of $\Xq$, $\Yq$ and the c-number to vanish respectively:
\begin{eqnarray}
  \Xq&:& \;\; \frac{\wq - i\kq}{2} x_\ssq^*  + \frac{\dot{y_\ssq}^*}{2}  + i \kq  \frac{x_\ssq - iy_\ssq}{2} = 0, \nonumber \\
  \Yq&:& \;\; \frac{\wq - i\kq}{2} y_\ssq^* - \frac{\dot{x_\ssq}^*}{2} - \kq \frac{x_\ssq-iy_\ssq}{2} + \ed \sin(\wdr t) = 0, \nonumber \\
  \text{c-number} &:& \;\; \frac{\wq - i \kq}{4} \left[ \left( x_\ssq^* \right)^2  + \left( y_\ssq^* \right)^2 \right] + \ed y_\ssq^* \sin(\wdr t) \nonumber \\ && \;\; + \frac{\dot{y_\ssq}^*}{2} x_\ssq^* + i \kq |\bar{a}|^2 - \dot{S}^* = 0. \label{Eq:CondsZeroComplex}
\end{eqnarray}
The third equation gives a prescription for $S$ as soon as $x_\ssq$ and $y_\ssq$ are found. The first two equations can be rearranged to give:
\begin{eqnarray}
  \dot{x_\ssq}^* &=& \left( \wq - i \kq \right) y_\ssq^* - \kq \left( x_\ssq - i y_\ssq \right) + 2 \ed \sin(\wdr t), \label{Eq:EOMx*Andy*} \\
  \dot{y_\ssq}^* &=& -\left( \wq - i \kq \right) x_\ssq^*  - i \kq \left(x_\ssq - iy_\ssq\right).  \nonumber
\end{eqnarray}

Let us search for real solutions for the classical quadratures. If $x_\ssq(t)$ and $y_\ssq(t)$ were real, the equations would be
\begin{eqnarray}
  \dot{x_\ssq} &=& - \kq x_\ssq + \wq y_\ssq + 2 \ed \sin(\wdr t), \label{Eq:EOMxAndy} \\
  \dot{y_\ssq} &=& - \kq y_\ssq - \wq x_\ssq.  \nonumber
\end{eqnarray}
This results in the second order differential equation for $x_\ssq$:
\begin{eqnarray}
  \ddot{x_\ssq} + 2 \kq \dot{x_\ssq} + (\kq^2 + \wq^2) x_\ssq = 2 \ed \wdr \cos( \wdr t ).
\end{eqnarray}
This is the equation of an oscillator of natural frequency $\wq = \sqrt{\kq^2 + \wq^2}$, decay rate $2 \kq$, driven by the periodic forcing term $2 \ed \wdr \cos(\wdr t)$.  The particular solution is $x_\ssq(t) = \eta_{x_\ssq} e^{-i\wdr t} + \eta_{x_\ssq}^* e^{i \wdr t}$, with
\begin{equation}
\eta_{x_\ssq} = \frac{\ed (\wdr + i \kq)}{\wq^2 - (\wdr + i \kq)^2},
\end{equation} 
while $y_\ssq(t) = \eta_{\ssq,y} e^{-i\wdr t} + \eta_{\ssq,y}^* e^{i \wdr t}$ with
\begin{equation}
\eta_{\ssq,y} = -i  \frac{\ed \wq}{\wq^2 - (\wdr + i \kq)^2}.
\end{equation}

We have found real $x_\ssq(t)$ and $y_\ssq(t)$ describing the steady state of the equations~(\ref{Eq:EOMxAndy}). Since the two quadratures are real, the equation for $S$, Eq.~(\ref{Eq:CondsZeroComplex}), becomes
\begin{eqnarray}
&&  \frac{\wq}{4} \left[ x_\ssq^2  + y_\ssq^2 \right] + \ed y_\ssq^* \sin(\wdr t) + \frac{\dot{y_\ssq}}{2} x_\ssq  - \dot{S}^* = 0, \nonumber \\ \;
\end{eqnarray}
implying that $S$ is real, and therefore the transformation matrix $\hat{V}(t)$ is unitary and therefore $W(t) = V^\dagger(t)$.

Finally, we have arrived at the following master equation for $\hat{\rho}'(t)$:
\begin{eqnarray}
\dot{\hat{\rho}}'(t) &=&
- i \left[ \frac{\wq - i \kq}{4} \left( \Xq^2 + \Yq^2 \right) \hat{\rho}' - \hat{\rho}' \frac{\wq - i \kq}{4} \left( \Xq^2 + \Yq^2 \right) \right] \nonumber \\
  &&  + 2 \kq \bq   \hat{\rho}'\bq^\dagger \nonumber \\
&=& - i \left[ \wq \bq^\dagger \bq, \hat{\rho}'(t) \right] + 2 \kq \mathcal{D}\left[ \bq \right] \hat{\rho}'(t).
\end{eqnarray}
All the complexity of solution to the classical driven-dissipative harmonic oscillator is now encapsulated in the unitary transformation $\hat{V}(t)$ that relates the lab frame density matrix $\hat{\rho}(t)$ to the displaced density matrix $\hat{\rho}'(t)$.

\section{Hierarchical equations for time-dependent generators}
 \label{Sec:Hierarchy}

In this appendix, we derive the hierarchical equations for the generator of our perturbation theory, $\hat{G}(t)$, for time-dependent Hamiltonians. The problem is to find the $\hat{G}(t)$ such that all number non-conserving terms up to $O(\epsilon^n)$ for $n \geq 1$ are removed from the left hand side of Eq.~(\ref{Eq:FloquetUGHUG}). 

More explicitly, we may reexpress the system Hamiltonian of Eq.~(\ref{Eq:DispH}) formally as a series in powers of the anharmonicity parameter $\epsilon$:
\begin{equation}
  \hat{\mathcal{H}}_{\text{s}}(t) = \hat{\mathcal{H}}_{\text{2}} - \epsilon \hat{\mathcal{H}}_4(t)  + \epsilon^2 \hat{\mathcal{H}}_6(t) + \ldots, \label{Eq:HstExpansion}
\end{equation}
where all of $\hat{\mathcal{H}}_{2n}$ are known, and separable into number-conserving and -nonconserving contributions:
\begin{equation}
  \hat{\mathcal{H}}_{2n}(t) = \hat{\mathcal{S}}_{2n}(t) + \hat{\mathcal{N}}_{2n}(t), \label{Eq:Sep}
\end{equation}
for any $n \geq 2$, where it is implicitly assumed that both terms on the right hand side of Eq.~(\ref{Eq:Sep}) are in normal-ordered form, and we define number-conserving terms strictly speaking as terms which are polynomials in the number operators $\nq$ and $\nc$. For example, a term of the form $\bq^\dagger \bq^\dagger \bc \bc$ belongs to $\hat{\mathcal{N}}_4(t)$.

The unknown generator of the unitary transformation, $\hat{G}(t)$, may also be expressed as a series in $\epsilon$:
\begin{eqnarray}
  \hat{G}(t) = \epsilon \hat{G}_4(t) + \epsilon^2 \hat{G}_6(t) + \ldots, \label{Eq:GtExpansion}
\end{eqnarray}
where $\hat{G}_{2n}(t)$ are unknown. The requirement that number-nonconserving terms in $\hat{\mathcal{H}}_\text{s}(t) - i \partial_t$ be removed translates to a hierarchical set of differential equations for the operator-valued coefficients in the Taylor series of the generator, $\hat{G}_{2n}(t)$.

This can be achieved by using the Baker-Campbell-Haussdorff expansion on Eq.~(\ref{Eq:FloquetUGHUG}) to find $\hat{G}(t)$ order by order by imposing that all number non-conserving terms up to some order $\epsilon^n$ disappear from the system Hamiltonian. The first step in this iterative process is to cancel all number-nonconserving terms which are order-$\epsilon$. We insert into Eq.~(\ref{Eq:FloquetUGHUG}) the Taylor series for $\hat{G}(t)$ and $\HO_\text{s}(t)$, then expand the resulting expression to linear order in $\epsilon$, to find that
\begin{eqnarray}
  &&e^{-\hat{G}(t)} \left[\hat{\mathcal{H}}_\text{s}(t) - i\partial_t \right] e^{\hat{G}(t)} = \\
  &&\;\;\;\;\;\;\;\;\; \hat{\mathcal{H}}_2 + \epsilon\left\{ - \hat{\mathcal{H}}_4(t) + \left[ \hat{\mathcal{H}}_2, \hat{G}_4 \right] - i \dot{\hat{G}}_4(t) \right\} + O(\epsilon^2). \nonumber
\end{eqnarray}
The curly brace on the right hand side contains all contributions of order $\epsilon$. Requiring that number-nonconserving terms be canceled at order $\epsilon$ amounts to an ordinary differential equation for $\hat{G}_4(t)$. Separating number-nonconserving terms of $\HO_4(t)$ as in Eq.~(\ref{Eq:Sep}), we have
\begin{eqnarray}
  - i \dot{\hat{G}}_4(t) + \left[ \hat{\mathcal{H}}_2, \hat{G}_4(t) \right] = \hat{\mathcal{N}}_4(t).\label{Eq:G4tODE}
\end{eqnarray}
With this condition satisfied, the effective Hamiltonian takes the form:
\begin{eqnarray}
  \hat{\mathcal{H}}_{\text{s},\text{eff}}(t) = \hat{\mathcal{H}}_2 - \epsilon \hat{\mathcal{S}}_4(t) + O(\epsilon^2).  \label{Eq:Heff4}
\end{eqnarray}
Even though Eq.~(\ref{Eq:G4tODE}) is an operator-valued ordinary differential equation, we may expand both $\hat{\mathcal{N}}_4(t)$ and $\hat{G}_4(t)$ over normal-ordered products of creation and annihilation operators, which we referred to in Part I as ``monomials'', and thereby obtain a solvable system of \textit{uncoupled} ordinary differential equations for the complex-valued coefficients of $\hat{G}_4(t)$. We select the following initial condition at $t=0$:
\begin{equation}
  \left[ \hat{\mathcal{H}}_2, \hat{G}_4(0) \right] = \hat{\mathcal{N}}_4(0)\label{Eq:G4tIC}
\end{equation}
such that the unitary generated by $G_4(0)$ removes $\hat{\mathcal{N}}_4(0)$ from the system Hamiltonian at $t=0$. 

Higher-order number non-conserving terms can be recursively canceled. If the time dependence of $\hat{G}_4(t)$ is known, then $\hat{G}_6(t)$ can be obtained upon requiring that all order-$\epsilon^2$ number non-conserving terms in Eq.~(\ref{Eq:FloquetUGHUG}) are vanishing. This condition reads
\begin{eqnarray}
-i\dot{\hat{G}}_6(t) + [\hat{\mathcal{H}}_2,\hat{G}_6(t)]+\hat{\mathcal{N}}_6(t)-[\hat{\mathcal{S}}_4(t),\hat{G}_4(t)]\nonumber \\ -\half\mathcal{N}\left(\left[\left[\hat{\mathcal{H}}_2,\hat{G}_4\right],\hat{G}_4\right]\right)=0, \label{Eq:G6tODE}
\end{eqnarray}
where $\mathcal{N}(\hat{O})$ for a normal-ordered operator $\hat{O}$ selects only those terms in $\hat{O}$ which are number non-conserving. There is an analogous initial condition for $\hat{G}_6(t)$ at $t=0$:
\begin{eqnarray}
  &&[\hat{\mathcal{H}}_2,\hat{G}_6(0)]+\hat{\mathcal{N}}_6(0)\nonumber -[\hat{\mathcal{S}}_4(0),\hat{G}_4(0)] \\ && \;\;\;\;\;\;\;\;\;\;\;\;\;\;\; -\half\mathcal{N}\left([\hat{\mathcal{N}}_4(0),\hat{G}_4(0)]\right)=0. \label{Eq:G6tIC}
\end{eqnarray}
The effective order-$\epsilon^2$ Hamiltonian depends only on $\hat{G}_4(t)$
\begin{eqnarray}
\hat{\mathcal{H}}_{\text{s,eff}}&=&\hat{\mathcal{H}}_2-\epsilon\hat{\mathcal{S}}_4+\epsilon^2\hat{\mathcal{S}}_6 \nonumber \\
&&\;\;-\frac{\epsilon^2}{2}\mathcal{S}\left(\left[\left[\hat{\mathcal{H}}_2,\hat{G}_4\right],\hat{G}_4\right]\right)
+ O(\epsilon^3),
\label{Eq:Heff6}
\end{eqnarray}
where we have analogously defined $\mathcal{S}(\hat{O})$ to denote the number-conserving terms of some normal-ordered operator $\hat{O}$. 

Transition frequency corrections due to the Josephson anharmonicity are obtained immediately from $\hat{\mathcal{H}}_{\text{s},\text{eff}}$, since it is diagonal in the Fock representation. We stress that, while energy corrections at order $\epsilon$ arise from the number-conserving terms in the Hamiltonian, corrections to energies at order $\epsilon^2$ and higher can arise from number non-conserving terms, as well, as illustrated by last term of Eq.~(\ref{Eq:Heff6}). 

To summarize, due to the time dependence of the system Hamiltonian, it follows that the terms in the expansion of the generator $\hat{G}(t)$ must satisfy operator-valued ODEs. Equations~(\ref{Eq:G4tODE}) and~(\ref{Eq:G6tODE}) are such equations for the first two terms in the expansion, $\hat{G}_4(t)$ and $\hat{G}_6(t)$, respectively. These equations can be reduced to systems of uncoupled ordinary differential equations by expansion over normal-ordered monomials in the bosonic operators. The procedure can be iterated to obtain equations for $\hat{G}_{2n}(t)$ for $n=4,5,\ldots$. The bookkeeping of terms becomes difficult as the degree of the monomials increases. We have performed the normal-ordering of the operators, as well as solutions to the resulting differential equations for the generators $\hat{G}_4(t)$ using computer algebra techniques.

\section{Qubit coupled to an infinite waveguide}
\label{Ap:SWCorr-DrQu}

In this appendix we present the treatment of a weakly driven weakly anharmonic qubit coupled to an infinite waveguide, in which we take as dominant source of quantum noise the noise on the flux quadrature. This is the situation of a frequency-tuned transmon qubit \cite{koch_et_al_2007}. Below, we are interested in the effect of flux noise solely, which dominates over charge noise in frequency-tuned transmons. 

\begin{figure}[b!]
  \includegraphics[width=0.6\linewidth]{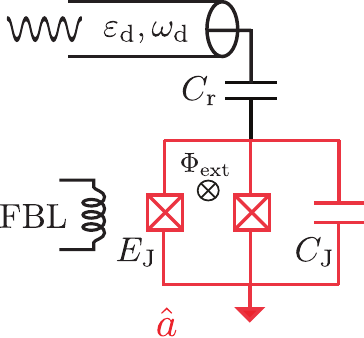}
  \caption{An EME is derived here for a transmon qubit (mode $\bq$) coupled to an infinite waveguide. A flux bias line used to tune the frequency of the transmon qubit by adjusting the magnetic flux through its SQUID loop. Flux noise is the dominant source of noise for this setup.}
\end{figure}

\subsection{Effective Master Equation}
The system circuit, which is shown schematically in Fig.~\ref{Fig:Circuits}a), is described by the following Hamiltonian 
\begin{eqnarray}
  \HO &=& \HO_\ssq + \HO_\text{d}(t) + \HO_\text{sb} + \HO_\text{b}.
\end{eqnarray} 
$\HO_\ssq$ is the qubit Hamiltonian, obtained from circuit quantization of the circuit in Fig.~\ref{Fig:Circuits}a) and upon expanding the Josephson potential to linear order in $\epsilon$: 
\begin{equation}
  \HO_{\ssq} = \wq \left( \bq^\dag \bq + \half \right) -
  \frac{\epsilon \wq}{48} \left(  \bq + \bq^\dag  \right)^4,
\end{equation}
whereas the Hamiltonians describing the drive on the charge quadrature, the system-bath coupling and the bath modes are
\begin{eqnarray}
\HO_\text{d}(t)  &=& -i \ed (\bq-\bq^{\dag})\sin(\wdr t), \nonumber \\
\HO_\text{sb} &=& (\bq + \bq^\dag) \sum_k g_{k} (\Bop + \Bop^\dag), \nonumber \\
\HO_\text{b}  &=& \sum_k \om_k \Bop^\dag \Bop.
\end{eqnarray} 

In order to prepare a simultaneous expansion in the anharmonicity parameter $\epsilon$ and in the drive strength $\ed$, we first perform a displacement transformation
\begin{eqnarray}
  \HO(t) - i\partial_t \to \hat{U}^\dag(t)\left[ \HO(t) - i\partial_t \right]\hat{U}(t), \nonumber \\
  \hat{U}(t) = e^{ i \Xq \frac{y_\ssq(t)}{2}} e^{ - i \Yq \frac{x_\ssq(t)}{2}} e^{ - i S_\ssq(t) },  \label{Eq:UtMain}
\end{eqnarray}
parametrized by three real functions of time $x_\ssq(t)$, $y_\ssq(t)$ and $S_\ssq(t)$. This transformation will remove the drive term $\HO_\text{d}(t)$ from the linear part of the dynamical equations [see App.~\ref{Ap:Shift} and App.~\ref{Ap:ShiftME}], provided that $x_\ssq$ and $y_\ssq$ obey the classical equations of motion for a driven harmonic oscillator,
\begin{eqnarray}
  \dot{x}_\ssq &=& \wq y_\ssq + 2 \ed \sin(\wdr t) - \kq x_\ssq, \nonumber \\
  \dot{y}_\ssq &=& -\wq x_\ssq - \kq y_\ssq,
\end{eqnarray}
and that $S_\ssq(t)$ is the associated action. The particular solution is $x_\ssq(t) = \eta_{\ssq,x} e^{-i\wdr t} + \eta_{\ssq,x}^* e^{i \wdr t}$, with
\begin{equation}
\eta_{\ssq,x} = \frac{\ed (\wdr + i \kq)}{\wq^2 - (\wdr + i \kq)^2},
\end{equation} 
while $y_\ssq(t) = \eta_{\ssq,y} e^{-i\wdr t} + \eta_{\ssq,y}^* e^{i \wdr t}$ with
\begin{equation}
\eta_{\ssq,y} = -i  \frac{\ed \wq}{\wq^2 - (\wdr + i \kq)^2}.
\end{equation}
Based on these, the remaining function $S_\ssq$ can be determined, which is handled in the Appendix. With this transformation, the resulting Hamiltonian takes the form
\begin{eqnarray}
  &&\HO_\ssq + \HO_\text{d}(t) \to \HO_\ssq(t), \label{Eq:Q1M0Disp} \\
  &&\HO_\ssq(t) = \wq \left( \bq^\dag \bq + \half \right) -
  \frac{\epsilon \wq}{48} \left(  \bq + \eta_{\ssq,x}e^{-i\wdr t} + \text{H.c.}  \right)^4, \nonumber
\end{eqnarray}
Note that this calculation of the displacement parameters is done without invoking the rotating wave approximation. This allows us to accurately account for the effect of all number non-conserving terms in the Josephson nonlinearity. Moreover, through the explicit dependence on the relaxation rate, this displacement transformation takes into account the effect of the bosonic bath to lowest order in the anharmonicity.

\begin{table}[b!]
  \begin{tabular}{|l|l|}
    \hline
    Operator  &  Coefficient \\
    \hline
    $\bq$ &  
    $-\frac{\wq \eta_{\ssq,x}^2 \left(-e^{-2 i t \wdr}+e^{2 i t \wq}\right)}{8 \left(\wdr+\wq\right)}$ \\
    &
    $-\frac{\wq  \eta_{\ssq,x}^{*2} \left(e^{2 i t \wdr}-e^{2 i t \wq}\right)}{8 \left(\wdr-\wq\right)}$ \\
    &
    $+\frac{1}{8} \left[\left(\eta_{\ssq,x}^2 + \eta_{\ssq,x}^{*2} \right)  e^{2 i t \wq}+2 |\eta_{\ssq,x}|^2 + 1\right]$ \\
    \hline     
    $\bq^\dag$ & c.c. \\
    \hline
    $\bq^2$       &   
    $-\frac{\wq \eta_{\ssq,x} \left(-e^{-i t \wdr}+e^{3 i t \wq}\right)}{4 \left(\wdr+3 \wq\right)}$ \\
    &
    $-\frac{\wq \eta_{\ssq,x}^* \left(e^{i t \wdr}-e^{3 i t \wq}\right)}{4 \left(\wdr-3 \wq\right)}$ \\
    &
    $-\frac{\wq \eta_{\ssq,x}  \left(e^{-i t \wdr}-e^{i t \wq}\right)}{4 \left(\wdr+\wq\right)}$ \\
    & 
    $+\frac{\wq \eta_{\ssq,x}^* \left(e^{i t \wdr}-e^{i t \wq}\right)}{4 \left(\wdr-\wq\right)} $ \\
    & 
    $+\frac{1}{12}
    \left(-3 e^{i t \wq} + e^{3 i t \wq}\right) \left(\eta_{\ssq,x} + \eta_{\ssq,x}^* \right)$ \\
    \hline
    $\left(\bq^\dag\right)^2$ & c.c. \\
    \hline
    $\bq^\dag \bq$ & 
    $+\frac{\wq \left[\eta_{\ssq,x}\left(e^{-i t \wq} - e^{-i t \wdr}\right)+ \text{c.c.}\right]}{2 \left(\wdr-\wq\right)}$ \\ 
    &$+\frac{\wq \left[\eta_{\ssq,x} \left(e^{-i t \wdr} - e^{i t \wq} \right) + \text{c.c.}\right]}{2 \left(\wdr+\wq\right)}$ \\
    & $+ \cos(\wq t) \left(\eta_{\ssq,x} + \eta_{\ssq,x}^*\right) $\\
    \hline
    $\left(\bq^\dag\right)^2 \bq + \text{H.c.}$ & $\frac{1}{8}$ \\
    \hline
    $\left(\bq^\dag\right)^3 + \text{H.c.}$ & $-\frac{1}{48}$ \\
    \hline
  \end{tabular}
  \caption{\label{Tab:QubitOnlyQuadratureTf} The terms of $\left[ \Xq, \hat{G}_4\right]$. The left column shows each operator entering the sum, and the right column shows its coefficient. The explicit derivations are provided in App.~\ref{Ap:1stOrderSW}.} 
\end{table}

In order to separate order-$\epsilon$ corrections to relaxation rates from the frequency corrections, we transform the displaced Hamiltonian, Eq.~(\ref{Eq:Q1M0Disp}), to perturbatively remove the number non-conserving terms. Specifically, we aim to find a unitary transformation defined by antihermitian operator $\hat{G}_4(t)$ such that 
\begin{eqnarray}
 && e^{-\epsilon \hat{G}_4(t)} \left[ \hat{\mathcal{H}}_\ssq(t) - i \partial_t \right] e^{\epsilon \hat{G}_4(t)} =  
  \HO_{\ssq,\text{eff}}(t) - i \partial_t + O(\epsilon^2).
  \nonumber \\ \;\label{Eq:SingleQubitCondition}
\end{eqnarray}
The effective Hamiltonian~$\HO_{\ssq,\text{eff}}(t)$ contains only number-conserving terms:
\begin{equation}
  \HO_{\ssq,\text{eff}}(t) = \wq \hat{H}_\ssq - \frac{\epsilon \wq}{48} \hat{S}_4(t),
\end{equation}
with:
\begin{eqnarray}
  \hat{S}_4(t) &=& 6 \hat{H}_\ssq^2 +  12 x_\ssq^2(t) \hat{H}_\ssq + x_\ssq^4(t) + \frac{6}{4}, \nonumber \\
  \hat{H}_\ssq &=& \hat{n}_\ssq + \half. \label{Eq:S4q}
\end{eqnarray}
\\
Defining $\hat{N}_4(t)$ to contain all the  number-nonconserving terms to quartic order in creation and annihilation operators, \textit{i.e.}
\begin{equation}
  -\frac{\epsilon \wq}{48} \hat{N}_4(t) \equiv \HO_\ssq(t) - \HO_{\ssq,\text{eff}}(t),
\end{equation}
the condition~(\ref{Eq:SingleQubitCondition}) is equivalent to
 \begin{equation}
  - i \dot{\hat{G}}_4(t) + \left[ \wq \hat{H}_{\ssq}, \hat{G}_4(t)\right]  = \frac{\wq \hat{N}_4(t)}{48}. \label{Eq:ODEG4}
\end{equation}
We derive the generator $\hat{G}_4(t)$ from this equation in App.~\ref{Ap:1stOrderSW}. In this section we will make use only of the resulting commutators of $\hat{G}_4(t)$ with system quadratures etc. which will yield the EME.

A compact form for the effective Hamiltonian is
\begin{eqnarray}
   \hat{\mathcal{H}}_{\ssq,\text{eff}}(t) &=&  \left\{ 1 - \frac{\epsilon}{8} - \frac{\epsilon}{2}\left[|\eta_{\ssq,x}|^2 + \Real{\eta_{\ssq,x}^2 e^{2it\wdr}} \right] \right\} \wq \hat{n}_\ssq \nonumber \\
  &&- \frac{\epsilon}{8}\wq \hat{n}_\ssq^2 + O(\epsilon^2),\label{Eq:HqEff} \nonumber
\end{eqnarray}
where in the last line we have neglected c-number contributions. Time-dependent contributions coming from the drive through $x_\ssq(t)$ are retained in $\hat{\mathcal{H}}_{\ssq,\text{eff}}(t)$. Equation~(\ref{Eq:HqEff}) contains the state-dependent renormalization of the qubit transition frequencies coming from the self-Kerr interaction.

We now focus on the effect of the unitary $e^{-\epsilon \hat{G}_4(t)}$ on the relaxation processes. Recall that $\hat{G}_4(t)$ can be calculated explicitly, and we provide the solution in App.~\ref{Ap:1stOrderSW}. The relaxation processes induced by the nonlinearity can be obtained by calculating the renormalized system-bath Hamiltonian, $e^{-\epsilon \hat{G}_4(t)} \HO_{\text{sb}} e^{\epsilon \hat{G}_4(t)}$. This unitary acts only upon the qubit quadrature, as
\begin{equation}
  e^{-\epsilon \hat{G}_4(t)} \Xq e^{\epsilon \hat{G}_4(t)}  = \Xq + \epsilon \left[ \Xq, \hat{G}_4\right] + O(\epsilon^2).
\end{equation}
We have listed in Table~\ref{Tab:QubitOnlyQuadratureTf} all order-$\epsilon$ terms arising from this transformation.

We next express the renormalized qubit quadrature in the interaction picture with respect to $\HO_{\text{s},\text{eff}}(t) + \HO_\text{b}$ as in the main text. This amounts to a sum of operators corresponding to transitions between states of the $\HO_{\text{s},\text{eff}}$, multiplied by phase factors oscillating at the corresponding transition frequency (a proof of this point can be found in App.~\ref{App:EME}):
\begin{eqnarray}
&&  e^{i \int_0^t dt' \HO_{\text{s},\text{eff}}(t')}  \left\{ \Xq + \epsilon \left[ \Xq, \hat{G}_4\right] \right\} e^{-i \int_0^t dt' \HO_{\text{s},\text{eff}}(t')} \nonumber \\ 
&&\;\;\;\;\;\;\;\;\;\equiv  \sum_j \hat{C}(\om_j) e^{i \om_j t},
\end{eqnarray}
where $j$ indexes a discrete set of frequencies $\{\om_1,\om_2,...\}$ which are linear combinations of $\om_{\text{d}},\wq$. $\hat{C}(\om_j)$ are operators at most linear in $\epsilon$, which will enter the dissipators of the EME, according to the prescription:  
\begin{eqnarray}
  \hat{C}(\om_j) e^{i\om_j t} \to 2 \kappa(\om_j) \mathcal{D}\left[ \hat{C}(\om_j) \right],
\end{eqnarray}
where $2\kappa(\om_j) = \SFN(\om_j)$, and $\SFN$ differs from the expression provided in the main text, Eq.~(\ref{Eq:SFN}), by replacing the charge $\hat{Y}_\text{b}$ quadratures with the phase $\hat{X}_\text{b}$ quadratures of the bath. 

Up to the leading order in $\epsilon$, this leads us to an EME (see App.~\ref{App:EME}):
\begin{eqnarray}
  \dot{\hat{\rho}}_\ssq(t)=-i\left[\hat{\mathcal{H}}_{\ssq,\text{eff}},\hat{\rho}_\ssq(t)\right]+ \sum_{j} 2 \kappa(\om_j)\mathcal{D}[\hat{C}(\om_j)]\hat{\rho}_\ssq(t). \nonumber \\ \label{Eq:Q1M0EME}
\end{eqnarray}
In the above, the operator entering the dissipator $\hat{C}(\om_j)$ has the following dominant contribution at the qubit frequency $\om_j = \wq$:
\begin{eqnarray} \label{Eq:CQPTQ1M0}
    \hat{C}(\wq) &=& \left[1 + \frac{\epsilon}{8} \left(1 + \hat{n}_\ssq + 2  |\eta_{\ssq,x}|^2 \right)\right] \bq \\
&& + \frac{\epsilon \wdr}{8} \left( \frac{\eta_{\ssq,x}^2}{\wdr-\wq} + \frac{\eta_{\ssq,x}^{*2}}{\wdr+\wq} \right) \bq^\dag \nonumber \\ 
    && - \frac{\epsilon \wdr}{4} \left( \frac{\eta_{\ssq,x}}{\wdr + \wq}  + \frac{\eta_{\ssq,x}^*}{\wdr - \wq} \right) \bq^2 \nonumber \\
    && + \frac{\epsilon \wdr}{12} \left( \frac{\eta_{\ssq,x}^*}{\wdr + 3 \wq} + \frac{\eta_{\ssq,x}}{\wdr - 3 \wq}  \right) \bq^{\dag 2} \nonumber \\
    && +\frac{\epsilon \wdr}{2} \left( \frac{\eta_{\ssq,x}}{\wdr - \wq} + \frac{\eta_{\ssq,x}^*}{\wdr + \wq}  \right) \bq^\dag \bq. 
 \nonumber
\end{eqnarray}
We note that setting the drive to zero, amounting to $\eta_{\ssq,x} \to 0$, leads to the expression found in Part I. At nonzero drive, there exists a relaxation-induced dephasing term $\propto \epsilon \eta_{\ssq,x}$, as well as an upward excitation term $\propto \epsilon \eta_{\ssq,x}^2$. Since the former is lower order in $\eta_{\ssq,x}$ compared to the latter, we keep track of dephasing terms in addition to single-photon terms for completeness. In addition to those contributions, two-photon transitions appear in this dissipator at the same frequency $\wq$. 

Finally, let us note that at nonzero temperature there would appear the Hermitian conjugate dissipator, 
\begin{equation}
\hat{C}(-\wq) = \hat{C}(\wq)^\dag. \label{Eq:CMinusWq}
\end{equation} 
Just as drive induced upward transitions in $\hat{C}(\wq)$, drive in the presence of finite temperature will allow for \textit{downward} transitions in $\hat{C}(-\wq)$.

The contributions in~(\ref{Eq:Q1M0EME}) are the dominant single-photon and dephasing contributions. Additionally, there are single-photon dissipators and dephasing dissipators, at frequencies distinct from $\wq$. As above, for generality, we list all the possible dissipators, including those at negative frequency which vanish at zero temperature.
\textalert{Dissipators with a single frequency denominator arise from a single oscillatory term in Table~\ref{Tab:QubitOnlyQuadratureTf}. All terms in the tables have been changed from $\hat{X}$ to $\hat{Y}$ quadrature corrections by multiplying them with a $\pm i$ factor. Therefore the dissipators will remain the same if they come from a single oscillatory term. There are a two dissipators listed below, each of which comes from a sum of two oscillatory terms oscillating at the same frequency. For these, I compared that the relative sign of the two terms of the sum remains the same by comparing the factors these terms have acquired in going from Table~\ref{Tab:QubitOnlyQuadratureTf} in version 20 ($\hat{X}$ coupling to bath) to the same Table in version 21 ($\hat{Y}$ coupling to bath).}
\begin{eqnarray} \label{Eq:AddtlTerms}
  &&+\SFN(\wq + 2 \wdr) \mathcal{D}\left[\epsilon \frac{\wq \eta_{\ssq,x}^2}{8(\wdr+\wq)} \bq \right] \nonumber \\ 
  &&+ \SFN(-\wq - 2 \wdr) \mathcal{D}\left[\epsilon \frac{\wq \eta_{\ssq,x}^{*2}}{8(\wdr+\wq)} \bq^\dag \right] \\
  + &&\SFN(\wq - 2 \wdr) \mathcal{D}\left[\epsilon \frac{\wq \eta_{\ssq,x}^{*2}}{8(\wdr-\wq)} \bq \right] \nonumber \\
  + &&\SFN(-\wq + 2 \wdr) \mathcal{D}\left[\epsilon \frac{\wq \eta_{\ssq,x}^2}{8(\wdr-\wq)} \bq^\dag \right] \nonumber \\
  + &&\SFN(\wdr) \mathcal{D}\left[ \epsilon \frac{\wq^2 \eta_{\ssq,x}}{-\wdr^2 + \wq^2 } \bq^\dag \bq \right] \nonumber \\
  + &&\SFN(-\wdr) \mathcal{D}\left[ \epsilon \frac{\wq^2 \eta_{\ssq,x}^*}{-\wdr^2 + \wq^2 } \bq^\dag \bq \right]
.  \nonumber 
\end{eqnarray}
The terms of Eq.~(\ref{Eq:AddtlTerms}) containing $\bq^\dag$ represent drive-induced upward transitions at zero temperature. Because dissipators are quadratic in their argument, the terms of Eq.~(\ref{Eq:AddtlTerms}) lead to order-$\epsilon^2$ contributions in the EME (in addition to order-$\epsilon$ contributions coming from Eq.~(\ref{Eq:Q1M0EME})).

In addition, there appear two- and three-photon relaxation processes, associated with collapse operators $\bq^2$ and $\bq^3$. For each of these processes, the corresponding dissipator and relaxation rate can be obtained analogously:
\begin{eqnarray}
  \label{Eq:CubicDissQ1M0}  
  + \SFN(\wdr + 2 \wq)\mathcal{D}\left[\frac{\epsilon}{4}\left( \frac{\wq}{\wdr + 3\wq} - \frac{\wq}{\wdr + \wq} \right)\eta_{\ssq,x} \bq^2\right] \nonumber \\
  + \SFN(-\wdr - 2 \wq)\mathcal{D}\left[\frac{\epsilon}{4}\left( \frac{\wq}{\wdr + 3\wq} - \frac{\wq}{\wdr + \wq} \right)\eta_{\ssq,x}^* \left(\bq^\dag\right)^2\right] \nonumber \\
  \SFN(-\wdr + 2 \wq)\mathcal{D}\left[-\frac{\epsilon}{4}\left(\frac{\wq}{\wdr - 3\wq}-\frac{\wq}{\wdr - \wq} \right) \eta_{\ssq,x}^* \bq^2\right] \nonumber \\
  \SFN(\wdr - 2 \wq)\mathcal{D}\left[\frac{\epsilon}{4}\left(-\frac{\wq}{\wdr - 3\wq}+\frac{\wq}{\wdr - \wq} \right) \eta_{\ssq,x} \left(\bq^\dag\right)^2\right] \nonumber \\
 + \SFN(3\wq) \mathcal{D}\left[ -\frac{\epsilon}{48} \bq^3\right] +  \SFN(-3\wq) \mathcal{D}\left[ -\frac{\epsilon}{48} \left(\bq^\dag\right)^3\right]. \nonumber \\
\end{eqnarray}
Note that the dissipators appear in pairs of two terms, the first of which corresponds to a either a one-, two- or three-photon relaxation process or dephasing, while the second corresponds to the Hermitian conjugate process at the negative transition frequency. At zero temperature, one of the two terms vanishes since the spectral function  $\SFN(\omega)\propto \Theta(\omega)$, where $\Theta$ is the Heaviside function (see App. \ref{App:EME}), \textit{i.e.} it is nonzero only for non-negative frequency.  The exception occurs for resonant situations where the drive frequency $\wdr$ and the oscillator frequency $\wq$ are commensurate and the spectral function is evaluated at zero frequency, which we will generally avoid in our numerics. 

\begin{figure}[b!]
  \includegraphics[width=\linewidth]{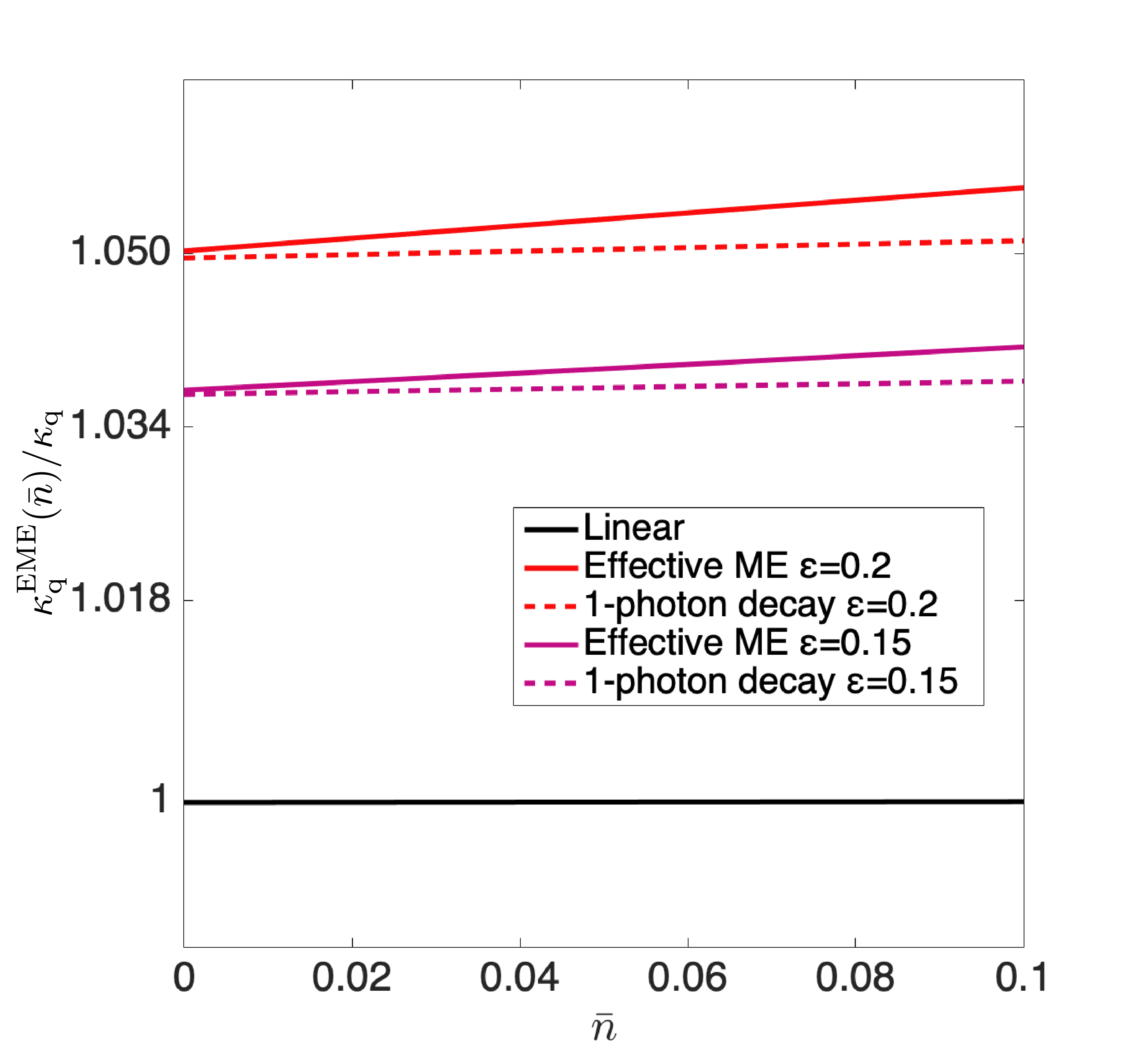}
  \caption{\label{Fig:EMEQ1M0}Results from EME solution for a qubit coupled to an infinite waveguide. The drive strength is represented on the horizontal axis in units of $\bar{n}$, which represents the steady-state mean photon number for the case of a linear oscillator. The vertical axis represents the relaxation rate $\kappa_\ssq^{\text{EME}}$ of the nonlinear oscillator as a function of $\bar{n}$, rescaled by the linear oscillator $\kappa_\ssq$. Solving the EME with $\epsilon = 0$ amounts to simulating a driven-dissipative linear oscillator, and the relaxation rate remains unchanged when the drive is applied (horizontal black curve). The solid red and purple curves show a renormalization of the nonlinear oscillator relaxation rate, for $\epsilon = 0.2$ and $\epsilon  = 0.15$, respectively. The dashed lines represent an estimate of the relaxation rate obtained from single-photon relaxation process in Eq.~(\ref{Eq:CQPTQ1M0}). This is an underestimate to the actual relaxation rate: in fact, additional drive-induced processes in Eqs.~(\ref{Eq:CQPTQ1M0}),~(\ref{Eq:AddtlTerms}) and ~(\ref{Eq:CubicDissQ1M0}), among which we mention single-photon excitation, dephasing, and multiphoton transitions, are responsible for the depletion of the first excited state and, consequently, an enhancement of the relaxation rate.}
\end{figure}

The EME for a qubit coupled to an infinite waveguide, Eq.~(\ref{Eq:Q1M0EME}), is specified by the dissipators in Eqs.~(\ref{Eq:CQPTQ1M0}),~(\ref{Eq:CMinusWq}),~(\ref{Eq:AddtlTerms}),~and(\ref{Eq:CubicDissQ1M0}). Figure~\ref{Fig:EMEQ1M0} shows results obtained from the numerical solution of the EME. In particular, we find that as the drive power is increased, there is an increase in the qubit relaxation rate. The qubit relaxation rate is obtained from the EME-generated time dependence of the qubit photon number, $\langle \bq^\dag \bq\rangle(t)$. The result is obtained by performing a least-squares fit of this time dependence to the photon number of a linear oscillator under the same conditions, with the relaxation rate $\kappa_\ssq^{\text{EME}}$ and oscillator frequency $\wq^\text{EME}$ as fit parameters. Figure~\ref{Fig:EMEQ1M0} shows that the relaxation rate of the nonlinear oscillator increases as a function of drive power and $\epsilon$. In the regime of weak drives, we find that this increase is linear in both $\epsilon$ and $\bar{n}$, with an increase of a few percent when the drive strength corresponds to a mean steady state population of one photon in the driven linear oscillator (see Fig.~\ref{Fig:EMEQ1M0}). The parameters chosen for the simulation are as follows: the $Q-$factor for the linear oscillator is $Q=10^2$, and a drive frequency $\wdr = 1.66 \wq$. In general, the renormalization of the qubit relaxation rate is a rescaling of the linear oscillator value by a factor larger than one which is linearly increasing with $\epsilon$ and $\bar{n}$. Therefore the result quoted here is not sensitive to the order of magnitude of $Q$. We assume that the bath spectrum is flat, such that the spectral function takes the form $\SFN(\om)= 2\kappa$ for all $\om$.

Finally, we can derive state-dependent relaxation rates by rederiving the EME in a Fock-state representation [for the detailed derivation, and comparison to Eq.~(\ref{Eq:Q1M0EME}) see App.~\ref{App:EME}]:
\textalert{The change for $\hat{Y}$ seems to occur in only one place. Now the overall (minding the spectral function taking the corrected energy as argument) rate decreases with drive, and, as opposed to the case studied up until version 20 with the $\hat{X}$ quadrature coupling to the bath, there is no longer a competition with the contribution from the spectral function.}
\begin{eqnarray}
    \dot{\hat{\rho}}_\ssq(t)&=&-i\left[\hat{\mathcal{H}}_{\ssq,\text{eff}},\hat{\rho}_\ssq(t)\right]\nonumber \\&&+\sum_{n\geq 1} 2\kappa_{\ssq,n,\downarrow}\mathcal{D}[|n-1\rangle\langle n|]\hat{\rho}_\ssq(t) \nonumber \\
&&+\sum_{n \geq 1} 2 \kappa_{\ssq,n,\uparrow}\mathcal{D}[|n\rangle\langle n-1|]\hat{\rho}_\ssq(t), \nonumber \\
&& +\sum_{n \geq 0} 2 \kappa_{\ssq,n,\varphi}\mathcal{D}[|n\rangle\langle n|]\hat{\rho}_\ssq(t),
      \label{Eq:Q1M0EMEFock}
\end{eqnarray} 
where there is a state-dependent relaxation rate
\begin{eqnarray}
&&2\kappa_{\ssq,n,\downarrow} = \label{Eq:DownRate}\\ 
&&n \left[ 1 +\frac{\epsilon}{4} \left(n + 2 |\eta_{\ssq,x}|^2 \right) \right]  \SFN\left(\left[ 1 - \frac{\epsilon}{4}(n + 2 |\eta_{\ssq,x}|^2) \right] \wq \right). \nonumber
\end{eqnarray} 
Note that, in deriving this form, we have averaged the effective Hamiltonian~(\ref{Eq:HqEff}) over one period of the drive in order to express the transition frequencies. It is remarkable to note that after this procedure there is a symmetry between the factor that renormalizes the qubit relaxation rate between the states $|n\rangle$ and $|n-1\rangle$, that is $1 +(\epsilon/4) \left(n + 2 |\eta_{\ssq,x}|^2 \right)$, and the factor that renormalizes the corresponding transition frequency, namely $1-(\epsilon/4) \left(n + 2 |\eta_{\ssq,x}|^2 \right)$. For single-photon transitions, rates and transition frequencies have equal and opposite changes relative to the linear theory quantities.

The upward transition rate is $\kappa_{\ssq,n,\uparrow}$, which is quadratic in $\epsilon$. It is analogously derived from Eq.~(\ref{Eq:AddtlTerms}) (we are considering zero temperature and omitting all multi-photon processes for simplicity):
\begin{eqnarray}
 2\kappa_{\ssq,n,\uparrow} = \epsilon^2 n \frac{|\eta_{\ssq,x}|^4}{64}\Bigg[ 
    &&\SFN(-\wq + 2 \wdr)  \left|\frac{\wq}{\wdr-\wq} \right|^2    \\
&&\;\;\;\;\;  + \SFN(\wq) \left|\frac{2\wdr^2}{\wdr^2-\wq^2} \right|^2 \Bigg]. \nonumber
\end{eqnarray}
The state-dependent dephasing rate reads:
\begin{eqnarray}
   2\kappa_{\ssq,n,\varphi} = \frac{\epsilon^2 |\eta_{\ssq,x}|^2 n^2}{(\wdr^2 - \wq^2)^2} \Big[ \wdr^4 \SFN(\wq) + \wq^4  \SFN(\wdr) \Big].
\end{eqnarray}
Note that, in the expressions of the rates above we have dropped order-$\epsilon$ corrections in the argument of the spectral function whenever those corrections would be beyond our level of approximation. 

One important conclusion drawn from this first example is that the physics predicted by the EME depends strongly on the properties of the spectral function in the neighborhood of the bare qubit transition frequency, $\wq$. The relaxation rate~(\ref{Eq:DownRate}) can increase or decrease with respect to $\kappa_\ssq$ depending on the frequency dependence of the bath spectral function. To better understand this, let us perform a Taylor series expansion around $\wq$:
\begin{eqnarray}
  \SFN\left(\left[ 1 - \frac{\epsilon}{4}(n +2 |\eta_{\ssq,x}|^2) \right] \wq \right) = \nonumber \\ \SFN(\wq) - \frac{\partial \SFN}{\partial \om} \left( \wq \right) \frac{\epsilon \wq}{4}(n + 2 |\eta_{\ssq,x}|^2) + O(\epsilon^2).
\end{eqnarray}
Inserting this form back into~(\ref{Eq:DownRate}) and retaining up to order-$\epsilon$ contributions, we arrive at 
\begin{eqnarray}
2  \kappa_{\ssq,n,\downarrow} &=& n \SFN(\wq) \label{Eq:DownRateExp}\\ 
  &&+\frac{n \epsilon}{4}\left[ \SFN(\wq)   
  - \wq  \frac{\partial  \SFN}{\partial \om} \left( \wq \right)  \right] \left(n + 2 |\eta_{\ssq,x}|^2 \right). \nonumber 
\end{eqnarray}
The relaxation rate~(\ref{Eq:DownRateExp}) expanded to lowest order in $\epsilon$ illustrates that the correction due to nonlinearity contains contributions which are both drive-dependent and drive-independent. More importantly, however, the sign of the correction of the qubit relaxation rate depends on the details of the frequency dependence of the spectral function in the neighborhood of $\wq$, as illustrated by the factor in the bracket in the second line of~(\ref{Eq:DownRateExp}).

\textalert{The following text highlighted in red, and any conclusion arising from it, is no longer true with the $\hat{Y}$ coupling. To linear order, there is always a decrease.} \textalert{The rate increases with drive power for every level $n$ if:
\begin{eqnarray}
  \SFN(\wq) > 2 \wq \frac{\partial  \SFN}{\partial \om} \left( \wq \right).
\end{eqnarray}
We have shown that the sign of the drive-induced renormalization of relaxation rates depends, to lowest order in $\epsilon$, on the slope of the spectral function at the oscillator frequency. In particular, an ohmic spectral function $\SFN(\om)=\alpha\om$ violates the inequality above, and predicts an overall decrease of the relaxation rate.}

To summarize, in this section we have built a classification of all the possible system-bath interactions induced by the number-nonconserving terms contained in the Josephson nonlinearity, to linear order in $\epsilon$. Keeping only the most relevant contributions, corresponding to single-photon processes, we have derived the EME for a qubit coupled to an infinite waveguide, Eq.~(\ref{Eq:Q1M0EME}). We have shown that the qubit relaxation rate is dependent on the qubit state and on the drive power, as shown explicitly in the Fock-state representation of the EME, Eq.~(\ref{Eq:Q1M0EMEFock}). 

Finally, by solving the EME numerically, we have extracted the qubit relaxation rate as a function of drive strength, and have shown that this relaxation rate increases linearly as a function of the anharmonicity parameter $\epsilon$ and the drive strength parameter $\bar{n}$.

\subsection{First-order Schrieffer-Wolff perturbation theory}
\label{Ap:1stOrderSW}
In this subsection we explicitly derive the generator $\hat{G}_4(t)$ of the unitary transformation for a driven-dissipative weakly anharmonic qubit from Eq.~(\ref{Eq:ODEG4}). The analogous problem for the qubit coupled to a cavity is an immediate generalization of this, but requires handling a large number of terms, for which we have employed computer algebra.
 
Assume that the Hamiltonian for the driven-dissipative qubit takes the form
\begin{eqnarray}
\hat{\mathcal{H}}_\text{s} &=& \wq \hat{H}_{\ssq} - \frac{\epsilon \wq}{48} \hat{H}_4(t), \nonumber \\
\hat{H}_4(t) &=& \hat{S}_4(t) + \hat{N}_4(t), \nonumber \\
\hat{H}_{\ssq} &=& \half(\bq^\dag \bq + \bq \bq^\dag). \label{Eq:SingleQubitHamiltonian}
\end{eqnarray}
Note that, by means of the unitary transformation introduced in App.~\ref{Ap:Shift}, the time-dependence due to the drive has been placed in the quartic terms. While the expressions for $\hat{S}_4(t)$ and $\hat{N}_4(t)$ will be provided below in Sec.~\ref{SubApp:1stSWPT-DrDissQuPart}, the result of this section holds for generic expressions. Moreover, to model dissipation, one would add to Eq.~(\ref{Eq:SingleQubitHamiltonian}) a system-bath coupling and a bath Hamiltonian.

We aim to find a unitary transformation generated by an antihermitian operator $\hat{G}_4(t)$, such that
\begin{eqnarray}
 && e^{-\epsilon \hat{G}_4(t)} \left[ \mathcal{H}_\text{s}(t) - i \partial_t \right] e^{\epsilon \hat{G}_4(t)} =  \nonumber \\
&& \;\;\;\;\;\;\;\;\;\;\;\;\;\;\;\;\;\; \wq \hat{H}_{\ssq} - \frac{\epsilon \wq}{48} \hat{S}_4(t) - i \partial_t + O(\epsilon^2). \label{ApEq:SingleQubitCondition}
\end{eqnarray}
Explicitly, the unitary transformation will remove the nonsecular contributions $-\frac{\epsilon \wq}{48}\hat{N}_4(t)$ in the system Floquet Hamiltonian $\hat{\mathcal{H}}_\text{s}(t) - i \partial_t$. These contributions will reappear at a higher order $\mathcal{O}(\epsilon^2)$ in the transformed Hamiltonian. The condition to cancel the nonsecular terms determines the generator of the unitary transformation $\hat{G}_4(t)$.

We summarize in this paragraph the main result of the subsection. Condition~(\ref{ApEq:SingleQubitCondition}) becomes equivalent to the operator-valued differential equation in Eq.~(\ref{ApEq:ODEG4}), together with the initial condition in Eq.~(\ref{Eq:G4IC}). The $\hat{G}_4(t)$ that solves these equations is presented at the end of this subsection in Eq.~(\ref{Eq:SolG4}), and is determined solely by $\hat{N}_4(t)$, in its more explicit form in terms of harmonics at the drive frequency, Eqs.~(\ref{Eq:N4Expa}) and~(\ref{Eq:N4Coeffs}). The remainder of this subsection contains the derivation.

We now proceed to finding $\hat{G}_4(t)$ that cancels the nonsecular terms to lowest order, \textit{i.e.} satisfies condition~(\ref{ApEq:SingleQubitCondition}). The transformation of $\hat{\mathcal{H}}_\text{s}$ is
\begin{eqnarray}
  &&e^{-\epsilon \hat{G}_4(t)} \left\{ \wq \hat{H}_{\ssq} - \frac{\epsilon \wq}{48} \left[ \hat{S}_4(t) + \hat{N}_4(t) \right] \right\} e^{ \epsilon \hat{G}_4(t)} =  \nonumber \\
  && \;\;\;\;\;\;\;\;\; \wq \hat{H}_{\ssq} - \frac{\epsilon \wq}{48} \hat{S}_4(t)  \label{Eq:TfHs} \\
  && \;\;\;\;\;\;\;\;\;\;\;\; + \epsilon \wq \left\{ -\frac{\hat{N}_4(t)}{48} + \left[ \hat{H}_{\ssq}, \hat{G}_4(t) \right] \right\} + \mathcal{O}(\epsilon^2). \nonumber
\end{eqnarray}
Under this same unitary, the energy operator transforms according to
\begin{eqnarray}
  e^{-\epsilon \hat{G}_4(t)} (-i\partial_t) e^{\epsilon \hat{G}_4(t)} = - i \partial_t - i \epsilon \dot{\hat{G}}_4(t) + \mathcal{O}(\epsilon^2). \nonumber \\ \; \label{Eq:TfEn}
\end{eqnarray}
Collecting the transformed Hamiltonian~(\ref{Eq:TfHs}) and the transformed energy operator~(\ref{Eq:TfEn}), we find that under the unitary transformation the Floquet Hamiltonian yields:
\begin{eqnarray}
&&  e^{-\epsilon \hat{G}_4(t)} \left[ \hat{\mathcal{H}}_\text{s}(t) - i \partial_t \right] e^{\epsilon \hat{G}_4(t)} = \nonumber \\
&& \;\;\;\;\;\;  \wq \hat{H}_{\ssq}  - \frac{\epsilon \wq}{48} \hat{S}_4(t) - i\partial_t  \label{Eq:SingleQubitTf} \\
&& \;\;\;\;\;\; + \epsilon \wq \left\{ -\frac{\hat{N}_4(t)}{48} + \left[ \hat{H}_{\ssq}, \hat{G}_4(t)\right] - \frac{i}{\wq} \dot{\hat{G}}_4(t) \right\} + \mathcal{O}(\epsilon^2). \nonumber
\end{eqnarray}

Imposing the condition~(\ref{ApEq:SingleQubitCondition}) in the expression for the transformed Floquet Hamiltonian~(\ref{Eq:SingleQubitTf}), we extract a first-order linear differential equation for $\hat{G}_4(t)$:
\begin{equation}
  - i \dot{\hat{G}}_4(t) + \left[ \wq \hat{H}_{\ssq}, \hat{G}_4(t)\right] = \frac{\wq \hat{N}_4(t)}{48} . \label{ApEq:ODEG4}
\end{equation}
The initial condition for $\hat{G}_4(t)$ is set such that $\hat{G}_4(0)$ removes the nonsecular terms at $t=0$, $\hat{N}_4(0)$, \textit{i.e.} we require that
\begin{equation}
  -\frac{\hat{N}_4(0)}{48} + \left[ \hat{H}_{\ssq}, \hat{G}_4(0)\right]  = 0, \label{Eq:G4IC}
\end{equation}
which is an algebraic equation for $\hat{G}_4(0)$. This initial condition ensures that nonsecular terms are removed by the unitary transformation for all $t \geq 0$.

Equation~(\ref{ApEq:ODEG4}) can be solved analytically. $\hat{N}_4(t)$ can be expressed as a sum over normal-ordered ``monomials'', $\left(\bq^\dag \right)^m \bq^n$, with time-dependent coefficients
\begin{eqnarray}
  \hat{N}_4(t) = \sum_{m\neq n} n_{4|m,n}(t) \left(\bq^\dag \right)^m \bq^n, \label{Eq:N4Expa}
\end{eqnarray}
where the sum is over integer $m \neq n \geq 0$. The time dependence of the coefficients of $\hat{N}_4(t)$ reduces to a sum over harmonics of the drive frequency, through the complex-number coefficients $n_{4,m,n}(t)$:
\begin{equation}
  n_{4|m,n}(t) = \sum_{p \in \mathbb{Z}} n_{4|m,n,p} e^{ i p \wdr t}, \label{Eq:N4Coeffs}
\end{equation}
where $\mathbb{Z}$ denotes the set of integers. Let us also expand $\hat{G}_4(t)$ over the same set of normal-ordered (nonsecular) monomials
\begin{eqnarray}
  \hat{G}_4(t) = \sum_{m \neq n} g_{4|m,n}(t) \left(\bq^\dag \right)^m \bq^n. \label{Eq:G4Expa}
\end{eqnarray}

The operator-valued differential equation~(\ref{ApEq:ODEG4}) reduces to determining the complex-valued functions of time $g_{4|m,n}(t)$. Using the identity
\begin{equation}
  \left[\hat{H}_{\ssq}, \left(\bq^\dag \right)^m \bq^n \right] = (m-n) \left(\bq^\dag \right)^m \bq^n,
\end{equation}
we may use the expanded forms for $\hat{N}_4(t)$, Eq.~(\ref{Eq:N4Expa}), and for $\hat{G}_4(t)$, Eq.~(\ref{Eq:G4Expa}), into the operator differential equation~(\ref{ApEq:ODEG4}). Collecting the coefficients term-by-term, we arrive at
\begin{eqnarray}
  (m-n) \wq g_{4|m,n}(t) - i \dot{g}_{4|m,n}(t) = \frac{\wq}{48} n_{4|m,n}(t). \nonumber \\\; \label{Eq:ODEG4Coeff}
\end{eqnarray}
The generator $\hat{G}_4(t)$ is constructed from its coefficients $g_{4|m,n}(t)$, which obey the differential equation of an oscillator of natural frequency $(m-n)\wq$ forced by the time-dependent term $(\wq/48) n_{4|m,n}(t)$.

Firstly, the particular solution to the ordinary differential equation~(\ref{Eq:ODEG4Coeff}) is constructed by expanding again over the harmonics of the drive frequency:
\begin{equation}
  g^{\text{(p)}}_{4|m,n}(t) = \sum_{p \in \mathbb{Z}} g^{\text{(p)}}_{4|m,n,p} e^{ i p \wdr t}.
\end{equation}
This is an Ansatz that solves~(\ref{Eq:ODEG4Coeff}) provided that
\begin{equation}
  g^{\text{(p)}}_{4|m,n,p} = \frac{\wq}{48} \frac{n_{4|m,n,p}}{(m-n)\wq + p \wdr},
\end{equation}
for all integer $m \neq n \geq 0$ and integer $p$. 

Secondly, the solution to the homogeneous part of~(\ref{Eq:ODEG4Coeff}),
\begin{equation}
  (m-n) \wq g_{4|m,n}(t) - i \dot{g}_{4|m,n}(t) = 0. \nonumber \\\; \label{Eq:ODEG4CoeffH}  
\end{equation}
is
\begin{equation}
  g_{4|m,n}^{\text{(h)}}(t) = g_{4|m,n}^{\text{(h)}}(0) e^{- i (m -n ) \wq t}.
\end{equation}
The general solution to~(\ref{Eq:ODEG4Coeff}) is then a linear combination of the particular and homogeneous solutions, 
\begin{equation}
  g_{4|m,n}(t) = g^{\text{(p)}}_{4|m,n}(t) + g^{\text{(h)}}_{4|m,n}(t),
\end{equation}
which has to obey the initial condition that derives from~(\ref{Eq:G4IC}), namely:
\begin{eqnarray}
  g_{4|m,n}(0) = \frac{n_{4|m,n}(0)}{48(m-n)}.
\end{eqnarray}
This fixes the amplitude of the homogeneous solutions to
\begin{eqnarray}
  g_{4|m,n}^{\text{(h)}}(0) = \frac{n_{4|m,n}(0)}{48(m-n)} - \sum_{p \in \mathbb{Z}} \frac{\wq}{48}\frac{n_{4|m,n,p}}{(m-n)\wq + p \wdr}
  \nonumber \\
  = \frac{\sum_{p \in \mathbb{Z}} n_{4|m,n,p}}{48(m-n)} - \sum_{p \in \mathbb{Z}} \frac{\wq}{48}\frac{n_{4|m,n,p}}{(m-n)\wq + p \wdr} \nonumber \\   = \frac{\wq}{48} \sum_{p \in \mathbb{Z}} n_{4|m,n,p} \left[  \frac{1}{(m-n)\wq} - \frac{1}{(m-n)\wq + p\wdr}  \right]. \nonumber \\ \;
\end{eqnarray}

Then the solution to Eq.~(\ref{ApEq:ODEG4}) obeying the initial condition~(\ref{Eq:G4IC}) can be written succinctly:
\begin{eqnarray}
  \hat{G}_4(t) &=& \sum_{m \neq n} g_{4|m,n}(t) \left( \bq^\dag \right)^m \bq^n,  \\
  g_{4|m,n}(t) &=& \frac{\wq}{48} \sum_{p \in \mathbb{Z}} \Bigg\{ \frac{n_{4|m,n,p} e^{-i(m-n)\wq t}}{(m-n) \wq} \nonumber \\
  &&+ \frac{n_{4|m,n,p} \left[e^{i p \wdr t} - e^{-i(m-n)\wq t}\right]}{(m-n)\wq + p \wdr} \Bigg\}, \nonumber \\
  && \text{ for }m \neq n. \label{Eq:SolG4} \nonumber
\end{eqnarray}
The coefficients $n_{4|m,n,p}$ are known and determine $\hat{G}_4(t)$. We turn to their explicit expressions in the next subsection, App.~\ref{SubApp:1stSWPT-DrDissQuPart}.

\subsection{Qubit coupled to infinite waveguide: Explicit solution}
\label{SubApp:1stSWPT-DrDissQuPart}
In this section we provide the explicit solution for $\hat{G}_4(t)$ for the driven weakly anharmonic oscillator Duffing oscillator
\begin{eqnarray}
  \hat{\mathcal{H}} = \hat{\mathcal{H}}_\text{s} + \hat{\mathcal{H}}_\text{d}(t),
\end{eqnarray}
with $\hat{\mathcal{H}}_\text{s}$ as defined in Eq.~(\ref{Eq:SingleQubitHamiltonian}), and
\begin{equation}
\hat{\mathcal{H}}_\text{d}\equiv \ed \left(-i\bq+i\bq^{\dag}\right)\sin(\wdr t).
\label{Eq:1stSWPT-DrDissQu-Def of Hd}
\end{equation}
The application of the displacement transformation leads to 
\begin{eqnarray}
\hat{\mathcal{H}}_{\ssq}(t)\to\hat{\mathcal{H}}_{\ssq}(t) &=&\wq \hat{H}_{\ssq} -\frac{\epsilon}{48}\wq \hat{H}_4(t),
\label{Eq:1stSWPT-DrDissQu-Ef Hq}
\end{eqnarray}
where
\begin{eqnarray}
\hat{H}_4(t) &\equiv& \left[\bq+\bq^{\dag}+ x_\ssq(t)\right]^4, \nonumber \\
\hat{H}_4(t) &=& \hat{S}_4(t) +\hat{N}_4(t), \nonumber \\
x_\ssq(t) &=& \eta_{\ssq,x} e^{-i \wdr t} +  \eta_{\ssq,x}^* e^{i \wdr t},
\end{eqnarray}
and $\eta_{\ssq,x}$ is
\begin{equation}
  \eta_{\ssq,x} = \frac{\ed (\wdr + i \kq)}{\wq^2 - (\wdr + i \kq)^2}
\end{equation}
Explicit forms of $\hat{S}_4(t)$ and $\hat{N}_4(t)$ are provided in the next paragraphs. 

\begin{widetext}
The number-conserving terms of the Hamiltonian, $\hat{S}_4(t)$, are
\begin{eqnarray}
\hat{S}_4(t) &=&  6 \bq^{\dag}\bq^{\dag} \bq \bq 
               + 12 \left[4 \eta_{\ssq,x}^2 \cos^2\left( \wdr t \right)+1\right] \bq^{\dag} \bq  
               + 16 \eta_{\ssq,x}^4 \cos^4\left(\wdr t \right)+24 \eta_\text{d}^2 \cos ^2\left(\wdr t \right)+3 \nonumber \\
             &=& +  6 \bq^\dag \bq \bq^\dag \bq + 12 \left[ 4 \eta_{\ssq,x}^2 \cos^2( \wdr t ) + \half \right] \bq^\dag \bq + \left[ 4 \eta_{\ssq,x}^2 \cos^2(\wdr t )  + 3 \right]^2 - 6  \nonumber \\
             &=& + 6 \bq^\dag \bq \bq^\dag \bq + 12 \left[ x_\ssq^2(t) + \half \right] \bq^\dag \bq + \left[ x_\ssq^2(t) + 3 \right]^2 - 6  \nonumber \\
             &=& +  6 \left(\hat{H}_{\ssq}^2 - \hat{H}_{\ssq} + \frac{1}{4}\right) + 12 \left[ x_\ssq^2(t) + \half \right] \left( \hat{H}_{\ssq} - \half \right) + \left[ x_\ssq^2(t) + 3 \right]^2 - 6  \nonumber \\
             &=& + 6 \hat{H}_{\ssq}^2 + \left[ -6 + 12 x_\ssq^2(t) + 6 \right] \hat{H}_{\ssq} + \left[ x_\ssq^2(t) + 3 \right]^2 - 6 - 6 \left[ x_\ssq^2(t) + \half \right] + \frac{6}{4} \nonumber \\
             &=& +  6 \hat{H}_{\ssq}^2 +  12 x_\ssq^2(t) \hat{H}_{\ssq} + x_\ssq^4(t) + \frac{6}{4} .
\label{Eq:Lind&PT-AdDrDissQu-Def of S4}
\end{eqnarray}
$\hat{S}_4(t)$ is diagonal in the number basis of the qubit Hilbert space. We add it to the quadratic Hamiltonian as a correction. The effective Hamiltonian can be expressed compactly
\begin{eqnarray}
\hat{\mathcal{H}}_{\ssq,\text{eff}}&=& \wq\left\{1-\frac{\epsilon}{8}\left[\hat{H}_{\ssq}+ 2 x_\ssq^2(t)\right]\right\}\hat{H}_{\ssq}  +\mathcal{O}(\epsilon^2) \nonumber \\
                               &=& \wq \hat{H}_{\ssq} - \frac{\epsilon \wq}{16} \left\{ \hat{H}_{\ssq}+ 2 x_\ssq^2(t) , \hat{H}_{\ssq} \right\} + \mathcal{O}(\epsilon^2); \nonumber \\
\;
\label{Eq:1stSWPT-DrDissQu-H_SW}
\end{eqnarray} 
in the expression above, we have dropped the contribution from the time-dependent c-number term of $\hat{S}_4(t)$. There are $O(\epsilon^2)$ secular terms, which arise from higher-order terms in the expansion of the unitary transformation $e^{\epsilon \hat{G}_4}$. However, here we confine ourselves to the analysis of the linear terms only.

The eigenstates and eigenenergies of $\HO_{\text{s},\text{eff}}$ can be readily obtained. The instantaneous eigenstates are exactly the eigenstates of $\wq \hat{H}_{\ssq}$, $| n \rangle$ for any $n \geq 0$ integer. The instantaneous eigenenergies corresponding to these kets are
\begin{equation}
  E_n(t) =  \wq\left\{1-\frac{\epsilon}{8}\left[\left( n + \half \right)+ 2 x_\ssq^2(t)\right]\right\}\left( n + \half \right) + O(\epsilon^2).
\end{equation}

The nonsecular part of the quartic nonlinearity is
\begin{eqnarray}
\hat{N}_4&=&\bq^4+\left(\bq^{\dag}\right)^4 + 4\left[\bq^{\dag}\bq^3+\left(\bq^{\dag}\right)^3\bq\right]  + 4 X_{\eta}(t) \left[ \bq^3 + \left(\bq^\dag\right)^3 \right]  + 12 X_{\eta}(t)\left[\bq^{\dag}\bq^2+\left(\bq^{\dag}\right)^2\bq \right] \nonumber \\
&& +6\left[ x_\ssq^2(t) + 1 \right]\left[\bq^2+\left(\bq^{\dag}\right)^2\right] + 4 x_\ssq(t) \left[ x_\ssq^2(t) + 3 \right]\left( \bq + \bq^\dag \right). 
\label{Eq:1stSWPT-DrDissQu-N4}
\end{eqnarray}
This allows us to read off the coefficients $n_{4|m,n}(t)$ of Eq.~(\ref{Eq:N4Expa}) and $n_{4|m,n,p}$ of Eq.~(\ref{Eq:N4Coeffs}).

We may now obtain the generator of the Schrieffer-Wolff unitary transformation, to linear order in $\epsilon$, in the form:
\begin{eqnarray}
  \hat{G}_4(t) &=& \left[ g_{4|4,0}(t) \left(\bq^{\dag}\right)^4 + g_{4|0,4}(t) \bq^4 \right] + \left[ g_{4|3,1}(t) \left(\bq^{\dag}\right)^3\bq + g_{4|1,3}(t)\bq^{\dag}\bq^3\right] \nonumber \\ &&+ \left[ g_{4|3,0}(t) \left(\bq^{\dag}\right)^3 + g_{4|0,3}(t)\bq^3 \right] + \left[ g_{4|2,1}(t) \left(\bq^{\dag}\right)^2\bq + g_{4|1,2}(t) \bq^{\dag}\bq^2 \right] \nonumber \\
  && +\left[g_{4|2,0}(t)\left(\bq^{\dag}\right)^2 + g_{4|0,2}(t)\bq^2\right] + \left[g_{4|1,0}(t) \bq^{\dag} + g_{4|0,1}(t)\bq\right].
\label{Eq:1stSWPT-DrDissQu-G4}
\end{eqnarray}
Due to the antihermiticity of $\hat{G}_4$, $\hat{G}_4(t) = -\hat{G}_4^\dag(t)$, the following conditions must hold:
\begin{eqnarray}
  g_{4|0,4}(t) = - g_{4|4,0}^*(t) &,&\; g_{4|1,3}(t) = - g_{4|3,1}^*(t),\nonumber \\
  g_{4|0,3}(t) = - g_{4|3,0}^*(t) &,&\; g_{4|1,2}(t) = - g_{4|2,1}^*(t),\nonumber \\
  g_{4|0,2}(t) = - g_{4|2,0}^*(t) &,&\; g_{4|0,1}(t) = - g_{4|1,0}^*(t).\nonumber \\
\end{eqnarray}
The expression of $\hat{G}_4(t)$ is determined by the time-dependent complex coefficients:
\begin{eqnarray}
  g_{4|4,0}(t) &=& \frac{1}{192}, \nonumber \\ 
  g_{4|3,1}(t) &=& \frac{1}{24}, \nonumber \\
  g_{4|3,0}(t) &=& +\frac{1}{18} \eta_{\ssq,x} e^{-3i t \wq} 
  +\frac{\wq}{48} \left[\frac{4\eta_{\ssq,x}  \left(e^{-i t \wdr}-e^{-3 i t  \wq}\right)}{-\wdr+3 \wq}
  +\frac{4\eta_{\ssq,x} \left( e^{i t \wdr}-e^{-3i t \wq} \right)}{\wdr+3 \wq}\right], \nonumber \\
  g_{4|2,1}(t) &=&    +\half \eta_{\ssq,x}   e^{-i t \wq} 
   +\frac{\wq}{48}\left[ \frac{12 \eta_{\ssq,x} \left(e^{-i t \wdr}-e^{-i t \wq}\right)}{-\wdr+\wq}
   +\frac{12\eta_{\ssq,x} \left(e^{i t \wdr}-e^{-i t \wq}\right)}{\wdr+\wq}\right], \nonumber \\
  g_{4|2,0}(t) &=& +\frac{1}{16} \left\{2\eta_{\ssq,x}^2  e^{-2 i t \wq} + 2 \eta_{\ssq,x}^2 +1\right\}
   +\frac{\wq}{48}\left[ \frac{6\eta_{\ssq,x}^2  \left(e^{-2 i t \wdr}-e^{-2 i t \wq}\right)}{-2\wdr+2\wq}
   +\frac{6 \eta_{\ssq,x}^2  \left(e^{2 i t \wdr}-e^{-2 i t \wq}\right)}{2\wdr+2\wq} \right], \nonumber \\
  g_{4|1,0}(t) &=& +\frac{1}{12}  \left[6 \left(1 + 2 \eta_{\ssq,x}^2 \right) \eta_{\ssq,x}  +  2\eta_{\ssq,x}^3 \right] e^{-i t \wq}\nonumber \\
    &&+ \frac{\wq}{48}\left[ \frac{12\eta_{\ssq,x} \left(\eta_{\ssq,x}^2+1\right) \left(e^{-i t \wdr} - e^{-i t \wq}\right)}{-\wdr+\wq}
    +\frac{12 \eta_{\ssq,x} \left(\eta_{\ssq,x}^2+1\right) \left(e^{i t \wdr}-e^{-i t \wq}\right)}{\wdr+\wq} \right] \nonumber \\
    &&+ \frac{\wq}{48} \left[ \frac{4\eta_{\ssq,x}^3 \left(e^{-3 i t \wdr} -e^{-i t \wq} \right)}{-3 \wdr + \wq}
    +\frac{4\eta_{\ssq,x}^3 \left(e^{3 i t \wdr} - e^{-i t \wq} \right)}{3 \wdr+\wq} \right].
\label{Eq:1stSWPT-DrDissQu-G4}
\end{eqnarray}
Table~\ref{Tab:QubitOnlyQuadratureTf} of the main text summarizes the terms entering the transformation of the qubit quadrature, according to the equation
\begin{eqnarray}
  e^{-\epsilon \hat{G}_4}\Xq e^{\epsilon \hat{G}_4} = \Xq + \epsilon \left[  \Xq, \hat{G}_4\right]. \label{Eq:QubitOnlyQuadratureTf}
\end{eqnarray}
\end{widetext}

\section{General derivation of the Effective Master Equation}
\label{App:EME}
This section contains a general derivation of the EME, with particular focus on the obtention of the corrected system-bath couplings, and the application of the Born-Markov and secular approximations. We start with the von Neumann equation for the density matrix defined over the tensor product Hilbert space of the system coupled to the environment:
\begin{equation}
  \dot{\hat{\rho}}_{\text{s}\otimes\text{b}}(t) = - i [ \HO(t), \hat{\rho}_{\text{s}\otimes\text{b}}(t)],
\end{equation}
where
\begin{equation}
  \HO(t) = \HO_\text{s}(t) + \HO_\text{b} + \HO_{\text{sb}}
\end{equation}
is the full system Hamiltonian. We are considering the situation where a displacement transformation has already been applied, so the drive term is absorbed in $\HO_\text{s}(t)$.

EMEs are obtained by performing a unitary transformation onto the system Hamiltonian, then obtaining the corrections that this unitary transformation induces onto the system-bath couplings. We therefore consider the transformed density matrix:
\begin{equation}
  \hat{\rho}_{\text{s}\otimes\text{b}}'(t) = e^{-\hat{G}(t)} \hat{\rho}_{\text{s}\otimes\text{b}}(t) e^{\hat{G}(t)},
\end{equation}
where $\hat{G}(t)$ is the antihermitian generator of the unitary transformation. The density matrix $\hat{\rho}_{\text{s}\otimes\text{b}}'(t)$ obeys the following von Neumann equation:
\begin{equation}
  \dot{\hat{\rho}}_{\text{s}\otimes\text{b}}'(t) = - i \left[ e^{-\hat{G}(t)} [ \HO(t) - i \partial_t ] e^{\hat{G}(t)},  \hat{\rho}_{\text{s}\otimes\text{b}}'(t)   \right].
\end{equation}
The generator $\hat{G}(t)$ is needed to eliminate the number non-conserving terms in the system Hamiltonian $\HO_\text{s}(t)$, amounting to condition~(\ref{Eq:FloquetUGHUG}) in the main text. The resulting effective von Neumann equation is
\begin{equation}
  \dot{\hat{\rho}}_{\text{s}\otimes\text{b}}'(t) = - i \left[ \HO_{\text{s},\text{eff}}(t) + \HO_{\text{b}} + \HO_{\text{sb},\text{eff}}(t),  \hat{\rho}_{\text{s}\otimes\text{b}}'(t)   \right],
\end{equation}
where
\begin{equation}
  \HO_{\text{sb},\text{eff}}(t) = \HO_{\text{sb}} + \epsilon \left[ \HO_{\text{sb}}, \hat{G}_4(t)\right] + O(\epsilon^2).
\end{equation}
After transforming to the interaction picture with respect to the effective system and bath Hamiltonian $\HO_{\text{s},\text{eff}}(t) + \HO_\text{b}$, and expanding the differential equation to second order in the pertubative system-bath coupling $\HO_{\text{sb},\text{eff}}(t)$, in what amounts to the Born-Markov approximation, and performing a trace over the bath degrees of freedom, we arrive at the Redfield equation \cite{redfield_1965, breuer_petruccione_2002} for the reduced density matrix   $\hat{\rho}_{\text{s},I}'(t) = \text{Tr}_\text{b}\left\{ \hat{\rho}_{\text{s}\otimes\text{b},I}'(t) \right\}$
:
\begin{eqnarray}
&&\dot{\hat{\rho}}_{\text{s},I}'(t)  \\ 
&&= \int_0^\infty ds \, \text{Tr}_\text{b}\Big\{ \HO_{\text{sb},\text{eff},I}(t-s) \hat{\rho}_{\text{s},I}'(t)\otimes \hat{\rho}_\text{b}(0) \HO_{\text{sb},\text{eff},I}(t) \nonumber \\ 
&&- \HO_{\text{sb},\text{eff},I}(t) \HO_{\text{sb},\text{eff},I}(t-s) \hat{\rho}_{\text{s},I}'(t)\otimes \hat{\rho}_\text{b}(0) \Big\} + \text{H.c.} \label{Eq:Redfield}  
\end{eqnarray}
where we have defined interaction picture operators with the subscript ``$I$'' as follows
\begin{eqnarray}
  \hat{O}_I(t) = e^{i \int_0^t dt' \left[\HO_{\text{s},\text{eff}}(t') + \HO_\text{b} \right]} \hat{O} e^{-i \int_0^t dt' \left[\HO_{\text{s},\text{eff}}(t') + \HO_\text{b} \right]}. \nonumber \\
\end{eqnarray}
Note that the time-ordering operator is absent from this expression since the effective Hamiltonians at different times, being diagonal in the number basis, commute. We have also assumed that at the initial time the bath is in thermal equilibrium at temperature $T$:
\begin{equation}
  \hat{\rho}_\text{b}(t) = \frac{1}{Z_\text{b}(k_\text{B}T)} e^{- \hat{\mathcal{H}}_\text{b}/ k_\text{B} T},
\end{equation}
where the normalization constant is the inverse partition function, such that $\text{Tr}_\text{b}\hat{\rho}_\text{b} = 1$.

Next, we need to formulate the Redfield equation for our particular problem. The first step is to simplify the expressions for the interaction-picture operators based on our expansion of the system Hamiltonian. To this end, we express separately in the effective Hamiltonian the quadratic, time-independent terms of $\hat{\mathcal{S}}_4$, and time-dependent terms of $\hat{\mathcal{S}}_4$, respectively, as follows:
\begin{eqnarray}
  \HO_{\text{s},\text{eff}}(t) = \HO_{2} - \epsilon \hat{\mathcal{S}}_{4,i} - \epsilon \hat{\mathcal{S}}_{4,d}(t),
\end{eqnarray}
where the subscripts $i$ and $d$ refer to time-independent and time-dependent contributions in $\hat{\mathcal{S}}_4$, respectively. We note that the time dependent terms are quadratic in $\eta_x$ and only contain harmonics at integer multiples of the drive frequency $e^{i n \wdr t}$, with $n \neq 0$, as can be easily verified by inspecting $\hat{\mathcal{S}}_4(t)$ in Eq.~(\ref{Eq:S4Text}). 

For concreteness, we provide here the expressions for $\hat{\mathcal{S}}_{4,d}(t)$ and  $\hat{\mathcal{S}}_{4,i}(t)$ as obtained from Eq.~(\ref{Eq:S4Text}):
\begin{eqnarray}
\hat{\mathcal{S}}_{4,d}(t) &=&  \lambda_{\ssq,d}(t) \nq  + \lambda_{\ssc,d}(t) \nc  , \nonumber \\ 
 \hat{\mathcal{S}}_{4,i}(t) &=&  \lambda_{\ssq,i} \nq  + \lambda_{\ssc,i} \nc + \chi_\ssqc \nq \nc + \alpha_\ssq \hat{\bar{n}}^2_\ssq  + \alpha_\ssc \hat{\bar{n}}^2_\ssc \label{Eq:S4Ap}, \nonumber \\ \;
\end{eqnarray}
with
\begin{eqnarray}
  \lambda_{\ssq,d}(t)&=& \frac{\bwq}{2}  \uqq^2 \eta_x^2 \cos(2\wdr t), \nonumber \\
  \lambda_{\ssc,d}(t)&=& \frac{ \bwq}{2} \uqc^2 \eta_x^2 \cos(2\wdr t),  \nonumber \\
  \lambda_{\ssq,i}&=& \frac{ \bwq}{8}  \uqq^2 \left[4\eta_x^2 +\uqq^2 + 2\uqc^2\right], \nonumber \\
  \lambda_{\ssc,i}&=& \frac{ \bwq}{8} \uqc^2 \left[4\eta_x^2  +\uqc^2+2\uqq^2\right],  \nonumber \\
  \chi_\ssqc &=& \frac{ \bwq}{4} \uqc^2 \uqq^2, \;\; \alpha_\ssq = \frac{ \bwq}{8} \uqq^4, \; \; \alpha_\ssc = \frac{ \bwq}{8} \uqc^4.
\end{eqnarray}

We may now factorize the unitary that takes us to the interaction picture as
\begin{eqnarray}
   e^{-i \int_0^t dt' \HO_{\text{s},\text{eff}}(t')} =    e^{-i \left(\HO_{2} - \epsilon \hat{\mathcal{S}}_{4,i}\right)t}    e^{-i \int_0^t dt' \epsilon\hat{\mathcal{S}}_{4,d}(t')} \nonumber \\
   =  \left[ 1 - \epsilon \hat{I}_4(t) \right] e^{-i \left(\HO_{2} - \epsilon \hat{\mathcal{S}}_{4,i}\right)t} + O(\epsilon^2), \nonumber \\ \label{Eq:TaylorExpansionI4}
\end{eqnarray}
where $\hat{I}_4(t)= i \int_0^t dt' \hat{\mathcal{S}}_{4,d}(t')$, and it is linear in the number operators:
\begin{eqnarray}
  \hat{I}_4(t) = i \frac{\bwq}{4\wdr} \eta_x^2  \sin(2\wdr t) \left( \uqq^2 \nq + \uqc^2 \nc \right).
\end{eqnarray} 
Notice that in the second row of Eq.~(\ref{Eq:TaylorExpansionI4}) we have perfomed a Taylor expansion of the second exponential. Its exponent is $\hat{I}_4(t)$ and it is a bounded function of time. On the other hand, the exponent that was not expanded is linear in $t$ and would lead to unbounded expressions in a Taylor expansion. However, this term can be handled fully and will yield the order-$\epsilon$ corrections from the Kerr theory to the eigenfrequencies of the system Hamiltonian. 
 
The system-bath coupling in the interaction picture becomes (recall that the bare cavity mode couples to the bath via the charge quadrature, $\bYc$):
\begin{eqnarray}
\HO_{\text{sb},\text{eff},I}(t) = \hat{Y}_{\text{b},I}(t)\otimes \nonumber \\ e^{i \left(\HO_{2} - \epsilon \hat{\mathcal{S}}_{4,i}\right)t} \left\{ \bYc + \epsilon \left[ \bYc,  \hat{I}_4 (t) + \hat{G}_4(t) \right] \right\}  e^{-i \left(\HO_{2} - \epsilon \hat{\mathcal{S}}_{4,i}\right)t},  \nonumber \\ \label{Eq:Ap:XcTf}
\end{eqnarray}
with 
\begin{equation}
  \hat{Y}_{\text{b},I} (t) = \sum_k g_k (-i B_k e^{- i \om_k t} + \text{H.c.}),
\end{equation}
in accordance with our notations in the main text. 

Our handling of the quartic terms of the Josephson nonlinearity has led to a renormalized system quadrature coupling to the bath. Let us define this quadrature as:
\begin{eqnarray}
  \hat{C}(t) = \bYc + \epsilon \left[ \bYc,  \hat{I}_4 (t) + \hat{G}_4(t) \right]. \label{Eq:DefC}
\end{eqnarray} 
This is a Hermitian operator:
\begin{equation}
  \hat{C}(t) = \hat{C}^\dag(t). \label{Eq:CtHerm}
\end{equation}

We further express the interaction-picture system-bath coupling of Eq.~(\ref{Eq:Ap:XcTf}) as a sum of harmonics upon decomposing the operator $\HO_{2} - \epsilon \hat{\mathcal{S}}_{4,i}$ into its (Fock-space) eigenstates. This amounts to 
\begin{eqnarray}
&& e^{i \left(\HO_{2} - \epsilon \hat{\mathcal{S}}_{4,i}\right)t} \hat{C}(t) e^{-i \left(\HO_{2} - \epsilon \hat{\mathcal{S}}_{4,i}\right)t} \nonumber \\ 
&&= \sum_{\bar{n}_\ssq\bar{n}_\ssc\bar{m}_\ssq\bar{m}_\ssc} e^{i(\om_{\bar{n}_\ssq\bar{n}_\ssc} - \om_{\bar{m}_\ssq\bar{m}_\ssc})t}  | \bar{n}_\ssq\bar{n}_\ssc \rangle \langle \bar{n}_\ssq\bar{n}_\ssc | \hat{C}(t) | \bar{m}_\ssq\bar{m}_\ssc \rangle \langle \bar{m}_\ssq\bar{m}_\ssc |  \nonumber \\
&&\equiv  \sum_j \hat{C}(\om_j) e^{i \om_j t}  \label{Eq:Ap:IPXC}
\end{eqnarray}
In the last row of Eq.~(\ref{Eq:Ap:IPXC}), we have introduced a sum over a set of distinct frequencies $\{ \om_j \}$, which are obtained from the transition frequencies $\om_{\bar{n}_\ssq,\bar{n}_\ssc} - \om_{\bar{m}_\ssq,\bar{m}_\ssc}$, plus linear combinations of $\wq$, $\wc$ and $\wdr$ arising from the phase factors present in $\hat{C}(t)$. These phase factors can be traced back to the commutator with the time-dependent generator $\hat{I}_4(t) + \hat{G}_4(t)$ in Eq.~(\ref{Eq:DefC}). The energies $\om_{\bar{n}_\ssq,\bar{n}_\ssc}$ are the eigenvalues of the time-independent part of the effective Hamiltonian, to wit:
\begin{eqnarray}
  \om_{\bar{n}_\ssq,\bar{n}_\ssc} &=& \bar{n}_\ssq \wq + \bar{n}_\ssc \wc  \label{Eq:NuNqNc}\\
  && - \epsilon( \lambda_{\ssq,i} {\bar{n}}_\ssq  + \lambda_{\ssc,i} {\bar{n}}_\ssc + \chi_\ssqc {\bar{n}}_\ssq {\bar{n}}_\ssc + \alpha_\ssq {\bar{n}}^2_\ssq  + \alpha_\ssc {\bar{n}}^2_\ssc). \nonumber
\end{eqnarray}
More explicitly, the term of the sum introduced in the last row of Eq.~(\ref{Eq:Ap:IPXC}) indicates that $\hat{C}(\om_j)$ is the coefficient of the harmonic $e^{i \om_j t}$ of $\hat{C}(t)$. Note that the Hermiticity of $\hat{C}(t)$ together with the expansion over Fock states~(\ref{Eq:Ap:IPXC}) imply that
\begin{equation}
  \hat{C}^\dag(\om_j) = \hat{C}(-\om_j),
\end{equation}
and that the set of frequencies $\{\om_j | j =0,1,2,... \}$ must in fact be symmetric: \textit{i.e.} for every frequency present in the set, the negative frequency is also present in the set.

These expressions enter the system-bath Hamiltonian in the interaction picture. From Eq.~(\ref{Eq:Ap:XcTf}), we may derive
\begin{equation}
  \hat{\mathcal{H}}_{\text{sb},\text{eff},I}(t) = \hat{Y}_{\text{b},I}(t) \otimes \sum_{j} \hat{C}(\om_j) e^{i \om_j t}.
\end{equation} 

The Redfield equation~(\ref{Eq:Redfield}) becomes
\begin{eqnarray}
  \dot{\hat{\rho}}_{\text{s},I}' = \int_0^\infty ds \, && \text{Tr}_\text{b} \left\{ \hat{Y}_\text{b}(t) \hat{Y}_\text{b}(t-s) \rho_\text{b}(0) \right\}  \nonumber \\
  && \sum_{jj'} e^{i\om_j (t-s)} e^{i \om_{j'} t} \nonumber \\
  && \times \Big[ \hat{C}(\om_j) \rho_{\text{s},I}' (t) \hat{C}(\om_{j'}) - \hat{C}(\om_{j'}) \hat{C}(\om_j) \hat{\rho}_{\text{s},I}'(t) \Big] 
  \nonumber \\
  + \text{H.c.} \label{Eq:Redfield2}
\end{eqnarray}
To bring this in a more compact form, let us define the unilateral power spectral density:
\begin{eqnarray}
  \sfn (\om) = \int_0^\infty d\tau \, e^{-i\om \tau} \text{Tr} \left[(1/Z_\text{b}) e^{- \hat{\mathcal{H}}_\text{b}/ k_\text{B} T} \hat{Y}_{\text{b}}(\tau) \, \hat{Y}_{\text{b}}(0) \right]. \nonumber \\ 
\end{eqnarray}
Assuming that the bath density matrix at the initial time $t=0$ corresponds to thermal equilibrium,
\begin{eqnarray}
  \hat{\rho}_\text{b}(0) = (1/Z_\text{b}) e^{- \hat{\mathcal{H}}_\text{b}/ k_\text{B} T},
\end{eqnarray}
we simplify~(\ref{Eq:Redfield2}) as follows:
\begin{eqnarray}
  \dot{\hat{\rho}}_{\text{s},I}' &=& \sum_{jj'} e^{i(\om_j +\om_{j'})t} \sfn(\om_j) \\
  && \times \Big[ \hat{C}(\om_j) \rho_{\text{s},I}' (t) \hat{C}(\om_{j'}) - \hat{C}(\om_{j'}) \hat{C}(\om_j) \hat{\rho}_{\text{s},I}'(t) \Big] 
  \nonumber \\
  + \text{H.c.} \label{Eq:Redfield20}
\end{eqnarray}

Now let us denote 
\begin{equation}
  \sfn(\om) \equiv \frac{\SFN(\om) + i \PFN(\om)}{2},
\end{equation}
where both $\SFN(\om)$ and $P(\om)$ are real functions of frequency. Recall that we introduced in the main text the bilateral power spectral density:
\begin{eqnarray}
  \SFN (\om) = \int_{-\infty}^\infty d\tau \, e^{-i\om \tau} \text{Tr} \left[(1/Z_\text{b}) e^{- \hat{\mathcal{H}}_\text{b}/ k_\text{B} T} \hat{Y}_{\text{b}}(\tau) \, \hat{Y}_{\text{b}}(0) \right]. \nonumber \\ 
\end{eqnarray}
One can check the following identity:
\begin{equation}
  \sfn(\om) + \sfn(\om)^* = \SFN(\om),
\end{equation}
which follows from the Hermiticity of $X_\text{b}(\tau)$ at all $\tau$. We can reexpress the master equation as follows:
\begin{eqnarray}
  \dot{\hat{\rho}}_{\text{s},I}' &=& + \sum_{jj'} e^{i(\om_j +\om_{j'})t} \frac{i}{2}\PFN(\om_j) \nonumber \\
  && \times \Big[ \hat{C}(\om_j) \rho_{\text{s},I}' (t) \hat{C}(\om_{j'}) - \hat{C}(\om_{j'}) \hat{C}(\om_j) \hat{\rho}_{\text{s},I}'(t) \Big] 
  + \text{H.c.} \nonumber \\
  && + \sum_{jj'} e^{i(\om_j +\om_{j'})t} \half\SFN(\om_j) \nonumber \\
  && \times \Big[ \hat{C}(\om_j) \rho_{\text{s},I}' (t) \hat{C}(\om_{j'}) - \hat{C}(\om_{j'}) \hat{C}(\om_j) \hat{\rho}_{\text{s},I}'(t) \Big] 
  + \text{H.c.} \nonumber \\
  &=& + \sum_{jj'} e^{i(\om_j +\om_{j'})t} \frac{i}{2}\PFN(\om_j) \nonumber \\
  && \times \Big[  - \hat{C}(\om_{j'}) \hat{C}(\om_j) \hat{\rho}_{\text{s},I}'(t)  +   \hat{\rho}_{\text{s},I}'(t) \hat{C}(\om_j)^\dag \hat{C}(\om_{j'})^\dag   \Big] 
  \nonumber \\
  && + \sum_{jj'} e^{i(\om_j +\om_{j'})t} \half\SFN(\om_j) \nonumber \\
  && \times \Big[ \hat{C}(\om_j) \rho_{\text{s},I}' (t) \hat{C}(\om_{j'}) - \hat{C}(\om_{j'}) \hat{C}(\om_j) \hat{\rho}_{\text{s},I}'(t) \Big] 
  + \text{H.c.} \nonumber \\
\label{Eq:Redfield21}  
\end{eqnarray}
Above, in simplifying the terms containing the imaginary part of the spectral function, $\PFN(t)$, we have removed the terms of the form $\hat{C}(\om_j) \rho_{\text{s},I}' (t) \hat{C}(\om_{j'})$ by adding the Hermitian conjugate, then renaming the summation indices $\om_j \leftrightarrow - \om_{j'}$, by virtue of the fact that the set $\{ \om_j \}$ is symmetric. 

It is common to perform a rotating wave approximation at the level of Eq.~(\ref{Eq:Redfield21}) which assumes that the smallest nonzero $|\om_j + \om_{j'}|$ is large compared to the typical relaxation rate of the system, and thus the contribution from terms oscillating at this frequency averages to zero. Retaining only those terms in Eq.~(\ref{Eq:Redfield21}) which have no oscillatory phase factor,
\begin{eqnarray}
  \dot{\hat{\rho}}_{\text{s},I}' &=&  
  + \sum_{j} \frac{i}{2}\PFN(\om_j) \nonumber \\
  && \times \Big[  - \hat{C}(\om_{j})^\dag \hat{C}(\om_j) \hat{\rho}_{\text{s},I}'(t)  +   \hat{\rho}_{\text{s},I}'(t) \hat{C}(\om_j)^\dag \hat{C}(\om_{j})   \Big] \nonumber \\
&&+\sum_{j} \half \SFN(\om_j) \nonumber \\
&&  \times \Big[ \hat{C}(\om_j) \hat{\rho}_{\text{s},I}'(t) \hat{C}^\dag(\om_{j}) - \hat{C}^\dag(\om_{j})  \hat{C}(\om_j) \hat{\rho}_{\text{s},I}' (t) \Big] \nonumber \\
  &&+ \text{H.c.} \nonumber \\
  &=&  - i \left[ \HO_{\text{Lamb}}, \hat{\rho}_{\text{s},I}'(t) \right] + \sum_{j} \SFN(\om_j) \mathcal{D}[\hat{C}(\om_j)] \hat{\rho}_{\text{s},I}'(t). \nonumber \\
 \label{Eq:Redfield5RWADi}  
\end{eqnarray}
We have denoted the Lamb shift Hamiltonian as
\begin{eqnarray}
  \HO_{\text{Lamb}} = \sum_j \half\PFN(\om_j) \hat{C}(\om_j)^\dag \hat{C}(\om_j).
\end{eqnarray}

Undoing the interaction-picture unitary transformation, one arrives at a master equation in Lindblad form: 
\begin{eqnarray}
  \dot{\hat{\rho}}_{\text{s}}'(t) = -i \left[ \HO_{\text{s},\text{eff}}(t) + \HO_\text{Lamb}, \hat{\rho}_\text{s}'(t) \right] \nonumber \\ \;\;\;\;\;+ \sum_{j} 2 \kappa(\om_j) \mathcal{D}\left[ \hat{C}(\om_j) \right] \hat{\rho}_\text{s}'(t), \label{Eq:Ap:EME}
\end{eqnarray} 
where $2\kappa(\om) = \SFN(\om)$. Equation~(\ref{Eq:Ap:EME}) is an EME to order $\epsilon$, within a Born-Markov approximation, as well as the rotating wave approximation introduced in the paragraph of Eq.~(\ref{Eq:Redfield5RWADi}). In the main text, we have neglected the Lamb shift contribution as we assume weak system-bath couplings $g_k$. These contributions can be reinstated should one require a calculation of bath-induced corrections on the system transition frequencies.  Equation~(\ref{Eq:Ap:EME}) yields the state-resolved EME of Eq.~(\ref{Eq:Q1M0EMEFock}) in the main text.

To obtain the more compact form of the EME, Eq.~(\ref{Eq:Q1M0EME}), one further approximation is in order. To make this approximation, we return to the definition of $\hat{C}(\om_j)$, implicit from Eq.~(\ref{Eq:Ap:IPXC}). The approximation that we make is that $\om_{\bar{n}_\ssq \bar{n}_\ssc} = \bar{n}_\ssq \wq + \bar{n}_\ssc \wc$, \textit{i.e.} we neglect the $\epsilon$-order corrections to the eigenenergies of the effective Hamiltonian, the second line of Eq.~(\ref{Eq:NuNqNc}). Inspection of~(\ref{Eq:NuNqNc}) shows that these corrections become large with increasing $\bar{n}_{\ssq,\ssc}$. However, this is not a problem, because it is the \textit{transition frequencies} that enter $\hat{C}(\om_j)$. Transition frequencies will suffer minor corrections from order-$\epsilon$ terms since $\hat{C}(t)$ connects at most Fock states whose photon numbers differ by three. To us, this approximation means that frequencies $\om_j$ are linear combinations of $\wq$ and $\wc$, consisting of any transition frequency of the linear system, plus integer multiples of $\wdr$:
\begin{eqnarray}
  \{ \om_j | j \text{ non-negative integer}\} = \nonumber \\ \{ \bar{d}_\ssq \wq + \bar{d}_\ssc \wc + d_\text{d} \wdr | \bar{d}_\ssq, \bar{d}_\ssc, d_\text{d} \text{ integers}\}.
\end{eqnarray}
The essential point here is that, in truncating the transition frequencies to zeroth order in epsilon, infinitely many transitions will occur at the same transition frequency, and consequently the transition operators can be summed over to obtain a single collapse or jump operator at the respective frequency. Hence, the dissipators of the EME will contain polynomial expressions in the creation and annihilation operators. In this way one obtains an EME of the form of Eq.~(\ref{Eq:Q1M0EME}) in the main text.

We should point out that the rotating wave approximation is not justified if the frequencies corresponding to distinct transitions can come close enough to each other (\textit{i.e.} $|\om_j + \om_{j'}|$ is small in the expressions above without $\om_j= -\om_{j'}$). This is the situation of the nonlinear transmon spectrum \cite{koch_et_al_2007}, whose high energy states form a continuum. We therefore quote the non-RWA EME, obtained by undoing the interaction picture on~(\ref{Eq:Redfield21}), as our more general result:
\begin{eqnarray}
  \dot{\hat{\rho}}_{\text{s}}' = -i \left[ \HO_{\text{s},\text{eff}}(t), \hat{\rho}_\text{s}'(t) \right] +\sum_{jj'} \sfn(\om_j) e^{i(\om_j+\om_{j'})t}  \nonumber \\
  \times \Big[ \hat{C}(\om_j) \hat{\rho}_\text{s}'(t) \hat{C}(\om_{j'}) - \hat{C}(\om_{j'})  \hat{C}(\om_j) \hat{\rho}_\text{s}' (t) \Big] \nonumber \\
  + \text{H.c.}  \nonumber \\
 \label{Eq:Redfield6}  
\end{eqnarray}
Without the RWA, the Lamb shift contribution is no longer in  commutator form as in Eq.~(\ref{Eq:Redfield5RWADi}). While more exact, this form is rather unwieldy. We have not used it for our numerics, primarily since we expect that for weak drives the density matrix will have nonzero weights primarily on states with low photon number, where transition frequencies are well separated.

We finish this section by providing expansions over bath modes of the bath spectral function. The first step is to evaluate the trace over the bath modes, which amounts to calculating
\begin{eqnarray}
  \text{Tr}_\text{b} \left\{ \hat{Y}_\text{b}(t) \hat{Y}_\text{b}(t-s) \rho_\text{b}(0) \right\} = \nonumber\\ 
  \sum_{kl} g_k g_l \text{Tr}_\text{b} \Big\{ \left(-i \Bop e^{-i \om_k t} + \text{H.c.} \right) \nonumber \\
   \left( -i \hat{B}_l e^{- i \om_l (t-s)} + \text{H.c.}\right) \Big\}  \nonumber \\
   = \sum_{kl} g_k g_l \delta_{lk} (1+n_k) e^{- i \om_k t} e^{i \om_l (t-s)} \nonumber \\
   + \sum_{kl} g_k g_l \delta_{kl} n_k e^{i \om_k t} e^{-i \om_l (t-s)} \nonumber \\
   = \sum_k g_k^2 (1+n_k) e^{-i\om_k s} + \sum_k g_k^2 n_k e^{i \om_k s}, \label{Eq:BathTrace}
\end{eqnarray}
where we have assumed that $\text{Tr}_\text{b} \left\{ \Bop \hat{B}_l^\dag \right\} \equiv \delta_{lk} (1+n_k)$, and $\text{Tr}_\text{b} \left\{ \Bop^\dag \hat{B}_l \right\} \equiv \delta_{kl} n_k$; $n_k =\left[e^{\om_k/(k_B T)} - 1\right]^{-1}$ is the value of the Bose-Einstein distribution at energy $\om_k$ and temperature $T$. We have assumed that anomalous bath correlation functions, \text{i.e.} $\text{Tr}_\text{b} \left\{ \Bop \hat{B}_l \right\}$, are all vanishing. 

Expanding the bath quadrature $\hat{Y}_\text{b}$ over the modes $\Bop$, we need the Sokhotski-Plemelj formula, $\int_0^\infty ds e^{- i(\om - \om_0)s} = \pi \delta(\om - \om_0)  - i \mathcal{P} \frac{1}{\om - \om_0}$, where $\mathcal{P}$ denotes the Cauchy principal value, and we arrive at:
\begin{eqnarray}
  \SFN (\om) &=& \sum_k 2\pi g_k^2 \left[(1+n_k) \delta(\om+\om_k) + n_k \delta(\om-\om_k) \right] \nonumber \\
  \PFN (\om) &=& \mathcal{P}\sum_k 2 g_k^2 \left[(1+n_k) \frac{-1}{\om+\om_k} + n_k \frac{-1}{\om-\om_k} \right]. \nonumber \\
  \;
\end{eqnarray}

\section{Transformations of qubit and cavity quadratures}
\label{Ap:Tables}
In Sec.~\ref{Sec:PT}, we needed to calculate the effect of the unitary transformation onto the system quadratures coupling to baths. Due to space constraints, those results are listed in the following three tables of this appendix: Tables~\ref{Tab:CoeffsQuadQ}, \ref{Tab:CoeffsQuadC}, and \ref{Tab:CoeffsQuadQC} for qubit-only, cavity-only and mixed processes arising from $\left[\Yq,\hat{G}_4(t)\right]$. There are three more sets of terms arising from the commutator with the charge operator of the cavity normal mode $\left[\Yc,\hat{G}_4(t)\right]$. Those can be obtained by changing indeces in the expressions as outlined in the main text, and will not be reproduced here.

\begin{table}
  \begin{tabular}{|c|c|}
    \hline  
    Operator                 & Coefficient \\
    \hline\hline
    $\bq$   &   
    $\frac{i\bwq \uqq^2 \left[\left(\eta_x^2+\eta_x^{*2}\right) e^{2 i t \wq} + \uqq^2 + \uqc^2 + 2 |\eta_x|^2 \right]}{8\wq}$\\ &
    $-\frac{i\bwq \eta_x^2 \uqq^2 \left(-e^{-2it\wdr}+e^{2 i t \wq}\right)}{8(\wdr+\wq)}$ \\ &
    $-\frac{i\bwq \eta_x^{*2} \uqq^2  \left(e^{2 i t \wq}-e^{2 i t \wdr}\right)}{8(\wq-\wdr)} $ \\ 
    \hline
    $\bq^\dag$   &  c.c. \\
    \hline
    $\bq^2$         &  
    $\frac{-i\bwq \eta_x^* \uqq^3  \left(e^{i t \wq}-e^{i t \wdr}\right)}{4(\wq-\wdr)}$ \\ & 
    $-\frac{i\bwq \eta_x^* \uqq^3 \left(e^{i t 3 \wq}-e^{i t \wdr}\right)}{4(3 \wq-\wdr)}$ \\ &
    $+\frac{-i\bwq \eta_x  \uqq^3  \left(-e^{-i t \wdr}+e^{i t \wq}\right)}{4(\wdr+\wq)}$ \\ &
    $-\frac{i\bwq \eta_x  \uqq^3  \left(-e^{-i t \wdr}+e^{i t 3 \wq}\right)}{4(\wdr+3 \wq)}$ \\ &
    $+\frac{i\bwq \left(\eta_x + \eta_x^* \right) \uqq^3 \left(3e^{i t \wq}+e^{3 i t \wq}\right)}{12\wq}$ \\
    \hline
    $\left(\bq^\dag\right)^2$ & c.c. \\
    \hline
    $\bq^\dag \bq$ & 
    $-\frac{i\bwq \uqq^3 \left(\eta_x^* e^{it\wdr} -\eta_x^*  e^{-i t \wq} +\eta_x  e^{it\wq} -  \eta_x  e^{-i t \wdr}\right)}{2(\wdr+\wq)}$ \\ &
    $-\frac{i\bwq \uqq^3  \left(-\eta_x^*  e^{i t \wdr}+ \eta_x^*  e^{i t \wq} - \eta_x  e^{-i t \wq} + \eta_x  e^{-i t\wdr}\right)}{2(\wq-\wdr)}$ \\ &
    $+\frac{i\bwq \left(\eta_x+\eta_x^*\right) \uqq^3  \left(-e^{-i t \wq}+e^{i t \wq}\right)}{2\wq}$ \\
    \hline
    $\bq^3$ & $\frac{i \uqq^4 \bwq}{16 \wq}$ \\ 
    \hline
    $\left(\bq^\dag\right)^3$ & c.c. \\
    \hline
    $\bq^\dag \bq \bq$ & $\frac{i \uqq^4 \bwq}{8\wq}$ \\
    \hline        
    $\bq^\dag \bq^\dag \bq$ & c.c. \\
    \hline
  \end{tabular}
  \caption{\label{Tab:CoeffsQuadQ}Operators acting on the qubit only, resulting from the expansion of $\left[\Yq,\hat{G}_4(t)\right]$. The left column contains the operator monomial, the right column contains its coefficient in the expansion.}
\end{table}

\begin{table}
  \begin{tabular}{|c|c|}
    \hline
    Operator & Coefficient \\
    \hline\hline
    $\bc$   &   
    $-\frac{i\bwq \eta_x^2 \uqq \uqc \left(-e^{-it2\wdr}+e^{i t \left(-\wq+\wc\right)}\right)}{4(-\wq+2 \wdr+\wc)}$ \\ 
& $-\frac{i\bwq \eta_x^{*2} \uqq \uqc \left(e^{i t \left(-\wq+\wc\right)}-e^{2 i t \wdr}\right)}{4(-\wq-2 \wdr+\wc)}$ \\ 
& $+\frac{i\bwq \uqq \uqc \left[2\Real{\eta_x^2} e^{i t \left(-\wq +\wc\right)} + \uqq^2 + \uqc^2 +2 |\eta_x|^2 \right]}{4(\wc-\wq)} $ \\ 
& $+\frac{i\bwq \uqq \uqc \left[2 \Real{\eta_x^2} e^{i t \left(\wq+\wc\right)}+  \uqq^2 + \uqc^2 + 2 |\eta_x|^2 \right]}{4(\wq+\wc)}$ \\ 
& $-\frac{i\bwq \eta_x^2 \uqq \uqc \left(-e^{-it2\wdr}+e^{i t \left(\wq+\wc\right)}\right)}{4(\wq+2 \wdr+\wc)}$ \\ 
& $-\frac{i\bwq \eta_x^{*2} \uqq \uqc \left(e^{i t \left(\wq+\wc\right)}-e^{2 i t \wdr}\right)}{4(\wq-2 \wdr+\wc)}$ \\
    \hline
    $\bc^\dag$   &  c.c. \\
    \hline
    $\bc^2$         &  
    $-\frac{i\bwq \eta_x^* \uqq \uqc^2 \left(e^{i t \left(-\wq+2 \wc\right)}-e^{i t \wdr}\right)}{4(-\wq-\wdr+2 \wc)}$ \\ &
    $-\frac{i\bwq \eta_x^* \uqq \uqc^2 \left(e^{i t \left(\wq+2 \wc\right)}-e^{i t \wdr}\right)}{4(\wq-\wdr+2 \wc)}$ \\ &
    $+\frac{i\bwq \eta_x  \uqq \uqc^2  \left(-e^{it(2\wc-\wq)}+e^{-i t \wdr}\right)
   } {4(-\wq+\wdr+2 \wc)} $ \\ &
     $-\frac{i\bwq \eta_x  \uqq \uqc^2 \left(-e^{-it\wdr}+e^{i t \left(\wq+2 \wc\right)}\right)}{4(\wq+\wdr+2 \wc)}$ \\ &
    $+\frac{i\bwq \Real{\eta_x} \uqq \uqc^2 e^{i t \left(2 \wc-\wq\right)}}{2(2 \wc-\wq)}$ \\ &
    $+\frac{i\bwq \Real{\eta_x}  \uqq \uqc^2 e^{i t \left(\wq+2 \wc\right)}}{2(\wq+2 \wc)}$ \\
    \hline
    $\left(\bc^\dag\right)^2$ & c.c. \\
    \hline
    $\bc^\dag \bc$ & 
    $-\frac{i\bwq  \uqq \uqc^2  \left(+ \eta_x^* e^{it\wdr} - \eta_x^* e^{-i t \wq} + \eta_x e^{i t \wq} - \eta_x e^{-i t \wdr}\right)}{8(\wq+\wdr)}$ \\
    & $-\frac{i\bwq \uqq \uqc^2 \left(- \eta_x^* e^{i t \wdr} + \eta_x^*  e^{it\wq} - \eta_x  e^{-i t\wq} + \eta_x e^{-i t\wdr} \right)}{8(\wq-\wdr)}$ \\ & $+\frac{i\bwq \Real{\eta_x} \uqq \uqc^2 e^{-i t \wq} \left(-1+e^{2 i t \wq}\right)}{4\wq}$ \\
    \hline
    $\bc^3$ & $\frac{i\bwq \uqq \uqc^3}{12(\wq+3 \wc)}+\frac{i\bwq \uqq \uqc^3}{12(3 \wc-\wq)}$ \\ 
    \hline
    $\left(\bc^\dag\right)^3$ & c.c. \\
    \hline
    $\bc^\dag \bc \bc$ & $\frac{i\bwq \uqq \uqc^3}{4(\wq+\wc)}+\frac{i\bwq \uqq \uqc^3}{4(\wc-\wq)}$ \\
    \hline        
    $\bc^\dag \bc^\dag \bc$ & c.c. \\
    \hline    
  \end{tabular}
  \caption{\label{Tab:CoeffsQuadC}Operators acting on the cavity only, resulting from the expansion of $\left[\Yq,\hat{G}_4(t)\right]$.}
\end{table}

\begin{table}
  \begin{tabular}{|c|c|}
    \hline
    Operator &  Coefficient \\
    \hline\hline
    $\bq \bc$   &  
    $-\frac{i\bwq \eta_x^* \uqc \uqq^2 \left(e^{i t \left(\wc+2 \wq\right)}-e^{i t \wdr}\right)}{2(\wc-\wdr+2 \wq)}$ \\ &
    $-\frac{i\bwq \eta_x  \uqc \uqq^2 \left(-e^{-it\wdr}+e^{i t \left(\wc+2 \wq\right)}\right)}{2(\wc+\wdr+2 \wq)}$ \\ &
    $-\frac{i\bwq \eta_x^* \uqc \uqq^2 \left(e^{i t \wc}-e^{i t \wdr}\right)}{2(\wc-\wdr)}$ \\
   &  $ -\frac{i\bwq \eta_x  \uqc \uqq^2 \left(-e^{-it\wdr}+e^{i t \wc}\right)}{2(\wc+\wdr)}$ \\ &
    $+\frac{i\bwq \Real{\eta_x} \uqc \uqq^2 e^{i t \left(\wc+2 \wq\right)}}{\wc+2 \wq}$ \\ &
    $+\frac{i\bwq \Real{\eta_x} \uqc \uqq^2 e^{i t \wc}}{\wc}$
    \\
    \hline
    $\bq \bc^\dag$   & 
    $\frac{i\bwq \eta_x^* \uqc \uqq^2 \left(e^{i t \left(-\wc+2 \wq\right)}-e^{i t \wdr}\right)}{2(\wc+\wdr-2 \wq)}$ \\ &
    $+\frac{i\bwq \eta_x  \uqc \uqq^2 \left(-e^{-it\wdr}+e^{i t \left(-\wc+2 \wq\right)}\right)}{2(\wc-\wdr-2 \wq)}$ \\ &
    $-\frac{i\bwq \eta_x^* \uqc \uqq^2 \left(-e^{-i t \wc}+e^{i t \wdr}\right)}{2(\wc+\wdr)}$ \\
     & $-\frac{i\bwq \eta_x  \uqc \uqq^2  \left(-e^{-it\wc}+e^{-i t \wdr}\right)}{2(\wc-\wdr)}$
    \\ & $-\frac{i\bwq \Real{\eta_x} \uqc \uqq^2 e^{-t \left(i \wc-2 i \wq\right)}}{\wc-2 \wq}$ \\ &
    $-\frac{i\bwq \Real{\eta_x} \uqc \uqq^2 e^{-i t \wc}}{\wc}$
    \\
    \hline
    $\bq \bc^\dag \bc$   &  $\frac{i\bwq \uqc^2 \uqq^2}{4\wq}$ \\
    \hline
    $\bc \bq^\dag \bq$   &  $-\frac{i\bwq \uqc \uqq^3}{2(\wq-\wc)}+\frac{i\bwq \uqc \uqq^3}{2(\wc+\wq)}$ \\ 
    \hline
    $\bq \bc^2$   &  $\frac{i\bwq \uqc^2 \uqq^2}{8(\wc+\wq)}+\frac{i\bwq \uqc^2 \uqq^2}{8\wc}$ \\
    \hline
    $\bq \left(\bc^\dag\right)^2$   & $-\frac{i\bwq \uqc^2 \uqq^2}{8\wc}-\frac{i\bwq \uqc^2 \uqq^2}{8(\wc-\wq)}$\\    \hline
    $\bq^2 \bc$   &  $\frac{i\bwq \uqc \uqq^3}{4(\wc+3 \wq)}+\frac{i\bwq \uqc \uqq^3}{4(\wc+\wq)}$\\    
    \hline
    $\bq^2 \bc^\dag$   &  $-\frac{i\bwq \uqc \uqq^3}{4(\wc-\wq)}-\frac{i\bwq \uqc \uqq^3}{4(\wc-3 \wq)}$\\    
    \hline
  \end{tabular}
  \caption{\label{Tab:CoeffsQuadQC}Correlated operators acting on both cavity and qubit, resulting from the expansion of $\left[\Yq,\hat{G}_4(t)\right]$. } 
\end{table}

\bibliographystyle{apsrev4-1}
\bibliography{SWPTBib}
\end{document}